\documentclass[pdfa,final]{aaltoseries}

\makeatletter
\@ifpackageloaded{inputenc}{%
  \inputencoding{utf8}}{%
  \usepackage[utf8]{inputenc}}
\hypersetup{hidelinks} 
\makeatother
\usepackage[finnish,english]{babel}   
\usepackage{setspace}  
\usepackage{afterpage} 
\usepackage{float}
\usepackage{svg}
\usepackage{esvect}
\usepackage{amsmath}
\usepackage{caption}
\usepackage{subcaption}

\microtypesetup{letterspace=25}       

\author{Pedro David Carneiro Neto}
\title{Deep Learning Based Analysis of Prostate Cancer from MP-MRI}

\begin{document}


\thispagestyle{empty}
\setcounter{page}{0}  

\newgeometry{left=23.2mm,right=23.2mm,top=13.5mm,bottom=18mm}

\pagecolor{aaltoBlack}\afterpage{\nopagecolor}
{\color{white}  

{\parindent0pt 
{\fontsize{11.9pt}{11.9pt}\bfseries\sffamily\lsstyle Master's Programme in Computer, Communication and Information Sciences}

\vspace{13.1mm}

\begin{spacing}{3.1}
{\fontsize{35}{35}\selectfont Deep Learning Based \\Analysis of Prostate Cancer from MP-MRI}
\end{spacing}

\vspace{2.2mm}

\begin{spacing}{1.24}
{\fontsize{14pt}{14pt}\bfseries\sffamily\lsstyle Classification, detection and segmentation of prostate cancer lesions from MP-MRI with deep learning based methods}
\end{spacing}

\vspace{7.2mm}

\rule{\textwidth}{1.25pt}

\vspace{8.5mm}

{\fontsize{13.9pt}{13.9pt}\bfseries\sffamily\lsstyle Pedro David Carneiro Neto}

\vfill

\begin{picture}(0,0)
\put(356,-7.8){\bfseries\sffamily\footnotesize\lsstyle MASTER'S}
\put(356,-17.4){\bfseries\sffamily\footnotesize\lsstyle THESIS}
\put(346,-26.5){\rule{.75pt}{25pt}}
\end{picture}

\AaltoLogoSmall{.66}{?}{white}

} 
} 


\newpage

\pagenumbering{roman}

\newgeometry{left=80.7mm,right=25mm,top=12.9mm,bottom=21mm}

\thispagestyle{empty}

{\parindent0pt 
\begin{spacing}{1.1}
\hspace{-39.1mm}{\fontsize{10.5pt}{10.5pt}\sffamily\lsstyle Aalto University}

\hspace{-39.1mm}{\fontsize{10.5pt}{10.5pt}\bfseries\sffamily\lsstyle MASTER'S THESIS} {\sffamily\lsstyle 2020}
\end{spacing}

\vspace{12.7mm}

\begin{spacing}{1.63}
{\fontsize{17.8pt}{17.8pt}\selectfont Deep Learning Based Analysis of Prostate Cancer from MP-MRI}
\end{spacing}

\vspace{10.5mm}

\begin{spacing}{1.2}
{\fontsize{13pt}{13pt}\selectfont Classification, detection and segmentation of prostate cancer lesions from MP-MRI with deep learning based methods}
\end{spacing}

\vspace{10.6mm}

{\fontsize{13.9pt}{13.9pt}\bfseries\sffamily\lsstyle Pedro David Carneiro Neto}

\vfill

{\fontsize{10.3pt}{10.3pt}\sffamily\lsstyle\raggedright
\begin{spacing}{1.06}

Thesis submitted in partial fulfillment of the requirements for the
degree of Master of Science in Technology.

Otaniemi, 25 May 2020

\begin{tabbing}
Supervisor:\hspace{6mm} \= professor Juho Kannala \\
Advisor 1: \> doctor Saad Ullah Akram\\
Advisor 2: \> doctor Harri Merisaari \\
Advisor 3: \> medical doctor Ivan Jambor\\
\end{tabbing}
\vspace{-4mm}
\end{spacing}
} 

\vspace{11.5mm}

\begin{spacing}{.9}
{\bfseries\sffamily\lsstyle Aalto University \\
School of Science \\
Master's Programme in Computer, \\Communication and Information Sciences}
\end{spacing}
} 


\newpage
\phantomsection
\addcontentsline{toc}{chapter}{Abstract}

\newgeometry{left=41.8mm,right=25mm,top=14.33mm,bottom=27mm}

\begin{spacing}{.88}

{\parindent0pt 
\AaltoLogoSmall{.625}{''}{aaltoBlack}

{\fontsize{13.9pt}{13.9pt}\selectfont
\vspace{-8.9mm}\hfill{\bfseries\sffamily\lsstyle Abstract}}

{\fontsize{9.48pt}{9.48pt}\selectfont
\vspace{.9mm}\hfill{\bfseries\sffamily\lsstyle Aalto University, P.O. Box 11000, FI-00076 Aalto~~\textcolor{aaltoGray}{www.aalto.fi}}}

\vspace{7.8mm}{\fontsize{10.5pt}{10.5pt}\bfseries\sffamily\lsstyle Author}\\
{\small Pedro David Carneiro Neto}

\vspace{-2.4mm}\rule{\textwidth}{.75pt}

{\fontsize{10.5pt}{10.5pt}\bfseries\sffamily\lsstyle Title}\\
\parbox[t]{\textwidth}{\raggedright\small Deep Learning Based Analysis of Prostate Cancer from MP-MRI}

\vspace{.5mm}\rule{\textwidth}{.75pt}

{\fontsize{10.5pt}{10.5pt}\bfseries\sffamily\lsstyle School}~~{\small School of Science}

\vspace{-2.4mm}\rule{\textwidth}{.75pt}

{\fontsize{10.5pt}{10.5pt}\bfseries\sffamily\lsstyle Master's programme}~~{\small Computer, Communication and Information Sciences}

\vspace{-2.4mm}\rule{\textwidth}{.75pt}

{\fontsize{10.5pt}{10.5pt}\bfseries\sffamily\lsstyle Major}~~{\small Computer Science - Big Data and Large-Scale Computing}\hfill{\fontsize{10.5pt}{10.5pt}\bfseries\sffamily\lsstyle Code}~~{\small SCI3042}

\vspace{-2.4mm}\rule{\textwidth}{.75pt}

{\fontsize{10.5pt}{10.5pt}\bfseries\sffamily\lsstyle Supervisor}~~{\small professor Juho Kannala}

\vspace{-2.4mm}\rule{\textwidth}{.75pt}

{\fontsize{10.5pt}{10.5pt}\bfseries\sffamily\lsstyle Advisor}~~{\small doctor Saad Ullah Akram; doctor Harri Merisaari; medical doctor Ivan Jambor}

\vspace{-2.4mm}\rule{\textwidth}{.75pt}

{\fontsize{10.5pt}{10.5pt}\bfseries\sffamily\lsstyle Level}~~{\small Master's thesis}\hfill{\fontsize{10.5pt}{10.5pt}\bfseries\sffamily\lsstyle Date}~~{\small 25 May 2020}\hfill{\fontsize{10.5pt}{10.5pt}\bfseries\sffamily\lsstyle Pages}~~{\small 74+4}\hfill{\fontsize{10.5pt}{10.5pt}\bfseries\sffamily\lsstyle Language}~~{\small English}

\vspace{-2.4mm}\rule{\textwidth}{.75pt}

\vspace{6mm}

} 
\end{spacing}
\begin{spacing}{1.05}

\noindent{\fontsize{10.5pt}{10.5pt}\bfseries\sffamily\lsstyle Abstract}
\vspace{.8mm}

{\small
The diagnosis of prostate cancer faces a problem with overdiagnosis that leads to damaging side effects due to unnecessary treatment. Research has shown that the use of multi-parametric magnetic resonance images to conduct biopsies can drastically help to mitigate the overdiagnosis, thus reducing the side effects on healthy patients. This study aims to investigate the use of deep learning techniques to explore computer-aid diagnosis based on MRI as input. Several diagnosis problems ranging from classification of lesions as being clinically significant or not to the detection and segmentation of lesions are addressed with deep learning based approaches. 

This thesis tackled two main problems regarding the diagnosis of prostate cancer. Firstly, a deep neural network architecture, XmasNet, was used to conduct two large experiments on the classification of lesions. Secondly, detection and segmentation experiments were conducted, first on the prostate and afterward on the prostate cancer lesions. The former experiments explored the lesions through a two-dimensional space, while the latter explored models to work with three-dimensional inputs. For this task, the 3D models explored were the 3D U-Net and a pretrained 3D ResNet-18. A rigorous analysis of all these problems was conducted with a total of two networks, two cropping techniques, two resampling techniques, two crop sizes, five input sizes and data augmentations experimented for lesion classification. While for segmentation two models, two input sizes and data augmentations were experimented. Moreover the experiments were conducted for both sequences independently, and within the lesion classification problem, the experiments were also conducted for both sequences simultaneously. However, while the binary classification of the clinical significance of lesions and the detection and segmentation of the prostate already achieve the desired results (0.870 AUC and 0.915 dice score respectively), the classification of the PIRADS score and the segmentation of lesions still have a large margin to improve (0.664 accuracy and 0.690 dice score respectively). It was also studied how some flaws in the dataset can be addressed to improve the results of all these problems. Further research on the problem is still needed, but nonetheless, this thesis established sufficient ground for future work to be conducted. 
}

\vfill

\end{spacing}
\begin{spacing}{.88}
{\parindent0pt 

\makebox[19mm][l]{\fontsize{10.5pt}{10.5pt}\bfseries\sffamily\lsstyle Keywords}\parbox[t]{123.6mm}{\raggedright\small deep learning, computer vision , prostate cancer, segmentation, classification, computer-aided diagnosis systems}

\vspace{.5mm}\rule{\textwidth}{.75pt}

\vspace{-2.4mm}\rule{\textwidth}{.75pt}

} 
\end{spacing}

\selectlanguage{english}  
\restoregeometry  


\newpage

\tableofcontents


\newpage

\pagenumbering{arabic}

\chapter{Introduction}

Prostate cancer is one of the biggest causes of cancer deaths in men around the world. It has also been consistently ranked as one of the most frequent cancers to be diagnosed in men by a myriad of organizations worldwide, with 84 different countries having it as the most diagnosed cancer in men \cite{Stewart2014}. Predictions for 2020 indicate that the number of new cases in the United States will be around 191,930, and 33,330 deaths caused by the disease \cite{newcases}. The disease's risk varies with some factors like the race, medical records of the family, age and daily diet choices \cite{Stewart2014}, and it only affects men since the prostate is a gland that belongs to the male reproductive system \cite{Romer1977}. Regarding these factors, it has been shown that the disease is not only more likely to be diagnosed in black men, but it also shows higher mortality rates \cite{Gallagher1998}.  Nevertheless, family medical record is still rather important because having a first-degree relative with prostate cancer doubles the chances of an individual to be also affected in the future \cite{10.1002/cncr.11262}.  Although responsible for enormous fatalities every year, it does not have a high mortality rate, especially if it is detected in early stages. Hence, the detection in these stages can prevent death and avoid other treatments, with respective side effects and harms, especially if the disease spreads to other body parts.  

It is estimated that two-thirds of the affected patients do not show any symptoms \cite{10.1002/cncr.11635}. Frequently, the initial phase of the diagnosis is the screening phase that can be performed through two exam. The first example exam is the measurement of the prostate-specific antigen (PSA), that uses blood tests to detect the levels of the antigen. If they are high, there is a chance of prostate cancer, even though the values may change with other diseases. Thus, it does not guarantee the presence of cancer \cite{TheAmericanCancerSociety2019}. The second screening technique is the digital rectal examination (DRE) \cite{PDQScreeningandPreventionEditorialBoard2002} that consists in the insertion of one finger in the rectum to directly feel anomalies in the prostate. However, these exams are controversial among the scientific community, for example, the PSA has been shown to increase the risks of overdiagnosis that usually implies overtreatment \cite{Gomella2011} and DRE considered to be ineffective \cite{Naji2018, Grossman2018}. Moreover, both tests are lacking evidence of benefits in reducing the mortality of the disease \cite{MHoffman2020, PDQScreeningandPreventionEditorialBoard2002}. Concerns related to the overtreatment (i.e. unnecessary biopsies) are supported by several harms that may result from the treatment (i.e. loss of sexual functions, blood in the urine) \cite{Roberts2016,El-ShaterBosaily2015}.

Since screening is not accurate enough to detect prostate cancer, these methods are often used as a procedure to decide between doing biopsies or not. High values generally require a biopsy, a process that consists of the direct removal of prostate samples using a special needle \cite{Patel2009}. The process is usually performed transrectally and implies some risks to the patient due to the invasiveness of the procedure, thus, since the mid-1980s, ultrasound images have been used as a guide, and by 2014 it was still the most common approach \cite{Bennett2016}. However, prostate cancer cannot be properly detected by ultrasound images since these have poor resolution while displaying soft tissue \cite{Bonekamp2011} leading to poor quality and rigour of biopsies. More recently, studies on multi-parametric magnetic resonance images (mp-MRI) have shown not only a better quality in identifying prostate cancer \cite{Bonekamp2011} by detecting cancer that was missed by traditional blind biopsies \cite{Sonn2013, Marks2013, Kuru2013}, but also decreasing the cases of overdiagnosis by 89.4\% \cite{Pokorny2014}. Thus, for the purpose of this thesis, mp-MRI is the medical imaging technique used and it is further detailed in Section \ref{Sec:mri}. 

Over the years, several researchers have attempted to build systems, both hybrid and fully automatic systems, that could help humans diagnosing the disease. The development of these systems required different techniques, where the most common were machine learning algorithms. Endeavouring to diagnose the disease by detecting and grading lesions, early research covered prostate segmentation, lesion detection, lesion classification, and lesion segmentation. Some of these tasks led to the creation of challenges and competitions as a way to motivate the research on those topics. ProstateX \cite{ProstateX} for clinical significance classification of lesions and the ProstateX2 \cite{ProstateX2} for lesion detection are some of those challenges, in which automated methods achieved promising results. 
In recent years, exponential growth in computer power, development of new machine learning algorithms and the integration of them with computer vision led to a significant improvement of computer-aided diagnosis systems (CAD). The whole new approach, called deep learning, required no handcrafted features and was the source of significant improvements, not only in other fields, but also in cancer screening, with some systems able to outperform radiologists \cite{McKinney2020}. Chapter \ref{Chap:backgroud} explores previous research on the topic with Section \ref{Sec:deep} focusing on details regarding deep learning and Section \ref{Sec:cad} diving deeper into computer-aided diagnosis systems. 

This thesis presents a research study on a recent dataset (IMPROD) \cite{JAronen} \cite{Merisaari2019} that contains samples of mp-MRI images, biomarkers, segmentation masks and lesion scores for 157 men with prostate cancer. The dataset is further explored in Chapter \ref{chap:dataset}. The two main problems addressed in this thesis are the classification of lesions, including PIRADS and clinical significance, and the detection and segmentation of the prostate cancer lesions. For these problems different experiments were performed on the dataset, firstly using a convolutional neural network (CNN), named XmasNet \cite{Liu2017}, to classify lesions regarding its clinical significance and a variation of this architecture to predict the PIRADS score and and for the other ones regarding the detection and segmentation of lesions on spatial images using two neural network architectures called 3D UNet \cite{Cicek2016} and a pretrained 3D ResNet-18 \cite{chen2019med3d}. The latter experiments were evaluated in two different problems in an attempt to explore their capabilities with prostate segmentation before applying it to lesions. Moreover, the architectures have shown results in a variety of different image segmentation problems. Experiments are respectively detailed in Section \ref{Sec:lesion_class_experiment} and Section \ref{Sec:lesion_segm_experiment} after introduction of the methods to be used in Chapter \ref{Chap:methods}. Finally in Chapter \ref{Chap:conclusion} results are discussed and the thesis concluded. 

\clearpage

\chapter{Background}
\label{Chap:backgroud}

Ever since computational methods could learn from data, prostate cancer has been studied, with more recent techniques being applied directly to medical images. The interaction of these methods with humans has been evolving, from methods that required human fine-tuning and corrections after the prediction, to algorithms that can predict with enough quality to be considered automatic. 

In this chapter, the current clinical practice for diagnosing prostate cancer is described. Moreover, the advantages and disadvantages of the current methodology are explored. 

Deep Learning techniques are also introduced and contextualized with the requirements of the problem, based on previous research on similar topics. Furthermore, it is explained how multi-parametric magnetic resonance images work, and why they should be used to detect, segment and classify prostate cancer lesions, due to the differences in the contrast of those regarding soft-tissue appearance against common CT images. 

Finally, previous research on computer-aided diagnosis systems for prostate cancer are discussed with previous deep learning-based CADs being mentioned and some details explored. 


\section{Prostate cancer: current clinical practice}

Manually diagnosing prostate cancer is a challenging process that requires a considerable amount of expertise, diligence, and potentially several test examinations. Thus, the process is rather expensive and time consuming that has a high rate of overdiagnosis. Treatments are also frequently associated with a myriad of secondary effects with distinct types of severity.

The process of diagnosing and treating prostate cancer is frequently divided into four main phases. First, the screening where an expert tries to assess by direct contact or blood analysis if cancer is present. Medical images are captured and analyzed to guide the third phase, the biopsy, which is used to confirm the presence and severity of cancer. The latter phase is rather invasive and may lead to potential harm. Finally, the treatment phase, where previously captured or new medical images can be used to guide the surgery. In the following paragraphs, these four phases are described and explained, and the potential automatization of those will be discussed, either to reduce the required resources, the required time, or to decrease the overdiagnosis. 

\subsection{Prostate cancer screening}

Prostate cancer screening is used to detect suspicions of prostate cancer, however, it is also used in individuals without symptoms, but that might have an undiagnosed tumor \cite{PDQScreeningandPreventionEditorialBoard2002}. Screening is usually the first step in the detection of prostate cancer, and it may improve the chances of early detection. The screening tools used by doctors are the digital rectal examination (DRE) and the analysis of the prostate-specific antigen (PSA) in the blood. And despite their frequent use, some experts disagree with their value, especially when thresholded with the potential harms of the overdiagnosis leading unnecessarily the patient to the subsequent phases \cite{Basch2012}. Considering the risks and the advantages of these techniques \cite{Wolf2010, Greene2013}, patients frequently are invited to decide jointly with the responsible doctor if they should perform the screening exams or not  \cite{prostate_guidelines_webmed}.

A DRE requires a doctor or a specialized person to introduce a gloved finger into the rectum to feel the prostate in an attempt to detect anomalies with its size. Its use is frequent in clinical practice as the first screening test, however, several studies, institutions, and experts have questioned the use of this test \cite{Grossman2018, Naji2018}, whereas others recommend this test to be used as the second line of test, one that should be used after PSA tests \cite{dre_secondary}. 

Testing for the presence of the prostate-specific antigen can be useful to detect anomalies in the prostate, therefore, it is used as a screening method. The PSA is produced in the prostate and can be detected with a blood test and a low quantity of this antigen in the blood indicates that the prostate is healthy. Yet, prostate cancer is not the only potential problem to raise the quantity of this antigen in the blood \cite{Catalona1994}, because prostatitis and benign prostatic hyperplasia can also be possible causes \cite{Velonas2013}. It is clear that PSA is not enough to diagnose prostate cancer, and that it might led to overdiagnose that is the cause of considerable harms due to further unneeded treatments. Thus, it is also recommended to not administer the test to young men due to the fact that it would potentially diagnose cancers that did not require immediate or future interventions \cite{Cooperberg2011}. 

\subsection{Image-guided prostate cancer biopsy}

Generally, image-guided prostate cancer biopsy is performed when there is at least a minor suspicion that prostate cancer may be present, for instance, high values in the prostate-specific antigen blood test, or anomalies found in the digital rectal examination. It is an invasive procedure that consists of the introduction of a special needle to remove samples from the patient's prostate for further tests and exams. It requires the introduction of the needle either in the urethra, the perineum, or transrectally, however, the latter is the most frequent\cite{Ghei2005}. 

There are two distinct types of prostate biopsies, one where ultrasound is used to guide the intervention (TRUS), and another where MRI images are used. Despite being widely used, and frequently mentioned as the most common method \cite{Stuckey, Bennett2016}, TRUS biopsies have, since 2005, lost ground to MRI-guided biopsies due to a better soft-tissue resolution, and higher potential to characterize prostate cancer \cite{Bonekamp2011}. In fact, the MRI-guided biopsies' potential was already shown in practice, and its ability to detect more cancers \cite{Marks2013} is verified by studies that confirm that it improved the detection of clinically significant prostate cancers by 17.7\% \cite{Pokorny2014}. Moreover, it also reduced the overdiagnosis by 89.4\% \cite{Pokorny2014}, and the correlation between the biopsy and the pathology is higher when using MRI-guided biopsies \cite{Hambrock2012}. 

As a consequence of the invasiveness of the procedure, some harm might be inflicted on the patient as a side effect. The most frequent side effects include blood in the urine, in 31\% of the cases, rectal pain, burning when urinating, poor erections and urinary infection potentially requiring hospitalization \cite{Bell2014, Roberts2017}. Due to this fact, in some cases, where there were previous negative biopsies, experts may recommend only the analysis of an MRI image if PSA values continue increasing \cite{Turkbey2018}.   

One of the areas that can benefit most of the automation is the analysis of medical images of prostate cancers. Firstly, the process of having a doctor annotate MRI images manually, either to guide a biopsy or to analyze the cancer lesions, is expensive and time consuming. To extend the implementation of this technique is crucial to reduce the costs, and here automation can assist if it succeeds to annotate the images with the same precision of the expert. Furthermore, if the annotation quality is increased by an automated tool, it might represent fewer biopsies, and fewer treatments, by reducing overdiagnosis. The reduction of these two interventions potentially could greatly reduce the risk of side effects and harm to the patient. 

\subsection{Prostate cancer treatment}

Treating prostate cancer is  challenging and often it requires a combination of surgical and non-surgical interventions \cite{Sartor2018}. Cancers in a more advanced state sometimes spread to other parts of the body near the prostate, endangering more organs and creating difficulties in the treatment. Therefore, these require special treatments such as hormonal therapy and chemotherapy, with minor exceptions, for example, when there is a limited amount of metastasis, sometimes radiation treatment is used \cite{Dhondt2019}. Other less advanced tumors can be treated with surgery, external beam radiation therapy or cryosurgery.  However, it is not wise to combine radiotherapy after a failed surgery since it may cause other problems (e.g. different cancers) \cite{Wallis2016, Mouraviev2006}. Both these procedures have similar prospects for the side effects after a five-year period \cite{Wallis2018}. 

Despite all the available treatments, radical prostatectomy is the main treatment for prostate cancer. Its effectiveness has been growing in recent years, with robotic-assisted procedures not only being available and common \cite{surgery} but also reducing the stay in the hospital \cite{Ilic2017}.  Both this procedure and radiotherapy have significant side effects that can inflict harm, to the patient, that may considerably affect his life. Within the side effects, we can include stress urinary incontinence and erectile dysfunction. The latter affects nearly all the patients that undergo treatment \cite{hopekinsdsd}.

\section{Deep Learning}
\label{Sec:deep}

\begin{figure}[h!]
    \centering
	\includegraphics[width=13cm]{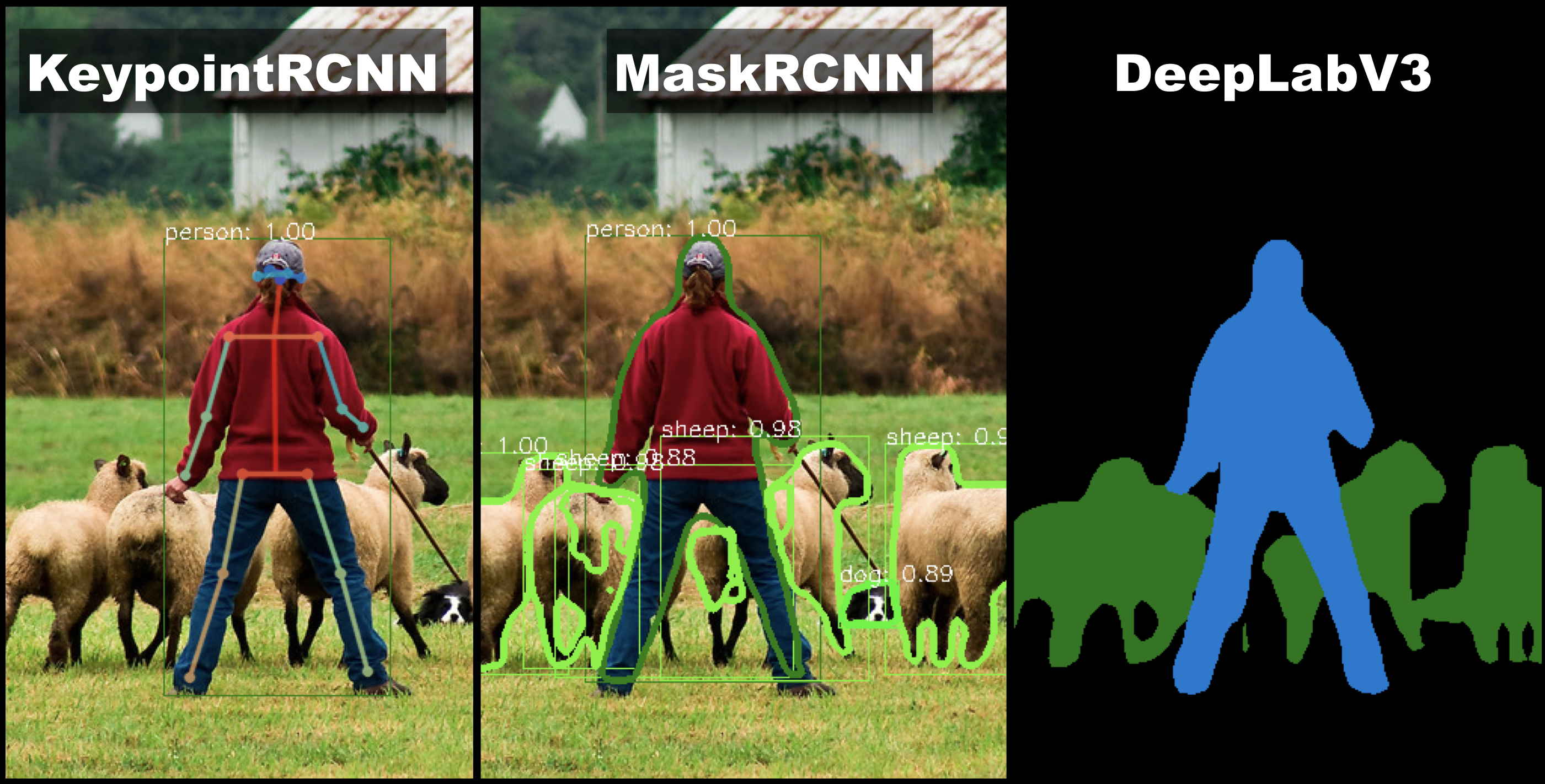} 
	\caption{Examples of use cases of deep learning with pose estimation, object detection and object segmentation with the respective models KeypointRCNN, MaskRCNN and DeepLabV3.\\ Image from the following blog post  - https://pytorch.org/blog/torchvision03/} 
    \label{example_segmentation_detection}
\end{figure}

Deep learning is the usage of deep neural networks (i.e. a neural network with several hidden layers stacked) to tackle problems ranging from natural language processing to computer vision and speech recognition \cite{Hinton}. It has been growing throughout the years, since the publication of the paper describing an architecture of a deep convolutional neural network called AlexNet that won the ImageNet challenge in 2012 by a considerable margin against traditional methods (e.g. previous state of the art algorithms) \cite{Krizhevsky}. The performance in the challenge of these networks is now superior to the performance of humans \cite{He}. 

Further research showed that deep learning was not only able to outperform previous models and techniques on classification problems, but it was also able to become the state of the art in a myriad of other problems such as the translation of a text to other languages, the detection \cite{Ren2017,Redmon2018} and the segmentation \cite{Ronneberger2015} of objects in images.  Figure \ref{example_segmentation_detection} shows examples of some computer vision tasks and models, such as pose estimation, object detection and segmentation.  These models can be used to detect a variety of objects from images, with slight variations, such as drawing a bounding box around an object or creating a segmentation mask for the object. There are plenty of different applications for these models, and they range from biomedical applications \cite{Ronneberger2015} to autonomous vehicles \cite{Li2019} and video description \cite{Aafaq2019}. 

From these methods, the segmentation is the one with more successful applications in biomedical domains, such as segmentation of cancer lesions \cite{Yousefikamal2019}, which is one of the main problems, applied to prostate cancer, that this thesis tries to solve. This section and respective subsections focus on the necessary deep learning foundations to be able to solve the proposed problems.

\subsection{Artificial neural networks }

Some computational algorithms were built based on and inspired by a specific biological system. One of these algorithms, called Artificial Neural Networks, was inspired by the nature neurons present in biological systems like the human brain.

\begin{figure}[h!]
    \centering
	\includegraphics[width=8cm]{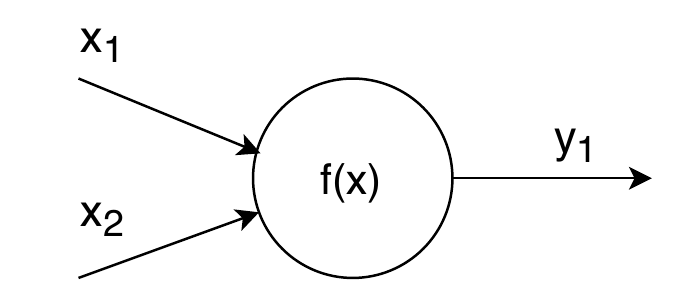} 
	\caption{An Artificial Neuron (also known as perceptron), basic unit of neural networks, with 2 input values, and one output value subject to the function f(x). } 
    \label{perceptron}
\end{figure}

Similarly to what it is possible to observe in nature, an artificial neural network has a basic unit, the artificial neuron. However, for historical reasons, the unit is sometimes called the perceptron and it is shown in Figure \ref{perceptron}. Several of these artificial neurons can be stacked either vertically or horizontally in order to create more complex compound units able to learn how to approximate nonlinear functions. 

To interact with the perceptron, it is necessary to feed it input values $\vv{x}$ to which it is applied a function $f(\vv{x})$ that originates output values $\vv{y}$. Either $\vv{x}$ and $\vv{y}$ can be represented with one or multiple values. As mentioned previously, these structures learn how to approximate nonlinear functions, however, $f(\vv{x})$, the applied function, is just a representation of a linear function $f(\vv{x}) = w^t\vv{x}+b$, where both $b$ and $w$ are learnt values, denominated respectively as bias and weights. Therefore as an attempt to approximate those nonlinear functions, researchers introduced nonlinearities after each function, called activation functions. Some of the most popular activations functions are the Tanh function, the Sigmoid function, Softmax function and the rectified linear unit (ReLU) \cite{10.5555/3104322.3104425}, and while those are broadly used some other activations, such as Mish \cite{misra2019mish} and Leaky rectified linear unit (Leaky ReLU) \cite{Maas2013}, have been used recently, showing promising results in some use cases.

\begin{equation} \label{equ:activation}
\begin{split}
\vv{y} & = \phi(f(\vv{x}))\\
& = \phi(w^t\vv{x}+b)
\end{split}
\end{equation}

As a result of including this activation in our artificial neuron, the output is now given by the expression shown in the equation \ref{equ:activation}.

\begin{equation} \label{eq:relu}
\begin{split}
\text{ReLU: \quad} \phi(x_i) & = 
  \begin{cases}
    0, & \text{if } x_i \leq 0 \\
    x_i, & \text{if } x_i > 0 
  \end{cases}
\end{split}
\end{equation}

\begin{equation} \label{eq:sigmoid}
\begin{split}
\text{Sigmoid: \quad} \phi(x_i) & = \frac{1}{1+e^{-x_i}} \\
\end{split}
\end{equation}

\begin{equation} \label{eq:softmax}
\begin{split}
\text{Softmax: \quad} \phi(x_i) & = \frac{x_i}{\sum_{j}e^x_j} \\
\end{split}
\end{equation}

Equations \ref{eq:relu}, \ref{eq:sigmoid} and \ref{eq:softmax} represent three of the most popular activation nonlinearities that are useful and important to the scope of the problems addressed in this thesis.  Each one of these activations has particularities, either in the range of the output, the computations needed to  computed and the derivative, or the circumstances where they are seen as particularly useful. ReLU, equation \ref{eq:relu}, has been used frequently as the activation that follows a convolutional layer (see subsection \ref{sub_sec:conv_nets}) since it has been shown that they usually improve deep neural networks training \cite{Lecun2015}. The computations needed to calculate this activation and its derivative are minimal, and the output ranges from $[0,\infty[$. While the ReLU range does not limit a maximum value to the output, both sigmoid and softmax functions limit the output to a value in the interval  $[0,1[$. The sigmoid activation, equation \ref{eq:sigmoid}, is particularly useful and popular for binary classification problems, for instance, when classifying if an image is from a dog or not, the output, between 0 and 1, can be interpreted as a probability of being a dog. On the other hand, the softmax activation, equation  \ref{eq:softmax}, is widely used on multiclass classification problems, due to two particular characteristics. First it is its range, and secondly the fact that it is a function of all the output values, in a way that the sum of the softmax of all the output values must sum to 1. Thus enabling it to be interpreted as a probability of belonging to that class, in other words it normalizes the outputs of one node based on the value of all nodes.

\begin{figure}[!h]
  \centering
  
  \begin{minipage}[b]{0.49\textwidth}
    \includegraphics[width=\textwidth]{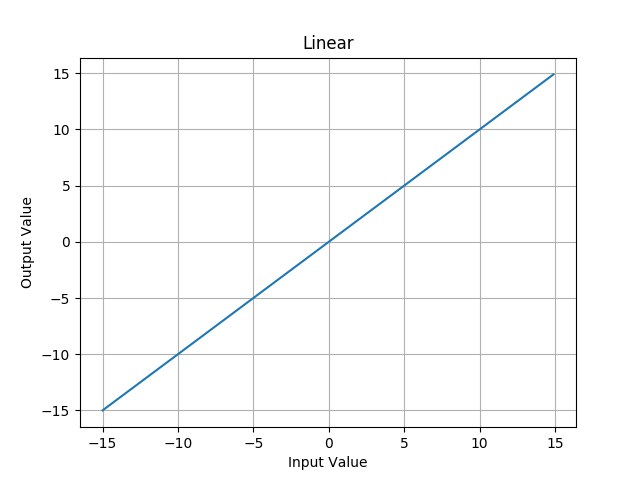}
  \end{minipage}
  \hfill
  \begin{minipage}[b]{0.49\textwidth}
    \includegraphics[width=\textwidth]{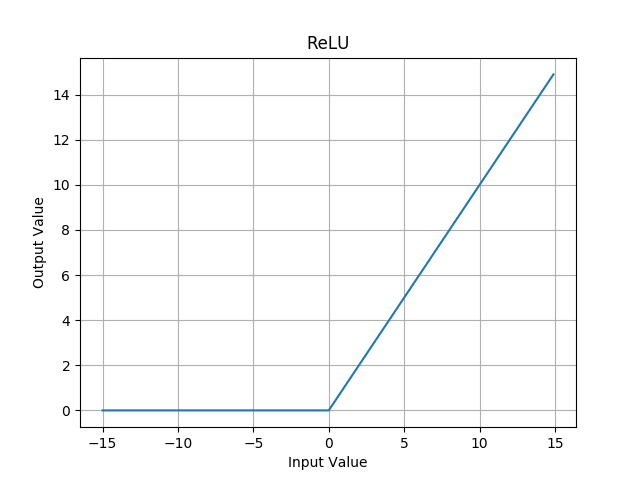}
  \end{minipage}
  \begin{minipage}[b]{0.49\textwidth}
    \includegraphics[width=\textwidth]{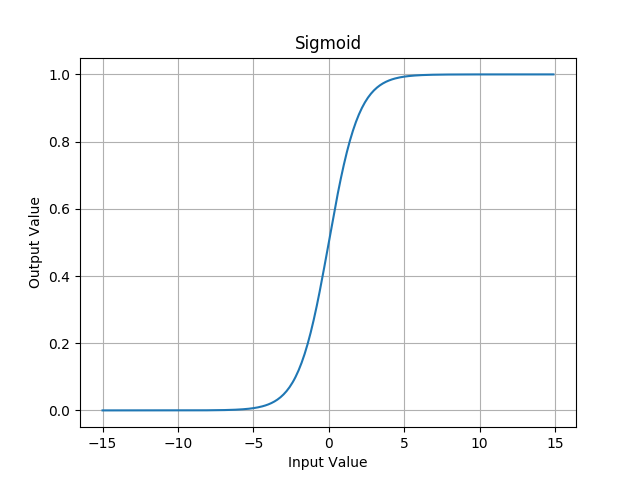}
  \end{minipage}
  \hfill
  \begin{minipage}[b]{0.49\textwidth}
    \includegraphics[width=\textwidth]{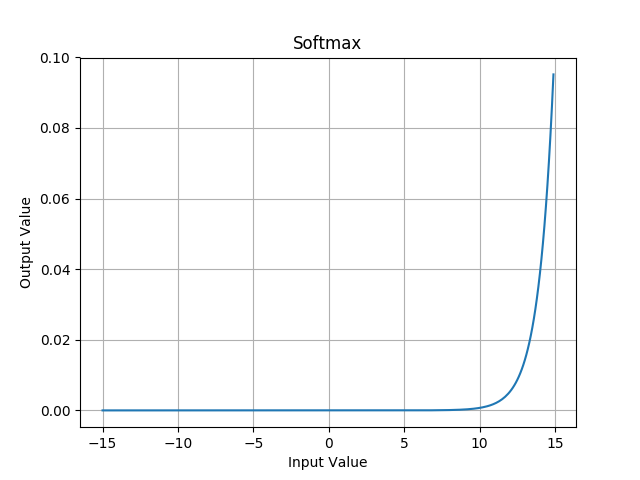}
    
  \end{minipage}
 \caption{Behavior of the ReLU, Sigmod and Softmax activation functions compared with the linear function. It is worth noting the scale of the y-axis in each plot.}
 \label{fig:activations_plots}
\end{figure}

In Figure \ref{fig:activations_plots} it is possible to observe the behavior of the three nonlinearities and compare them with the linear function. It is also worth noting that despite having the same scale on the x-axis in all the plots, each of them has a specific scale for the y-axis. This is of particular relevance when comparing ranges of functions, their progression and the impact that it may have on training (e.g. what if all the values are 0 before being given to a ReLU nonlinearity). However, despite allowing the network to learn more complex functions, the introduction of nonlinearities transforms the optimization problem in a non-convex optimization problem, meaning that it is now more complicated to optimize when compared to a convex problem \cite{Jain2017}.

\begin{figure}[h!]
    \centering
	\includegraphics[width=8cm]{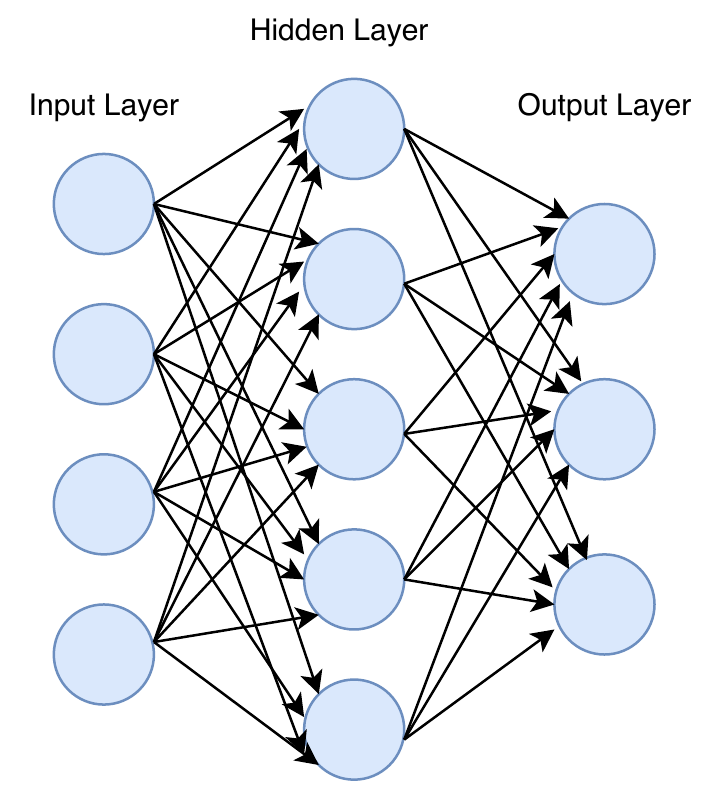} 
	\caption{A simple fully connected (also known as feed forward) artificial neural network with 4 input values, 1 hidden layer composed of 5 neurons and 3 output values.} 
    \label{ann}
\end{figure}

An artificial neural network is not only characterized by the number of input values, the number of hidden layers with the respective number of nodes and the number of output values, but also by how these layers and each node is stacked and connected to the others. One of the oldest and more popular neural networks is the feed-forward, and it contains several layers, the input, and output layers and also a stack of one or more hidden layers. However, each layer has also a stack of artificial neurons (or nodes), increasing the number of nodes increases the complexity of the learned function but also makes the training process more difficult. The same happens when the number of hidden layers is increased. The layers of this network are usually denominated as fully-connected layers, as a result of the fact that every node of the layer $n$ feeds its output to all the nodes of the layer $n+1$ as input. In other words, this means that a simple network as seen in Figure \ref{ann} contains, if all the bias are ignored, 35 parameters to be learned in the optimization process.

\subsection{Optimization of neural networks}

Due to its non-convex nature, artificial neural network optimization cannot be solved by an analytical mathematical expression. Instead, an algorithm that is a special case of reverse accumulation, designated back-propagation, must be used \cite[p.~217-218]{Goodfellow-et-al-2016}. This algorithm solves the values of the parameters (weights) of a neural network in an iterative two-step process, the forward and the backward step. Despite being independent algorithms, back-propagation often takes advantage of the gradient descent algorithm to update the weights and try to find a solution \cite[p.~200]{Goodfellow-et-al-2016}. Despite the fact that configuring the hyperparameters carefully usually allows the algorithm to get a solution that has the reqired performance in practice, there is no guarantee about finding an optimal solution or even a solution that is close to the optimal. Optimization of neural networks is usually called the training of the network, and only during training, the backward step is computed. 

\begin{table}[h!]
    \centering
\caption{Loss functions and the type of machine learning problems where they are frequently used.}
\label{tab:loss_functions}
\begin{center}
 \begin{tabular}{|c | c |} 
 \hline
 Name  & Problem \\ [0.5ex] 
 \hline
 Cross-Entropy loss& Classification \\ 
 \hline
 Logistic loss &  Classification \\
 \hline
 Dice loss &  Image Segmentation \\
 \hline
 Focal Loss\cite{Lin2017} &  Detection \\
 \hline
 Mean Square Error & Regression \\ 
 \hline
 
\end{tabular}
\end{center}
\end{table}

In order to know how to update the network, and to be able to integrate gradient descent, it is necessary to give an objective to the optimization process. Therefore, during training, besides the input, $\vv{x}$, and the output, $\vv{y}$, it is necessary to have a label representing the expected value for $\vv{y}$ \cite{Mohri, Russel2012}. Frequently this label is mentioned as ground truth. Some optimizations do not require labels (e.g. unsupervised learning) \cite{Hinton1999}, however, for the scope of this thesis, only supervised learning is covered. In supervised learning problems, during training, the objective is to obtain an output each iteration closer to the ground truth, and the closeness of both is computed using a loss function. Loss functions generate what can be seen as a numerical error that can be used to analyze the performance of the model. A proper optimization algorithm can, through the minimization of the error, update the weights of the network in a way that will better approximate the optimal function and generate results closer to the ground truth.

There is a myriad of different loss functions, each of them having different properties, advantages, and disadvantages, being arguably more appropriate to one type of problem than to others. Some of these can be seen in Table \ref{tab:loss_functions} with a reference to the type of machine learning problems where they are frequently used.

The forward step was previously described, and it is the process used to generate an output from an input. After computing an output during training, the next step is to compare it with the ground truth through the calculation of an error, using, as mentioned before, a loss function. The backward step then takes advantage of a dynamic programming technique to compute, with the chain rule, the gradients of the input with respect to the loss function. This process goes from the last layer to the first, computing each gradient in one whole iteration, and, for instance, does not repeat redundant computations in order to compute the gradients of some layer \cite[p.~200-220]{Goodfellow-et-al-2016}. The gradients are then fed to a gradient descent algorithm that will perform a gradient step, which is used to finally update the weights \cite[p.~200]{Goodfellow-et-al-2016}.  Different variations of gradient descent can be used, such as mini-batch gradient descent, which uses a small batch of the training set at each gradient step, or stochastic gradient descent, which uses only one sample per gradient step \cite[p.~303-307]{Taddy}. These variations frequently reduce computation cost (e.g. not all the data is loaded in the memory or in the GPU) and help to avoid local optimums, however, the convergence rate decreases \cite[p.~351-368]{PaulH.Sra2011}.

In the inference phase, no gradients are computed and new samples are given to the network. The network must be able to generalize and have error margins in never seen data close to the ones given in the training set, after training. When this does not happen it is said that the network overfitted the training data and strategies to avoid this problem range from reducing the complexity of the network to adding a regularization term to the loss function or increasing significantly the number of training samples \cite[p.~1-12]{Hawkins2004}.

\subsection{Convolutional neural networks}
\label{sub_sec:conv_nets}

The application of deep learning and deep artificial neural networks to computer vision tasks led to the detection of a problem with feed-forward models. Even small images have a large number of pixels, for example, a 128x128x3 (height x width x color channels) image contains $128*128*3 =  49,152$ pixels, which represent the number of nodes in the input layer. The effect of these pixels get worse when we consider that the next layer will connect all the nodes to these input nodes, and that complex tasks require more nodes. Thus, a model that receives this image has 100 nodes in the second layer and five outputs totaling $49,152 * 250 + 250 * 5 = 12,289,250$ parameters to be optimized. Optimization of such a large number of parameters is rather complicated, despite the fact that the network shown is not even a deep neural network. 

To construct feasible networks that work with images as input, researchers started replacing most of the fully-connected layers with convolutional layers. Networks with these layers are called convolutional neural networks (CNN). Inspired by biological receptive fields, the CNN architecture mimics, to some extent, this feature of the animal visual cortex \cite{Hubel1968, Fukushima1988}. The aforementioned receptive field is used by animals as detectors, to detect special characteristics in an image, such as edges. The approximation can be achieved with the convolutional operator \cite{Marr1980}.

\begin{figure}[h!]
    \centering
	\includegraphics[width=8cm]{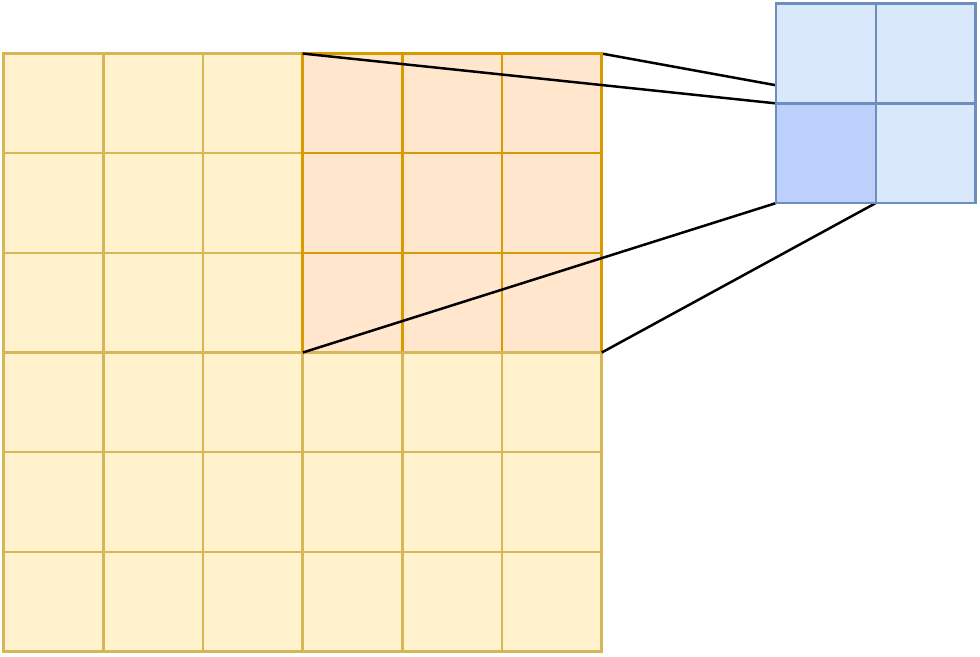} 
	\caption{Visual representation of a convolution happening in a 2 dimensional input of size 6x6. This convolution has dilatation of 0, stride of 3, padding of 0 and a kernel size of 3. The output is of size 2x2} 
    \label{2d_conv}
\end{figure}

A convolutional layer, the main building block of convolutional neural networks, is typically defined by a set of filters/kernels, which are the representation of their weights. Despite going through all the input, these filters are usually considerably smaller than the input size. In order to detect the spatial characteristics of the input, each kernel is used to compute the dot product between the kernel and the image, throughout the entire width and height, generating an activation matrix similar to what can be seen visually represented in Figure \ref{2d_conv}. This process can also be mentioned as a convolution operation. Furthermore, it also includes a third dimension, the channels where the size of that dimension represents the number of stacked filters to be used in the convolution. Using this dimension is what allows one convolutional layer to be able to learn several features in the spatial representation of the input in one single convolution. These characteristics might be useful for future layers to understand and learn more complex spatial shapes.

Besides the filters dimension and the number of channels, the output shape of these layers is controlled by other hyperparameters that require careful attention and tuning to approximate. The three hyperparameters are dilation, stride, and zero-padding.

\begin{itemize}
    \item  \textbf{Stride} - Represents the movement of the filter along with the image height and width, it affects the size of the output since a larger stride will represent a smaller sized output \cite{CS231n}. If the filter is seen as a sliding window that moves around the input, the stride is the step size, in terms of pixels, that the sliding window should move. If the stride is smaller than the size of the kernel, it means that there will be overlapping between activations in the output, otherwise, the same input pixel will not be in more than one activation. 
    
    \item  \textbf{Dilation} - The filter usually convolves on a specific size of consecutive pixels, however, sometimes it is useful to utilize a specific type of convolutions, called dilated convolutions \cite{Yu2015}. This parameter, dilation, represented by an integer, is the spacing between each pixel to be convolved, i.e. a regular convolution has dilation of 0 since there are 0 pixels of spacing between each pixel to be convolved in the input. 
    
    \item \textbf{Zero-padding} - Some use cases require that a padding of 0's is added to the input, this increases the size of the input and is usually used to control the spatial size of the output, for example, when it is required that the output size is the same of the input size. 
\end{itemize}

\begin{equation} \label{eq:output_size}
\begin{split}
 Dim(Y) & = \frac{ Dim(X) + 2 * P - D * (K-1) - 1}{S} + 1\\
\end{split}
\end{equation}

The output size $Dim(Y)$ can be written in function of the input size $Dim(X)$, the filter size $K$, the stride $S$, the dilation $D$ and the padding $P$ by the expression given in the equation \ref{eq:output_size} \cite{conv_size}.

In convolutional layers, the optimization process does not try to optimize the weight of every single neuron, what it does is optimize the existent values in the sliding window. There is an assumption that if the values of the kernel are able to acquire information at some spatial position, they should be able to do the same in the others. In other words, it means that if a convolutional layer is composed of five channels, each with a filter of size 4x4 and it gets as input a 128x128x3 image, the number of parameters of the layer will be $5*3*4*4=240$. This characteristic, that is mentioned as parameter sharing, contributes to two main aspects that might improve the computational performance and the results obtained. The first one is that it does contribute to the translation invariance characteristic that is frequently associated with CNN. Secondly, they require considerably fewer parameters, easing the optimization of the network, and allowing its design to be deep (i.e. include more layers) \cite{HabibiAghdam2017}. The optimization of convolutional neural networks can be done, similarly to feed-forward networks, with backpropagation and gradient descent \cite{LeCun1989}. 

Besides convolutional layers, convolutional neural networks frequently have max/min/average pooling layers that try to reduce the spatial representation of the image by using a sliding window and max, min or average operations to pick the output value. These layers are particularly useful to reduce the size of the input, however recently researchers have discarded them \cite{Springenberg2015} or reduced the size of their filters \cite{Graham2014}.

To finish a convolutional network architecture, usually, one or more fully connected layers are added to do the high-level reasoning. This works similarly to what was shown before regarding artificial neural networks, including the activations. Including several of these layers might have a negative impact on the computational cost, thus usually only one to three are added after several layers of convolutions and poolings. 

\begin{figure}[h!]
    \centering
	\includegraphics[width=8cm]{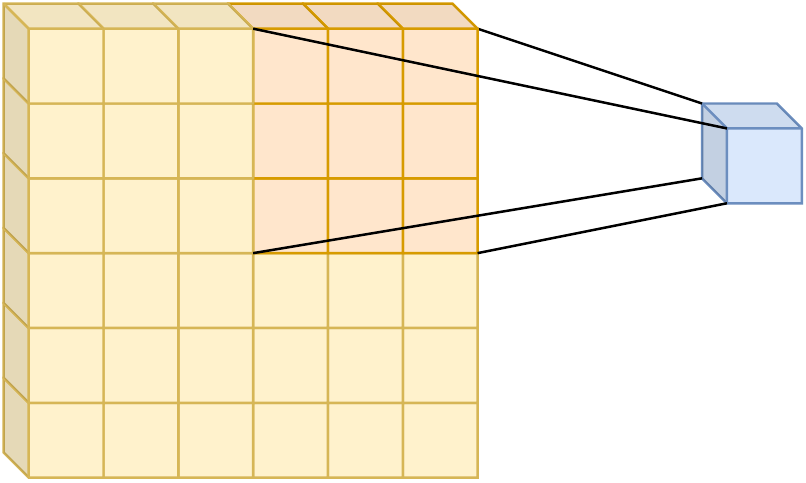} 
	\caption{Visual representation of a convolution happening in a 3 dimensional input of size 6x6x1. This visualization shows the capture of a single activation using a kernel size of 3x3x1. The full output is not displayed.} 
    \label{3d_conv}
\end{figure}

While this subsection focused on convolutional neural networks working in two dimensions, the architecture is not limited to it. Convolutions on three dimensions are possible, and they work similarly, however instead of a sliding window (2-d kernel), it is used a sliding cube (3-d kernel). In Figure \ref{3d_conv} it is possible to observe a simple case of how these convolutions work to generate one entry of the output matrix.

\subsection{Image segmentation}

Similar to other research fields, computer vision has also benefited from the rise of deep learning-based methods. Not only previous results were improved in a myriad of tasks and use cases, but also the process was simplified due to the fact that deep learning does not require manual feature extraction. One of the most frequent topics to be mentioned in research articles is image segmentation, that is other words can be described as the process of dividing an image, given as output, into segments. Some images can have their analysis simplified simply by changing their representation to something with more significance to the problem \cite{Shapiro2001,Lee2003}.

\begin{figure}[h!]
    \centering
	\includegraphics[width=12cm]{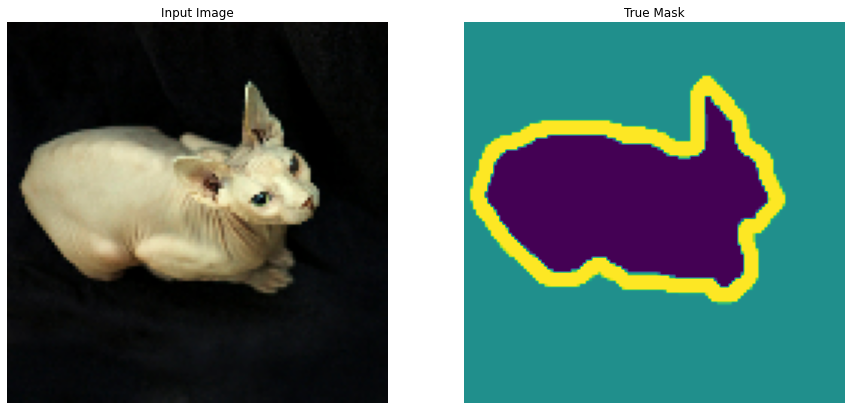} 
	\caption{Segmentation example, where the boundaries and the animal image are segmented as separated segments. On the left it is possible to observe the original image, on the right the 3 different segments, the background (blue), the boundaries (yellow) and the cat (purple). Image from https://www.tensorflow.org/tutorials/images/segmentation} 
    \label{segmentation_example}
\end{figure}

Segments can vary accordingly to the problem, however, frequently the main use of image segmentation is to find segments that contain some object, and to draw its boundaries. For instance, if a problem consists of segmenting the ball from a football game picture, the idea is that the output includes the same label in each pixel where the ball is represented and another label (e.g. a background label) in the other ones.  This representation not only gives the position of the ball, as it gives its boundaries too. Figure \ref{segmentation_example} shows an example of a segmentation problem, where the goal is not only to segment the cat, but to segment separately the boundaries between the cat and the background.  

The interception of image segmentation with other research areas also led to significant advances and some even more promising future prospects. One of the areas, where it has been applied to, was medical imaging \cite{Pham2000, Forouzanfar2010}, with several uses cases, such as detecting and segmenting tumors \cite{BenGeorge}, surgery planning and studying the anatomical structure to detect, and diagnose problems \cite{Kamalakannan2010}.

\subsection{Data augmentations for biomedical images}

Training a regular artificial neural network takes a considerable amount of time and data, and some specific areas, such as deep learning, where the neural networks also extract the features from the data, it is required even larger volumes of data. This is also true to avoid problems like overfitting, which might hurt the performance of the network when it is exposed to never seen samples, due to the fact that the complex model fitted almost perfectly the testing data, decreasing the generalization capabilities. Moreover, it is important that the data has representations of a great part of the input data distribution, so it can properly approximate the real distribution, in other words, when trying to segment prostate lesions it is important that the model trains on different tumors shapes, sizes, and aggressiveness. 

Constructing a large medical dataset is an almost impossible task due to diverse factors. To start with the labeling of the samples must be precise, accurate and correct, which requires expert knowledge. Nevertheless, not only the price of these radiology experts is rather expensive, the process of manually annotating the labels in a medical scan is time-consuming, and a complicate process. Hence, the datasets provided by these clinical experts are frequently small in size. Furthermore, recently the restrictions imposed by privacy laws created extensive problems for the investigation labs, health institutions or clinical trial lab to combine their data and joint efforts to create a large meaningful dataset for deep learning. Comparing to other Computer Visions datasets with millions of available samples for training, the frequently less than one thousand samples in medical images datasets shows the lack of proper data in these problems. Moreover, medical images are frequently represented with one color channel, which can be seen as one extra challenge to the learning process.  

There are, however, techniques that were created as an attempt to bypass this problematic lack of data. The technique with the most success cases is called data augmentation, or more specifically image augmentation when referring to image data. Since one of the advantages of having a big dataset is to have a high variation in the data which drives the network to focus on the important features instead of focusing on specific of each image, these techniques apply small random transformations to the data. The application of these augmentations can be done statically and applied to all the dataset, however, the best results have been attained with generators, that will apply on the fly the augmentations to an image randomly varying the magnitude and which transformations are applied. Generators usually help to further tackle the overfitting problem.

Several transformations that can be applied to the data, however, not all of them are appropriate to be used in medical images. Considering the impact that the features of medical images are important, and should be preserved, the most frequently used augmentations include the use of linear transformations. These can be divided into rigid and non-rigid transformations, with the latter not keeping the image shapes unchanged. Some of the rigid transformations are translation, rotation, flipping.

\begin{itemize}
    \item  \textbf{Translation} - Shifting of the image towards some direction, changing the absolute position of the image elements. 
    \item  \textbf{Rotation} - Rotating the training image by a randomized number of degrees. 
    \item \textbf{Flipping} - Due to the fact that anatomical structures are, in several cases, symmetric, this technique mirrors the image either vertically or horizontally. This also tackles the lack of variety of the dataset, in the sense that images givens could, without any particular medical reason, have the region of interest located more frequently in on side of the organ.
   
\end{itemize}

Examples of the non-rigid linear transformations are stretching and shearing. 

\begin{itemize}
    \item  \textbf{Stretching} - Zooming in and out with different ratios stretches the image, and changes both absolute and relative features of the image. However, this transformation must be used carefully since if its magnitude is too high it can hurt the model performance instead of improving it.
    \item \textbf{Shearing} - While stretching works in one direction, shearing is a similar transformation that consists of moving the top and the bottom, or the left and the right side of the image in different directions. It can be seen as stretching in two directions.
\end{itemize}

 Finally, in an attempt to improve the generalization of the model across different medical imaging capture machines, it might be useful to apply a transformation that will vary the intensity of the gray-scale pixels in the image. This also helps to optimize the model in a more robust way when the limited dataset had all the samples captured by the same machine.
   
\section{Multi-parametric MRI}
\label{Sec:mri}

The complexity of the task of detecting, segmenting, or diagnosing disease from medical images is due to several factors. First of all, not only it requires appropriate expert knowledge so the image is properly analyzed, but the results are affected by the type and quality of the image being visualized. In other words, not all techniques to capture an image are adequate for the visualization and accurate diagnosis of either prostate cancer lesions or the classification of the aggressiveness of those lesions.

For prostate cancer, two main imaging techniques have been used over the years. Ultrasound (US) scan was the main technique to diagnose prostate cancer, however, in recent years, Magnetic Resonance Imaging (MRI) started to have its use increased in several countries. One of the reasons for this increase is that ultrasound images have poor soft tissue resolution whereas magnetic resonance images show better resolutions \cite{Bonekamp2011}. Moreover, MRI has been used to help doctors assessing other details, such as the difficulty of the surgical procedure, and which direction should they follow when planning the surgery \cite{Tan2012}.  

To capture an image of the organs of the body, magnetic resonance images rely on radio waves and in strong magnetic fields. Since it does not use radiation or ionization it can be seen as a better solution when compared to alternatives like Computed Tomography and X-Rays. However, MRI requires longer capture times, which can become an obstacle when this technique becomes widely used on a higher number of patients. The mapping of the organs is based on the detection of water and fat in the body through the interaction and excitation of hydrogen atoms present in those biological tissues. Small changes to the configuration can generate different types of images with a huge variety of contrasts and use cases applied to different diagnoses, these different types are frequently called sequences. When a magnetic resonance image is composed of two or more sequences, it usually is called multi-parametric magnetic resonance image (mpMRI) \cite{Tahmassebi2018} \cite{Marino2018}. Some of the most common sequences are T1 weighted (T1W), T2 weighted (T2W),  diffusion-weighted image (DWI) and apparent diffusion coefficient (ADC). Each of these sequences captures different details, therefore they have different uses based on the problem to analyze and diagnose (e.g. DWI and T2W are frequently seen in prostate cancer studies).

\subsection{PI-RADS}

Prostate Imaging Reporting and Data System, also known as PI-RADS is a  set of standards to capture and report prostate cancer images with mpMRI and assess from them the risk of clinically significant cancer being present. The first version of these standards was especially focused on the clinical significance classification \cite{Barentsz2012}, whereas the second version, PI-RADSv2, focused also in creating global standards for MRI \cite{Weinreb2016}. Overall, both versions aimed to improve the quality of image capture and reporting. 

A plethora of studies have tried to assess the impact of these standards across the entire workflow of a prostate cancer diagnosis. Both versions have shown positive results in classifying clinically significant prostate cancer lesions \cite{Thompson2016,Vargas2016}, however, some limitations regarding the classification of small ($\leq$0.5mL) significant lesions (GS$\geq$4+3) have been noticed \cite{Vargas2016}. Furthermore, PI-RADS has proved to be useful in other applications as the detection of the extension of cancer outside of the prostate \cite{Schieda2015} which is an important step in the staging of the cancer. Predicting when the active surveillance termination period should be defined, based on the aggressiveness of lesions, is another of the uses of this set of standards and evaluation criteria \cite{Abdi2015}. 

\begin{table}[h!]
    \centering
\caption{PI-RADS scores \cite{Read_prostate}}
\label{tab:pirads_grades}
\begin{center}
 \begin{tabular}{|c | c|} 
 \hline
 PI-RADS & Probability of clinical significance \\ [0.5ex] 
 \hline
 1 & Very low \\ 
 \hline
 2 & Low \\
 \hline
 3 & Intermediate \\
 \hline
 4 & High \\ 
 \hline
 5 & Very High \\
 \hline
\end{tabular}
\end{center}
\end{table}

Table \ref{tab:pirads_grades} shows the scores that compose the PI-RADS score system and their meaning in terms of risk of the cancer being clinically significant. 

\subsection{T2W}

After excitation, tissues affected follow a relaxation process, on T2 weighted a specific relaxation process, by the name Spin-Spin relaxation, is used in order to generate the image. This process of capturing the images is based on the decay of the transverse component of the magnetization vector to an equilibrium state. The decay is exponential and the time necessary to reach the equilibrium given in function of a constant known as spin-spin relaxation time, or T2 \cite{Abragam1983, Claridge2016}. The constant T2 (given in the order of seconds for protons) varies, and each biological tissue has its own, for example, water-based tissues range from 40 to 200 ms whereas fat-based tissues range from 10 to 100 ms.

\begin{figure}[!tbp]
     \centering
     \begin{subfigure}[b]{0.495\textwidth}
         \centering
         \includegraphics[width=6.5cm]{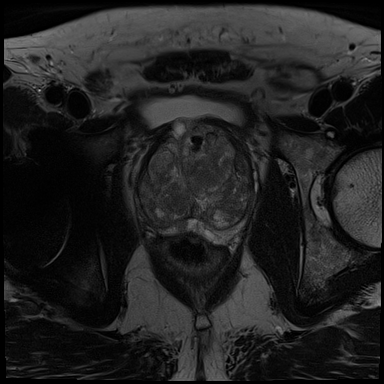}
         \caption{T2W sequence. ProstateX dataset \cite{ProstateX}.}
         \label{back_t2w}
     \end{subfigure}
     \hfill
     \begin{subfigure}[b]{0.495\textwidth}
         \centering
         \includegraphics[width=6.5cm]{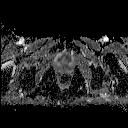}
         \caption{ADC sequence. IMPROD dataset \cite{JAronen, Merisaari2019}}
         \label{back_adc}
     \end{subfigure}
        \caption{Visualization of T2W and ADC slices, from a mp-MRI of the prostate, side by side. }
        \label{fig:three graphs}
\end{figure}


Variation in the capture time of different tissues generates different signal intensities, therefore it is possible to observe a contrast between some tissues and body organs based on what is in their composition. Some of the high signals in T2W images are given by water tissues \cite{UniversityofWisconsin}, such as inflammation, edema, tumors or infections \cite{Johnson} whereas bone, air, protein-rich fluids, and fibrosis originate low signals \cite{UniversityofWisconsin, Johnson}. The representation of a high signal is a white, brighter pixel while a low signal is represented by a darker pixel. In Figure \ref{back_t2w} it is shown an example of a T2W image of a prostate MRI.

\subsection{DWI-ADC}

The diffusion of water molecules in tissue is constrained and affected by the interaction of water molecules with different components of that tissue, such as fiber and proteins. Images are generated through the analysis of water diffusion patterns that are used to map the microarchitecture of biological tissues into an image with the appropriate contrast to detect and diagnose diseases and anomalies \cite{Dunn2015, Merboldt1985}. Images generated with this technique are called diffusion-weighted magnetic resonance images (DWI).

The composition of a tumor, frequently highly cellular, which originates more constraints to water diffusion allows this type of pathology to originate a high signal in DWI images \cite{Koh2007}. Hence, this sequence type is frequently used to detect, stage and monitor cancer tumors. 


Standard diffusion-weighted images have an inherent relation with T2 weighted images, which in some cases has a possible negative impact on the image capture. For instance, lesions that do not restrict water diffusion are not supposed to have a high signal, however, if their T2 relaxation time is long then the pixels will be bright \cite{Hammer}. To reduce this effect, a special type of DWI, called apparent diffusion-coefficient (ADC), is used. The capture of an ADC image requires several diffusion-weighted images to be captured with different weights, the rate of diffusion is then calculated from the change in the signal between weights. Despite the similarities and the relation between ADC and DWI, the second uses bright pixels to represent constraint movement, whereas ADC uses dark pixels for that representation \cite{Elster}. In Figure \ref{back_adc} it is possible to observe an example of a ADC image of a prostate MRI, and it is also possible to observe that this image has frequently lower resolution when compared to T2W images, as can be seen by comparing with Figure \ref{back_t2w}.

\section{CAD to prostate cancer}
\label{Sec:cad}

Several researchers have attempted to develop solutions that would help to diagnose prostate cancer. These solutions varied greatly in their methodology, not only in terms of deep learning models, but in the way that they saw the problem. For instance, some solutions tried to classify the clinical significance of lesions from diffusion-weighted images using a dataset with 427 patients \cite{Yoo2019}. Others directed their efforts to segmentation problems from these magnetic resonance images. 

Since it is an easier problem, considerable research was conducted for the prostate segmentation with varying results regarding different datasets and methodologies. For this problem, some researchers used only T2W sequences, either from public datasets such as PROMISE12 \cite{Karimi2019} or private \cite{To2018}. In an effort to further increase the performance of single sequence segmentation, some other researchers co-registered both DWI/ADC images with T2W images. While in some cases this was the only approach \cite{Feldman2019}, it is possible to see in others that the co-registration, in fact, increases the performance  when compared to single sequence inputs \cite{Schelb2019}.

Some authors decided to conduct more extensive experiments and efforts, endeavoring to detect and segment prostate cancer lesions from these images. The results reflected the hardness of the task at hand, and in some cases, neither of the independent sequences or a combination of both was enough to surpass a dice score of 0.60 \cite{Schelb2019}. However, there has been some progress with some solutions obtaining a dice score of 0.64 which is considerably closer to the reported dice score for two expert clinicians of 0.67 \cite{Feldman2019}.

A myriad of other techniques was attempted ranging from traditional machine learning with features extracted and treated manually by researchers, probabilistic methods, and others that required post-processing of the output by experts. Since the main focus of this thesis is the use of deep learning based methods, the focus of this section was in those methods previously developed by other researchers.

\clearpage

\chapter{Improd Dataset}
\label{chap:dataset}

Recent progress in machine learning and deep learning research has been backed up by the improvement of public datasets for a myriad of domains. These datasets allow a more objective evaluation of the methods since they are all tested on the same data. However, that is not the case with datasets containing prostate magnetic resonance images, since those do not combine high image and annotations quality and different types of annotations. Therefore, for this project, the existence of a private dataset, IMPROD, reuniting all this characteristic is of greater importance, even if the objective comparison with other work might be slightly affected. Moreover, not only the quality is important, but the size and some other statistical details of the dataset are important.

The analysis of a dataset includes several questions to be answered such as "What is the source of the dataset?", "How was the dataset annotated?", "Why is this dataset valuable for the problem that is being solved?" and "How do those annotations align with the experiments necessary to solve the problem?". Furthermore, since this dataset contains medical images it is also important to discuss the expertise of the annotator and the quality of the dataset. Moreover, studying if the data available is representative of its real-world distribution that would allow generalizing to the deployment of the model.  

This chapter discusses all the previously mentioned questions addressing, in particular, the  IMPROD dataset and its characteristics. Further important statistics to set up, interpret and evaluate experiments using the methods detailed in the previous chapter are analyzed thoughtfully to understand the weaknesses and strengths of this particular dataset in the context of this thesis. 

\section{Source of images and annotations}

The IMPROD (Improved Prostate Cancer Diagnosis - Combination of Magnetic Resonance Imaging and Biomarkers) dataset originated from a clinical trial conducted in a joint effort by the Turku University Hospital and the University of Turku. The trial included 175 patients between 40 to 85 years with suspicions of prostate cancer supported by screening results (i.e. abnormal DRE or 2,5-25ng/ml PSA in two measurements).

In order to maximize the quality, the dataset was, not only gathered using images captured with recent and high-quality magnetic resonance scanners (Magnetom Verio 3T, Erlangen, Germany), but it was also carefully annotated by experts in the field that belong to the institutions conducting the clinical trial. This maximization of the data quality due to the techniques used denotes an attempt to improve the outcome of machine learning and deep learning solutions in a myriad of prostate cancer-related problems.

From the possible image sequences captured by a magnetic resonance machine, the selected to compose the dataset are T2W and ADC sequences captured from five diffusion-weighted images with a $b$ value varying between 0 and 500. Both of them are accompanied by the respective prostate mask, and if there are one or more lesions, by the lesion masks. Distinct lesions have different independent masks, even if the patient and the magnetic resonance image are the same. Each sequence was manually segmented to construct the mask of the lesions and the prostate under the premise that despite requiring twice the necessary work, it generates better quality annotations. Concerning the dimensions of each sequence, T2W is the largest of them with 360x360 with a varying number of slices. ADC, on the other hand, is considerably smaller with only 128x128, also with a varying number of slices.

Besides images and corresponding masks, metadata of all lesions of every patient is also gathered to facilitate development of automated methods. It includes scores interpreted from magnetic resonance images, such as the Gleason Grade Group and both the PIRADS and Likert scores. Moreover, a Gleason score given is also provided allowing for potential studies on the overdiagnosis problem too. The analysis and the classification of lesions can be done using several approaches and interpretations of this data. In this project, since, lesions labels indicating clinical significance are not given directly in the dataset, they are inferred based on the provided details concerning each lesion. Similar to the aforementioned mask annotations, experts from the institutions involved in the clinical trial were responsible for given the scores of each lesion.

\begin{figure}[!tbp]
  
  \centering
  \begin{minipage}[b]{0.495\textwidth}
    \includegraphics[width=\textwidth]{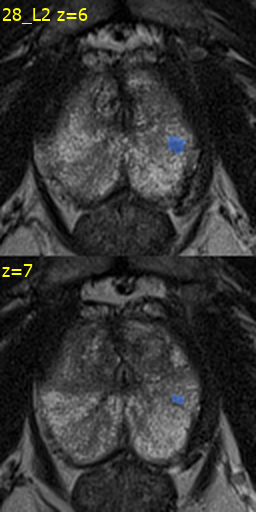}
  \end{minipage}
  \hfill
  \begin{minipage}[b]{0.495\textwidth}
    \includegraphics[width=\textwidth]{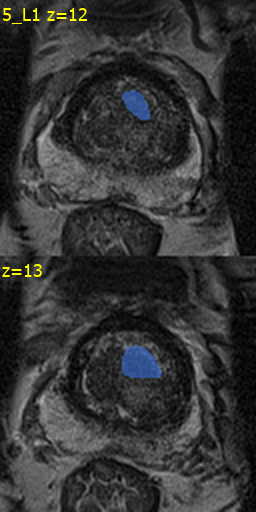}
    
  \end{minipage}
  \caption{Example extracted from the IMPROD dataset of lesions masks over a T2W multi-parametric magnetic resonance image. The figure on the left is the second lesion of the patient identified as 28, with upper and lower images representing respectively the consecutive slices 6 and 7. The figure on the right is the lesion number one of the patient identified as 5, with upper and lower images representing respectively the consecutive slices 12 and 13. Note the lesion size difference between patients and between slices of the same patient.}
  \label{T2W_lesion_mask}
\end{figure}

\section{Clinical Interpretation of the data }

Interpreting the data from a clinical perspective is crucial to understand and solve the problem. Notwithstanding the fact that deep learning architectures implicitly learn feature extraction, other details must be considered in order to properly set up the experiments. For instance, what the meaning and the reason behind the distinction of a lesion as clinically significant, or the three-dimensional representation of the data, and how masks are visualized.  

The images are divided into three dimensions, the first two, height and width, are common to regular two-dimensional images. The third dimension that gives extra spatial information is in this case the slices. This is captured by collecting prostate images in a different position with a given spacing between them, which is relevant to the way the images are processed. Thus, even though the same institution usually captures the images with the same technique, there has been an attempt to create a standard technique to collect these images, allowing them to combine in the future several datasets.

\begin{figure}[!tbp]
 
  \centering
  \begin{minipage}[b]{0.495\textwidth}
    \includegraphics[width=\textwidth]{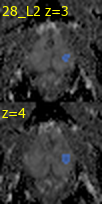}
  \end{minipage}
  \hfill
  \begin{minipage}[b]{0.495\textwidth}
    \includegraphics[width=\textwidth]{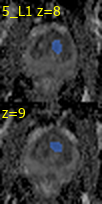}
    
  \end{minipage}
  \caption{Example extracted from the IMPROD dataset of lesions masks over a ADC multi-parametric magnetic resonance image. The figure on the left is the second lesion of the patient identified as 28, with upper and lower images representing respectively the consecutive slices 3 and 4. The figure on the right is the lesion number one of the patient identified as 5, with upper and lower images representing respectively the consecutive slices 8 and 9. Note the lesion size difference between patients and between slices of the same patient.}
  \label{ADC_lesion_mask}
\end{figure}

A clinically significant lesion can be defined through several scoring systems such as the Gleason grade groups (GGG), the PIRADS and Likert scores given from analysis of multi-parametric magnetic resonance images. Some datasets use the real Gleason score verified after the surgery. For this project, the Gleason grade group from the MRI (GS\_MRI) is used. Positive labels are given to lesions that belong to group two, whereas lesions from groups one and zero are classified with a negative label of zero.

\begin{figure}[h!]
    
    \centering
	\includegraphics[width=11.5cm]{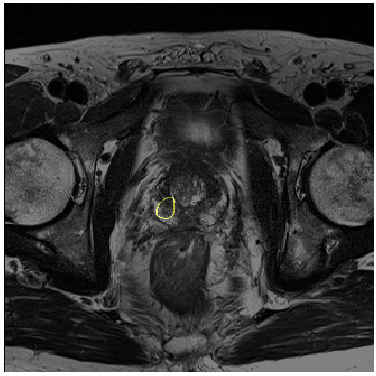} 
	\caption{Example of a lesion contour in a T2W sequence. This contour is drawn from the segmentation mask and it is only used from visualization purposes. Slice 13 from the patient with id 10.} 
    \label{countour_segmentation}
\end{figure}

Lesions are much less represented in the dataset than the prostate, thus generating less positive labels and an increasingly difficult task due to a more skewed dataset. Figure \ref{T2W_lesion_mask} shows the visualization of lesions masks in T2W sequences. In the first analyzes of the image, it is perceptible that lesions do not have the same size. Comparing this image to the one in Figure \ref{ADC_lesion_mask} denotes two key details, firstly the resolution of the latter, representing ADC sequences, is inferior. Secondly, both figures include the same lesions, however, they are visible in different slices of each sequence, in other words, these sequences are not registred. Moreover, in Figure \ref{countour_segmentation} it is possible to see the contour of the lesion drawn from the segmentation mask.

\section{Dataset statistics}

The size and the quality of the annotations are undoubtedly important to the quality of the learning. However, some other key factors might influence the ability of the model to generalize to or work properly with all the classes and potential requirements of the problem. For instance, as previously mentioned, the classification of both clinically significant lesions or the PIRADS score is one of the problems that this thesis is trying to solve, and they can be affected by skewed datasets. In other words, a dataset where one label has a much significant representation than the others, or when one label does not have proper representation (e.g. 95\% of the dataset from the same class).

\begin{figure}[h!]
    
    \centering
	\includegraphics[width=11cm]{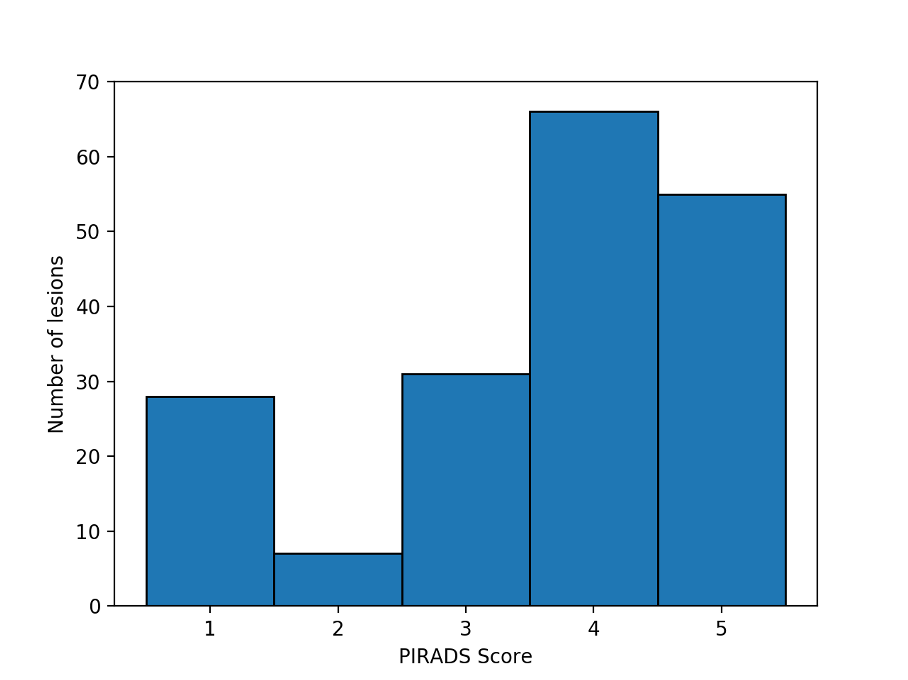} 
	\caption{Distribution of the PIRADS score across lesions. This histogram includes all the images and lesions in the dataset, even those that do not have individual lesion masks, such as most of the lesions with a PIRADS score of 1.} 
    \label{PIRADS_DISTRIBUTION}
\end{figure}

\begin{figure}[h!]
    
    \centering
	\includegraphics[width=11cm]{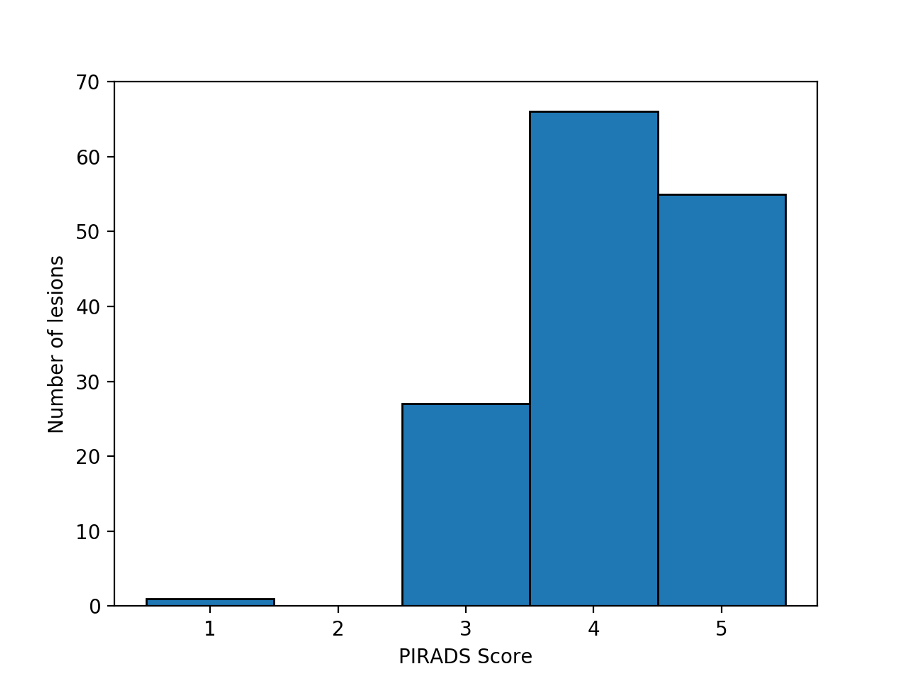} 
	\caption{Distribution of the PIRADS score across lesions. This histogram includes all the images and lesions in the dataset, even those that do not have individual lesion masks, such as most of the lesions with a PIRADS score of 1.} 
    \label{PIRADS_DISTRIBUTION_REMOVED}
\end{figure}

All the lesions in the dataset are classified by experts according to the PIRADS system, and one of the important points to the prediction of this score is to see if the dataset has a proper representation of all classes. Thus, in Figure \ref{PIRADS_DISTRIBUTION} a histogram with the distribution of the scores is shown. The most represented labels are those of the scores four and five, accounting for more than 60\% of the labels. On the other hand, label two has a poor representation of less than 10\% of the data. Despite being present in this histogram, not all the lesions that have a PIRADS score have an independent mask, so it is impossible to locate them in the magnetic resonance image. Thus, since the lesion cannot be located through the mask coordinates they are excluded from all the final classification problems. Furthermore, these lesions are lower grade lesions of labels one and two, which might become considerably problematic due to the lack of representation of these. The distribution of the PIRADS scores after the exclusion of lesions without mask can be seen in Figure \ref{PIRADS_DISTRIBUTION_REMOVED}. It is relevant to note that there is no lesion with a score two and only one with score one.

\begin{figure}[h!]
    \centering
	\includegraphics[width=11cm]{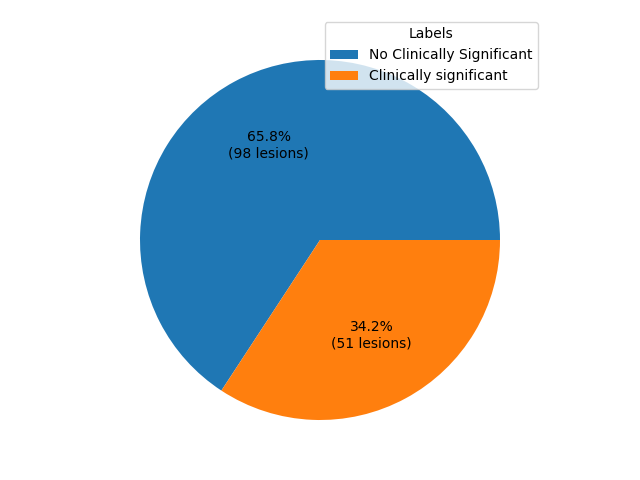} 
	\caption{Distribution of the Clinically Significant (positive label) and No Clinically Significant Lesions (negative label). The former is represented by the orange slice, while the latter is represented by the blue slice and represents most of the lesions. This pie plot only includes the lesions which have mask representations of it and can be used in classification problems. } 
    \label{Significance_DISTRIBUTION}
\end{figure}

After removing the lesions that do not have masks and cannot be located, it is possible to start studying the distribution of clinically significant lesions. The exclusion of the lesion leads to a considerable reduction in the number of available samples for classification problems with only 149 lesions to be used. In figure \ref{Significance_DISTRIBUTION} a pie chart shows the representation of the positive and negative labels in the dataset of lesions that can be cropped and classified by a machine learning model. In contrast to what is seen in Figure \ref{PIRADS_DISTRIBUTION}, since clinical significance is not, in the context of this thesis, related to the PIRADS score, the presence of no clinically significant lesions is much higher. The majority of the dataset, 65.8\%, has a negative label, where the remaining samples 34.2\% are clinically significant, thus, associated with a positive label.

Lesion characteristics have an impact on how they are classified, and some of those can be studied in order to optimize the solution search space or understand the problems related to one unfitting approach.  The size of the lesion is one of those elements that are relevant to decide how to approach the problem. Through this several analyses can be done, such as the height-width relationship, the area of lesions and the direct impact of those in the clinical significance of the lesions. All the images were resampled to a 224x224 height and width while keeping intact the number of slices.

\begin{figure}[h!]
    \centering
	\includegraphics[width=11cm]{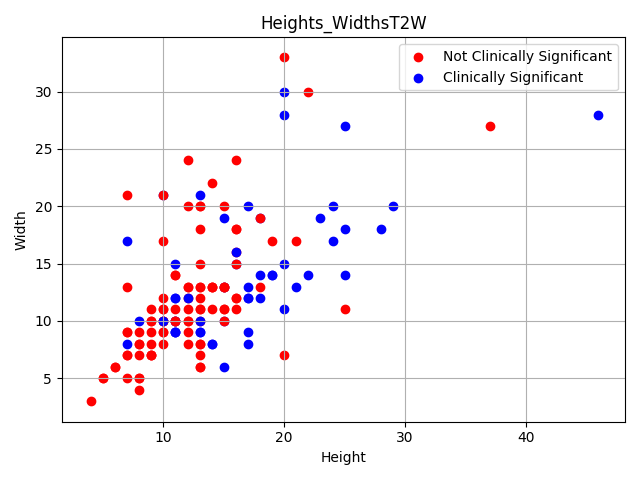} 
	\caption{Scatter plot displaying the relationship between the weight and the height of lesions in T2W sequences in the slice that maximizes the area of the lesion. Only lesions with masks are used to generate this plot that also indicates with blue if the lesion is clinically significant or with red if it is not clinically significant. This allows to understand the influence of the size in the significance. Lesions are measured after a resampling of the images from 360x360 to 224x224 keeping the same number of slices.} 
    \label{T2W_scatter_lesions}
\end{figure}

\begin{figure}[h!]
    \centering
	\includegraphics[width=11cm]{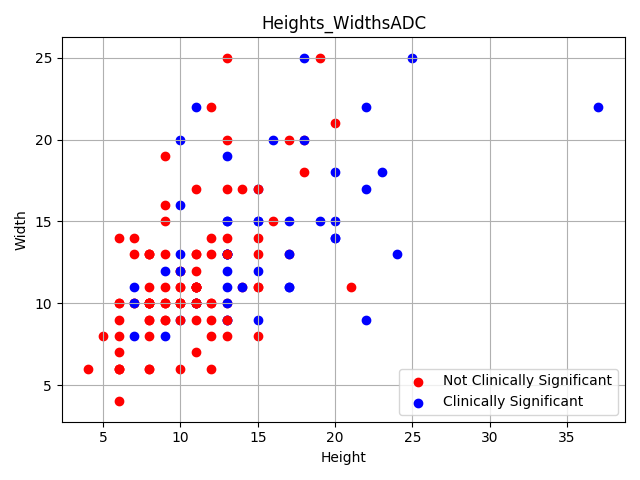} 
	\caption{Scatter plot displaying the relationship between the weight and the height of lesions in ADC sequences in the slice that maximizes the area of the lesion. Only lesions with masks are used to generate this plot that also indicates with blue if the lesion is clinically significant or with red if it is not clinically significant. This allows to understand the influence of the size in the significance. Lesions are measured after a resampling of the images from 128x128 to 224x224 keeping the same number of slices.} 
    \label{ADC_scatter_lesions}
\end{figure}

A potential relationship between the height and the size of the lesions is explored in T2-weighted, in Figure \ref{T2W_scatter_lesions}  and in apparent diffusion coefficient images, in Figure \ref{ADC_scatter_lesions}. Despite the original size difference, with the resampling both of the scatter plots show provide information to acquire similar knowledge and draw identical conclusions from them. T2W lesions width values are between 2$<W_{T2W}<$35 while for ADC the values are 2$<W_{ADC}<$25. The value fore the height, are 2$<H_{T2W}<$50 and 2$<H_{ADC}<$40 respectively. It is also worth noting, that the absolute difference between the height and the width is in the majority of the cases limited to ten, indicating that both dimensions have similar values. Moreover, a more careful analysis denotes that the size seems to have an influence on the classification of the lesions as clinically significant. It is shown that lesions with both dimensions with values below ten do not have, with a high degree of certainty, clinical significance. However, as the size increase, the impact in the classification seems to be blander, and although having a larger size denotes more likelihood of being clinically significant, it is not a sufficiently accurate indicator to draw any conclusion related to the significance. Both images show several lesions with high dimensions and negative labels. In \ref{T2W_scatter_lesions}  the resampling has an unnoticeable effect on the distribution of the sizes, but since ADC sequences were resampled to a higher resolution it is possible to notice in Figure \ref{ADC_scatter_lesions} small consistent gaps between the sizes.

Several strategies can be used in order to locate, crop and prepare a lesion to be classified. Yet, not all the strategies are adequate for this problem, and the selection of the strategy to be used is highly dependent on the range of sizes of lesions. For instance, if there is a broad range of sizes, a technique that just does a crop of a fixed size centered in the center of the lesion might be problematic. Smaller lesions crops will include an excessive amount of background, which might interfere with the learning since the region of interest is insignificant when compared with the surrounding. On the other hand, if a lesion is larger than the predefined crop size important parts for the classification might be cropped out, also hurting the performance of the model. Therefore, knowing the size of the lesions and how the cropping technique might affect the performance is valuable to understand the reasons behind the success or lack of it in further experiments.  

\begin{figure}[h!]
    \centering
	\includegraphics[width=11cm]{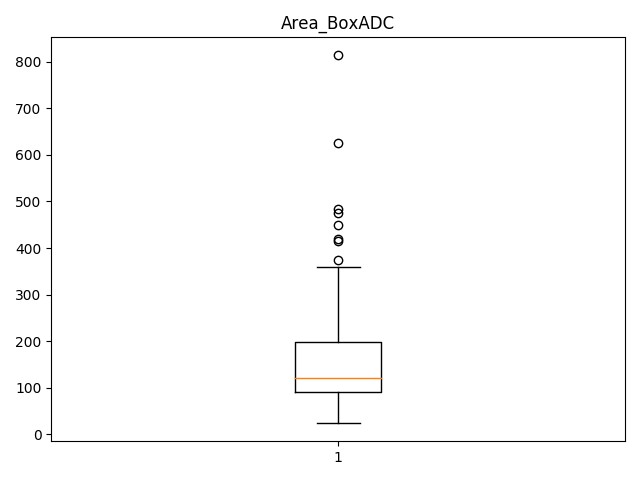} 
	\caption{Box plot of the area occupied by ADC lesions. This takes into account only the size of the lesion in the slice that maximizes its size, and only lesions with masks are used to generate this data. Lesions are measured after a resampling of the images from 128x128 to 224x224 keeping the same number of slices.} 
    \label{ADC_AREA_DISTRIBUTION}
\end{figure}

\begin{figure}[h!]
    \centering
	\includegraphics[width=11cm]{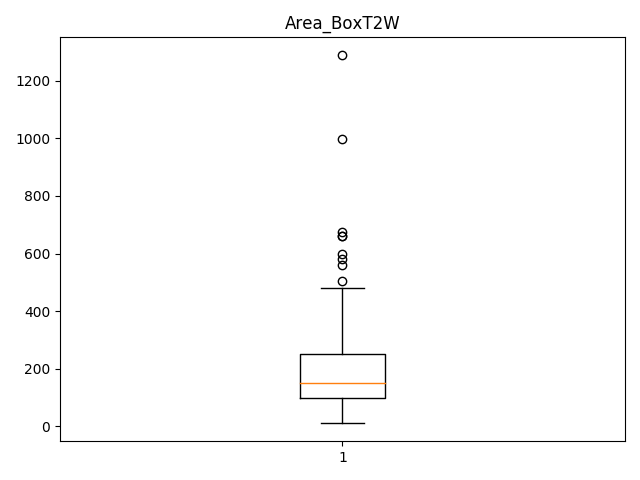} 
	\caption{Box plot of the area occupied by T2W lesions. This takes into account only the size of the lesion in the slice that maximizes its size, and only lesions with masks are used to generate this data. Lesions are measured after a resampling of the images from 360x360 to 224x224 keeping the same number of slices.} 
    \label{T2W_AREA_DISTRIBUTION}
\end{figure}

Implications of the size of lesions are broad and affect other tasks such as the segmentation of those lesions. To start with, a larger lesion area impacts directly the number of positive labels in each mask, for example, smaller lesions have fewer pixels thus fewer labels, and this interferes with how skewed the dataset is. Secondly, a considerable variation in the area occupied by lesions might affect negatively the performance, since models sometimes deduct correlations that might make the predictions more prone to a specific size. To analyze the occupied area of lesions, the area of one lesion was considered to be approximated by the multiplication of its width and height dimensions. However, it was only considered the slice where each lesion had the biggest approximated area since it is the same slice to be used in classification problems. In Figure \ref{ADC_AREA_DISTRIBUTION} and \ref{T2W_AREA_DISTRIBUTION}, a box plot of this area in ADC and T2W sequences is shown respectively.  And even though both show similar means, and as a consequence of higher width and height maximum values, T2W sequences are more prone to outliers with larger areas. This might imply problems when cropping the lesions.

\clearpage

\chapter{Methods}
\label{Chap:methods}

In the context of this thesis, there are two central research directions. The first of them is to approach the data to work in a classification problem, more specifically, the classification of lesions. Whereas the other focuses on detection and segmentation from the multi-parametric magnetic resonance images. These problems require distinct approaches and methodologies, as well as computing power, which shall be taken into consideration while designing the model.

The classification of lesions is a broad domain due to the most variate scores scales given to the lesions, hence this problem can be divided into two smaller problems that follow more narrow guidelines. To start with, a binary classification problem based on the prediction of the clinical significance of the lesions. And, another more complex problem aiming to predict the PIRADS score, thus a multi-class classification problem.  Both of these problems share common characteristics, however, dataset differences and other smaller details entitle each of them to be individually researched and explored.

Despite employing similar methods, detecting and segmenting lesions or the prostate are two distinct problems. On one hand, the same loss functions,  models or data processing techniques can be used. On the other hand,  the methods working in one task might perform poorly on the other one or even show strange behavior.

This Chapter's focus is to establish the research grounds used in the context of this thesis, as well as discuss other factors such as research questions, problems to be solved and implications of the chosen methods. Furthermore, it will be explained how the conducted research will proceed for the experiments and study alternative solutions and techniques.  

\section{Lesion classification}

Prostate cancer lesions vary in shape, size, and intensity, even their aggressiveness changes accordingly. Being able to define and categorize the aggressiveness of a lesion is one of the roles of experts when dealing with multi-parametric magnetic resonance images of the prostate. Moreover, the classification can be performed according to the guidelines of several score metrics and scales, requiring a deep knowledge of the problem and years of study in order to classify them correctly. Yet, it is a tedious, expensive and prone to errors task, since there is a considerable quantity of bad diagnosis, especially overdiagnosis with some lesions being classified as more aggressive than what they actually are.

In an attempt to solve the classification problem, two smaller tasks were undertaken and similar methods were employed. For both problems the input is the same, so the base model to tackle them was the same, with some minor differences and some other distinctions in the optimization and evaluation process.

\subsection{Data}

The original format of the data was not appropriate for the problems at hand. Therefore, some previous processing was required to generate appropriate input and labels that could be used in these tasks. Originally the data was available as the 3D representations of magnetic resonance images of the prostate and surrounding organs. However, the expected format to solve these problems was a 2D representation of the lesions, avoiding as much as possible unnecessary body elements.

\begin{figure}[!h]
  \centering
  \begin{minipage}[b]{0.49\textwidth}
    \includegraphics[width=\textwidth]{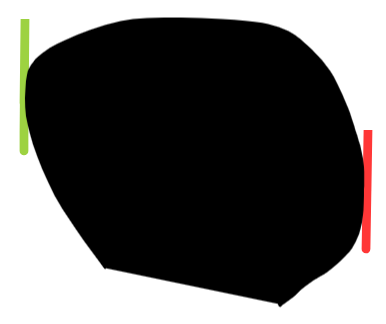}
    
  \end{minipage}
  \hfill
  \begin{minipage}[b]{0.49\textwidth}
    \includegraphics[width=\textwidth]{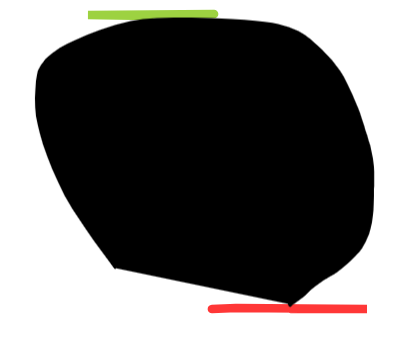}
  \end{minipage}
  
  \caption{Example of how the lesions are located and their area approximated. The lesion is represented by the black area. In the image in the left, the green line represents the left most point of the lesions while the red represents the point that is more to the right. In the image in the write the green represent the point that has the biggest height and the red line represents the point with lowest height. This points can be used to approximate the lesion area to a square, or to point the center of the lesion. }
  \label{width_height_crops}
  
\end{figure}

The first step to attain the desired data format was to spatially locate the lesions in the 3D space. Lesions were previously detected and marked by experts, therefore the use of the masks to locate them is possible. This also means that lesions that appear in the data but do not contain an individual mask cannot be part of it and need to be excluded. To efficiently locate the lesions, their domain area was approximated, and the initial and final points were marked independently in every slice as shown in 
Figure \ref{width_height_crops}. In the figure, it is possible to see at the green starting points and the red ending points if the lesion is interpreted from top left to bottom right. These points can be used to draw a bounding box without predefined size that contains the entire lesion and potentially some background too. Alternatively, they can be used to approximate center and draw a fixed size bounding box that might crop the lesion, or capture almost exclusively background. 

Following the location of the lesion at each slice, the approximated area of these lesions was calculated on a slice basis. The slices were then ordered based on the largest lesions size, and the one that maximized the size was chosen. A crop to the lesion was performed in that slice and the output of the cropping was saved as an image to input to the classification model. The label of the lesion was respectively associated with the new two-dimensional image. 

\subsection{XmasNet}

In all the machine learning tasks the model is as important as the data, and in this case, the same happens. Previous challenges such as ProstateX already tried to address the clinical significance classification problem. As a result, many submissions attained reasonable values at this task, therefore, one of the submissions was selected to be the base of the classification system developed in this thesis.

\begin{figure}[h!]
    \centering
	\includegraphics[width=13cm]{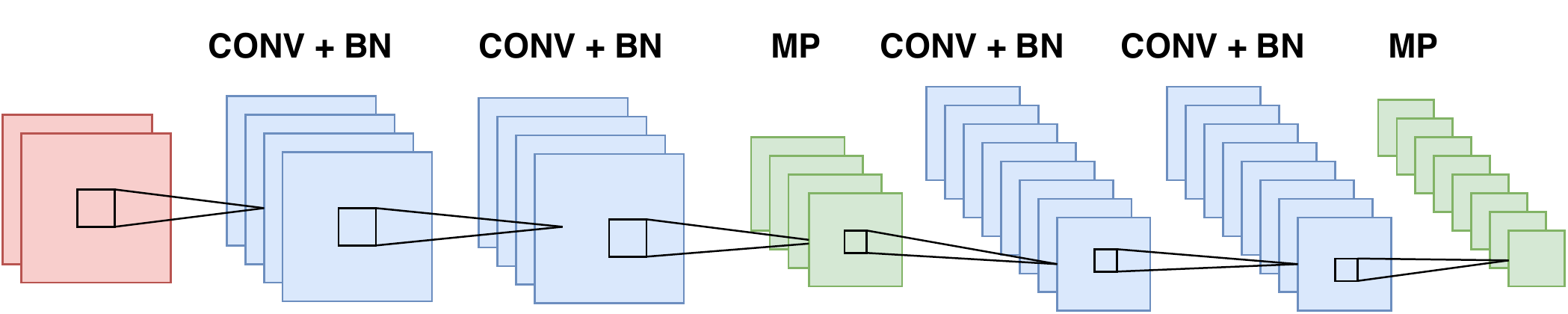} 
	\caption{XmasNet \cite{Liu2017} based network architecture with 2 input channels, and without fully-connected layers. Red squares represent each one channel of the input layer. Blue squares represent each 8 channels of a convolutional layer with a kernel size of 3, padding of 1 and stride of 1 plus a batch normalization layer right after. Green layers represent each max pooling layers with kernel size of 2 and stride of 2.   } 
    \label{xmas_base}
\end{figure}

The selected architecture, named XmasNet achieved the second place with an AUC of 0.84 in the ProstateX challenge surpassing traditional machine learning techniques \cite{Liu2017}. Additionally to the model architecture, the submission presented also a technique to preprocess the data and feed it to the neural network. This preprocessing is considerably different from that was used in this project, therefore the only inspiration is the model \cite{Liu2017}. 

Figure \ref{xmas_base} partially shows the model, more specifically, it shows the common layers to both classification problems. This part is responsible for extracting useful information from the lesions, in other words, the features of the lesions, regarding for instance, shape, size or intensity. The architecture despite being shown to have two input channels works also with one input channel, depending on how many sequences are used. The input is followed by a convolution that increases the number of channels to 32 and another convolution that keeps the number of channels. To these, a max-pooling layer follows, down-scaling the size by half, and feeding it to a convolutional layer that will double the number of channels. Another convolution and max-pooling layer follow, keeping the number of channels and halving the size again. It is also worth noting that every convolution has a kernel size of three and padding of one and they are followed by a batch-normalization layer and ReLU non-linearity. Max-pooling layers have both kernel and stride set to two.

\subsection{Clinical significance classification}

Predicting if a lesion has clinical significance or not is a typical machine learning binary classification problem. In this sort of problem, the model predicts if an input belongs to the positive or negative classes which in this case are being clinically significant or not respectively. The output of the network is one single value, between zero and one, the probability of belonging to the positive class. 

\begin{figure}[h!]
    \centering
	\includegraphics[width=11cm]{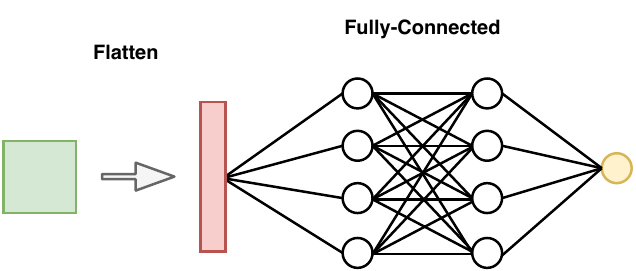} 
	\caption{Additional layers to be added to the network for a binary classification of the clinical significance of lesions. Green square represents the last max pooling layer. Arrow to the write represents the flatten of the square data to a vector that can be fed to fully convolutional layers. Yellow circle represents the output followed by a sigmoid activation.} 
    \label{xmas_bc}
\end{figure}

In order to tackle this task, to the partial convolutional model shown in Figure \ref{xmas_base}, some extra layers, as seen in Figure \ref{xmas_bc}, were appended. These layers are responsible for using the extracted features from the convolutional part and compute from them the prediction, and are specific for the binary classification problem. The output of the last max-pooling layer is flattened to one single dimension, besides the batch dimension, and is then fed as input to three fully-connected layers. The first layer receives the flat vector that has a length dependent on the size of the lesion image given to the network accordingly to the formula $ \frac{height * width}{4} * 64 = length$, for example, a 32x32 lesion image will originate a vector of length of $8^2*64 = 4096$. This first layer connects this input vector to 1024 nodes, while the second layer connects the 1024 node to the 256 following nodes. Finally, the third and last layer generates one output from all of the 256 nodes and applies a Sigmoid function in order to convert it to a probability. The first two layers are followed by ReLU non-linearity.

\begin{equation} \label{binary_cross_entropy_eq}
\begin{split}
BCE(y,t) & = -t*log(y) - (1 - t)*log(1 - y))\\
 & = \left\{\begin{matrix} & - log(y) & & if & t = 1 \\ & - log(1 - y) & & if & t = 0 \end{matrix}\right.
\end{split}
\end{equation}

To optimize the network and to approximate the output the binary cross-entropy loss was minimized. The expression for this loss function can be seen in the Equation \ref{binary_cross_entropy_eq} where $t$ stands for the expected class (one or zero), and $y$ is the output given by the network that ranges between zero and one. This loss function is a special case of the general cross-entropy loss function and is frequently used in binary classification problems. Moreover, the stochastic optimization was performed by the Adam optimizer with a learning rate of $7 * 10^{-4}$ and a weight decay of  $10^{-4}$. Each step computed the results and afterward the gradients of a mini-batch of size 8.

\subsection{PI-RADS classification}

Classifying a lesion as clinically significant or not is a useful task, however, it can frequently be seen as rather limiting and lacking detail. Therefore, in a clinical environment, the PIRADS score is a better classification system for the lesions, since it does try to measure the likelihood of a lesion to be from a clinical significance cancer or not. In practical terms, it means that the network had to be changed in order to produce five different outputs. Moreover, not only the outputs were required to be between zero and one, but their sum had to result in the value one.

\begin{figure}[h!]
    \centering
	\includegraphics[width=11cm]{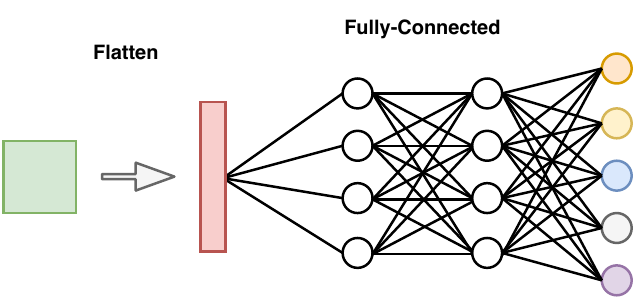} 
	\caption{Additional layers to be added to the network for a multi-class classification of the PIRADS score of lesions with five possible outputs. Green square represents the last max pooling layer. Arrow to the write represents the flatten of the square data to a vector that can be fed to fully convolutional layers. Yellow circles represents the outputs followed by a softmax activation. } 
    \label{xmas_mc}
\end{figure}

The model used in this task was a variation from the partial convolutional model displayed in Figure \ref{xmas_base}. To the last max-pooling layers, some other layers were added, such layers are seen in Figure \ref{xmas_mc}. For the most part, these layers work similarly to the ones approached in the previous problem, with the flattening of the matrix, the input dependent number of nodes and the node count of posterior layers. Nevertheless, the output was in this problem represented by five different nodes, where each node represents a class or a score from the PIRADS scale. Moreover, to the output a Softmax function is applied, so that they can be used to minimize the cross-entropy loss.

\begin{equation} \label{_cross_entropy_eq}
\begin{split}
CE(y,t) & = -log\left(\frac{e^{y_{t}}}{\sum_{j}^{C} e^{y_{j}}} \right )\\
 & = -y_{t} + log\left(\sum_{j}^{C} e^{y_{j}} \right )
\end{split}
\end{equation}

The optimization of the network was similar to the one described in the previous problem, although there were small differences. To start with, the learning size was decreased to a value of $5 * 10^{-5}$ to ensure smoother convergence, while the optimizer and the weight decay were kept the same. The loss function was also slightly different since it had to handle all the five inputs and the five classes. Cross-entropy loss was used and its equation is represented in Equation \ref{_cross_entropy_eq}, where $y$ represents the predictions vector and $t$ the ground truth class. Moreover, as seen in the equation $t$ can also be used as an index of the vector $y$.

\section{Detection and segmentation}

Analyzing a medical image is not an effortless simple task, not only it requires a considerable amount of training and expertise, but due to low resolution or the format of the images, some elements such as organs and their boundaries are difficult to locate. A machine learning segmentation model tries not only to detect these elements but to segment them and delineate their boundaries. In the multi-parametric magnetic resonance images, two segmentation tasks can be defined and tackled, the segmentation of the prostate and the segmentation of prostate cancer lesions. Methods for both these tasks are discussed in this section, and how the solution for each one was approached.

For the most part, the tasks are similar with the main goal being to predict a mask close to the ground truth mask. Therefore, similar methods were used in both cases, similar architectures trained and tested for both problems, and the same loss functions were also used. However, this section focus on similarities further details on how the problems differ are discussed in the following sections. 

\subsection{Loss function}

There are several loss functions that could have been applied to the problem at hand, however, only dice loss was picked. This loss function comes from the similarity measure Sørensen–Dice coefficient applied to a three-dimensional space.

\begin{equation} \label{_3d_dice_eq_binary}
\begin{split}
DICE & =\frac{2*TP}{2*TP+ FN + FP} \\
\end{split}
\end{equation}

 The formula to calculate the Dice coefficient or score is in Equation \ref{_3d_dice_eq_binary} in terms of binary data that composes the masks. The intuition behind this score is to measure the correctly predicted values divided by the sum of those values and all the values predicted wrongly. A perfect value of one in this function would require that no label of one is incorrectly predicted as zero (a false negative) and that no label of zero is wrongly predicted as one (a false positive). Although the binary formulation works perfectly in an evaluation setup, it is not appropriate for training.

\begin{equation} \label{_3d_dice_eq}
\begin{split}
DICE(Y,M) & =\frac{2 * Y * M}{Y * Y + M * M} \\
\end{split}
\end{equation}

During the training of the model, the mask predictions are supposed to be given in probabilities, in other words, each pixel in the three-dimensional space is supposed to have a probability of being classified with a positive label. While in the final output to test a prediction a threshold can be established to attribute a label of one after that specific value, during training the loss function must be able to work with the probabilities. Therefore, in order to use the Dice coefficient effectively in these problems, and to approximate the value of the real mask, the formulation to be used is shown in the Equation \ref{_3d_dice_eq}. For that formulation $Y$ represents the predicted mask while $M$ represents the target mask.

\begin{equation} \label{_3d_dice_loss_eq}
\begin{split}
DICE\_L(Y,M) & =1 - DICE(Y,M) \\
\end{split}
\end{equation}

Moreover, a loss function usually measures the distance between prediction and target and not the similarity, hence, the Dice Score in Equation \ref{_3d_dice_eq} can be further adapted to represent a function to be minimized. The adaptation is rather simple, and it does only require that to a value of one, the Dice Score is subtracted. The formulation for the loss is given in Equation \ref{_3d_dice_loss_eq}. Despite not being used directly as loss, the Dice Score is also a great tool to measure the similarity and evaluate the performance of the model further on.

\begin{equation} \label{combined_dice_eq}
\begin{split}
Comb(Y,M,n) =  \frac{1.5 *\sum_{i}^{n}{BCE(Y_i,M_i)}}{n} + DICE\_L(Y,M)
\end{split}
\end{equation}

Despite the wide use of Dice loss to optimize models in image segmentation tasks, there are other losses that can be explored and combinations of losses. For this thesis, a weighted combination of the dice loss with the binary cross-entropy loss was used. The formulation of this loss can be seen in Equation \ref{combined_dice_eq} and the weights of both are respectively 1 and 1.5. Uniting losses requires some thought and attention, for this particular task, the combination of these was conceived in order to try and improve the overlapping of the predicted and ground truth mask. 

\subsection{Models}

In terms of the architectures to tackle these problems, the first model to be used was a 3D U-Net, that was specifically designed for segmentation in medical imaging problems \cite{Cicek2016}. 

\begin{figure}[h!]
    \centering
	\includegraphics[width=13cm]{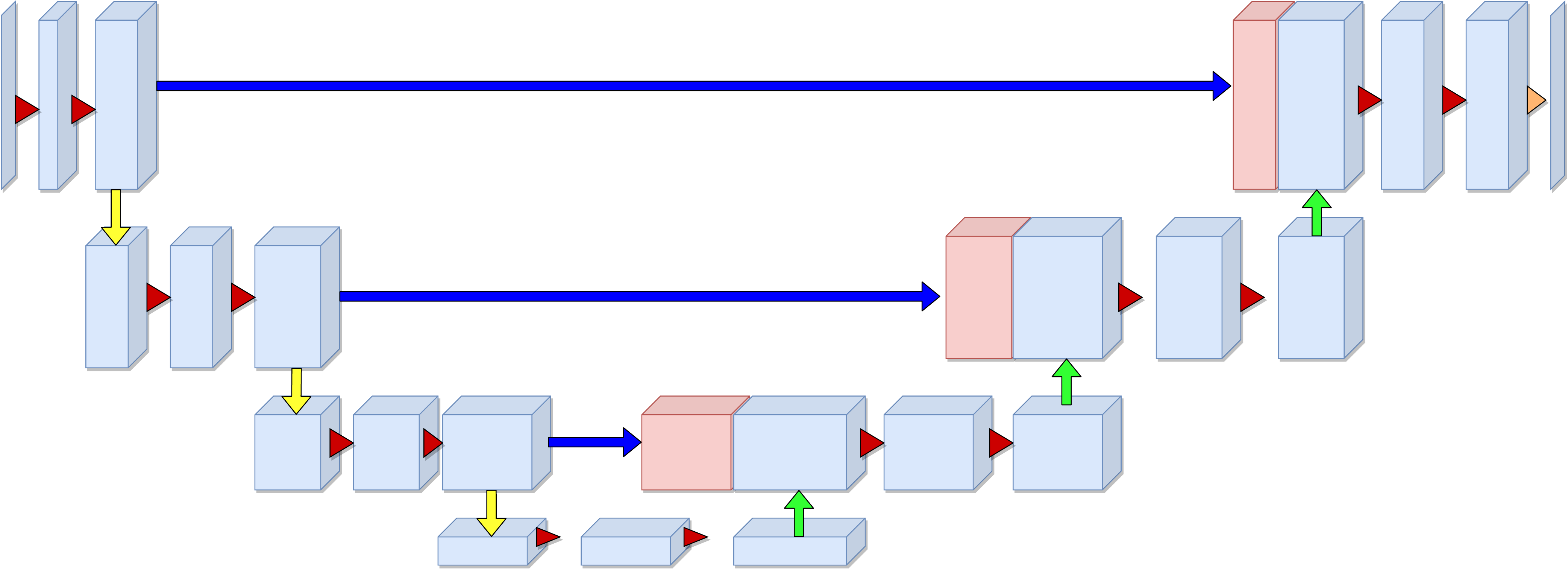} 
	\caption{3D U-Net \cite{Cicek2016} architecture. Blue arrows represent a concatenation operation (blue cubes in the tail of the arrow are seen as red cubes in the head of the arrow). Red arrows represent a 3D convolution followed by batch normalization and ReLU. Yellow arrows are max pooling operations while green arrows are up-convolutions. The final orange arrow represents simply a convolution to generate the output.  } 
    \label{u-net_architecture}
\end{figure}

 The 3D U-Net model architecture can be seen in Figure \ref{u-net_architecture}. It is characterized by two main features in its design. To start with the model has "deep levels" where after two convolutions operations across three dimensions, the input is downscaled to half of the size and the number of channels doubled. After reaching the desired depth, in this case after three downscaling operations, the input is upscaled until it returns to the original dimension. In spite of having, between upscaling operations, 3D convolutions that do not change the size, the first convolution is responsible for reducing the number of channels. Another particularity of this model is the shortcut connections between levels, where the output of one level in the downscaling side is concatenated to the input of the same level in the upsample side. These shortcut connections are especially important to ensure that no information is lost between all the convolutions, and to ensure that spatial information is preserved from the original image and through the entire downscale-upscale process.

\begin{table}[h!]
    \centering
\caption{Comparison of the number of parameters of different models}
\label{tab:parameters_models}
\begin{center}
 \begin{tabular}{|c c c |} 
 \hline
 Code & Model name & Parameters\\ [0.5ex] 
 \hline
 U & 3D U-Net & 19,069,955 \cite{Cicek2016}\\ 
 \hline
 R & 3D ResNet-18 &  32,990,000 \cite{chen2019med3d}  \\[1ex] 
 \hline
\end{tabular}
\end{center}
\end{table}

The second model used in these problems was a 3D ResNet-18 \cite{Hara} that is considerably more complex than 3D U-Net as can be seen in Table \ref{tab:parameters_models}. However, the complexity does not guarantee better learning due to the considerable difference between architectures and different types of connections. The construction of this model is based on building blocks, and in this thesis, the original model was slightly adapted to work with the dimensions of the input.

\begin{figure}[!h]
  \centering
  \begin{minipage}[b]{0.40\textwidth}
    \includegraphics[width=\textwidth]{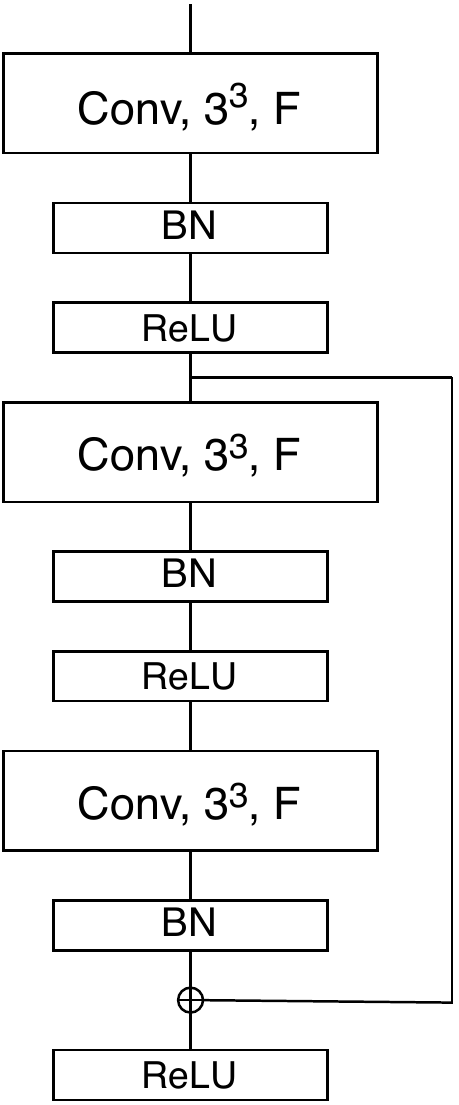}
    
  \end{minipage}
  \hfill
  \begin{minipage}[b]{0.40\textwidth}
    \includegraphics[width=\textwidth]{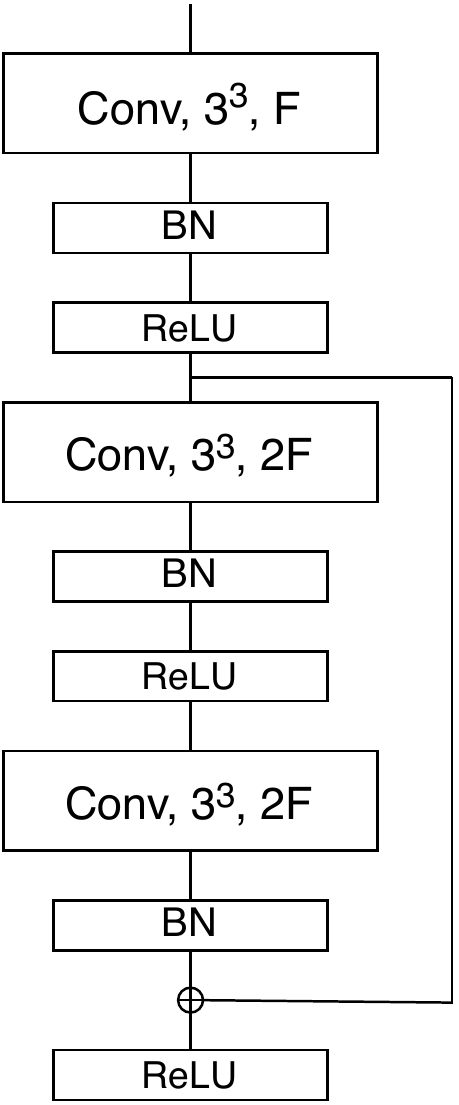}
  \end{minipage}
  
  \caption{Adapted  architecture of the basic block for three dimensional residual convolutional neural networks (3DResNet)\cite{Hara}. This block is composed by the following components: BN as Batch Normalization; ReLU as the Rectified Linear Unit nonlinearity; and Conv, $X^3$, F as a convolution with a kernel of size X across the three dimensions and that maps the input to an output with F channels.  Basic block on the left is the Block A while the one on the right is Block B. }
  \label{basic_block_architecture}
  
\end{figure}

The adapted building blocks, seen in Figure \ref{basic_block_architecture}, differently from the original version have a stride of size one, removing the downsampling properties of these layers. These building blocks have three convolutional layers each followed by a batch normalization layer and a ReLU activation. The difference between both blocks is that the block on the left (Block A) does not change the number of channels, whereas Block B doubles the number of channels in the first convolution. Moreover, both versions include a shortcut residual connection between the output of the first convolutional layer and the output of the third convolutional layer. 

\begin{figure}[h!]
    \centering
	\includegraphics[height=14cm]{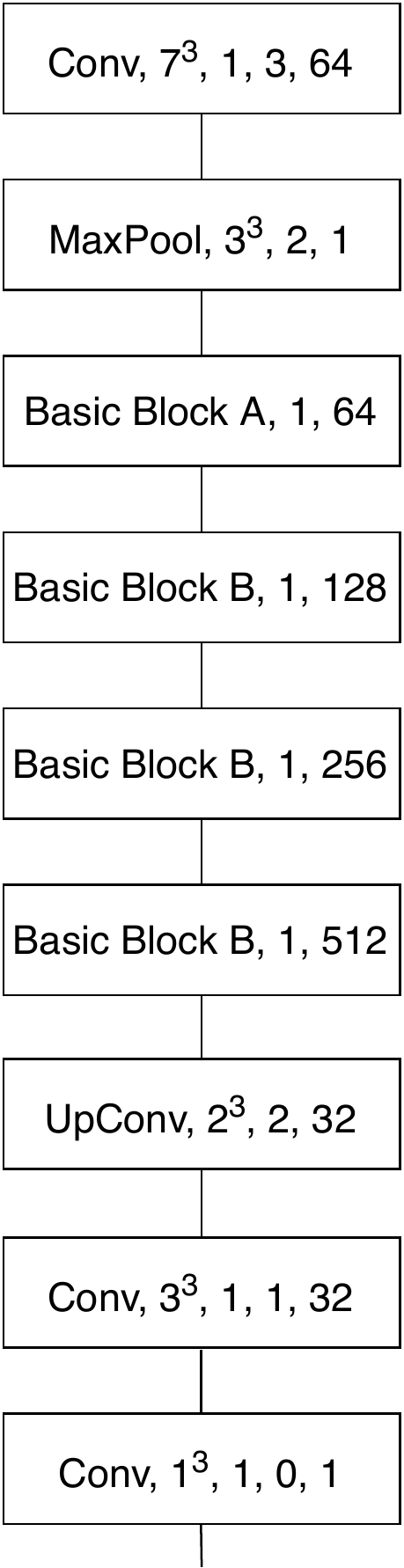} 
	\caption{Adapted architecture of a 3DResNet-18 \cite{Hara2} from the MedicalNet project, which trained several three dimensional neural networks in a combined dataset of various organs \cite{chen2019med3d}. The weights of the network are available and any adaptation of the network did not affect the original parameters. } 
    \label{3dresnet_architecture}
\end{figure}

The complete architecture is seen in Figure \ref{3dresnet_architecture}. It is composed of several instances of the basic blocks, as well as other layers. The networks start by convolving the input with a 7x7x7 kernel and mapping one channel to 64. The following max-pooling layer downsamples to half the size and then feeds it to a sequence of four basic blocks, with the first being of type A and the following of type B. The original size is then restored by an up-convolution and processed by two more convolutional layers. The final output has one channel subject to sigmoid at each pixel. The network used was already pretrained in another dataset for other organs \cite{chen2019med3d} and the weights were used as initial weights of this model. 

\subsection{Data augmentations}

Neither the prostate cancer lesion segmentation problem nor the prostate segmentation problem had abundant data and examples to learn from. As seen before, machine learning and especially deep learning algorithms rely considerably on the number of samples in the training set that represent potentially different situations and variability between samples. Moreover, variability in the data and a considerable number of distinct samples is a mechanism to avoid also overfitting and improve the generalization capabilities of the model at hand.

\begin{figure}[!h]
  \centering
  \begin{minipage}[b]{0.495\textwidth}
    \includegraphics[width=\textwidth]{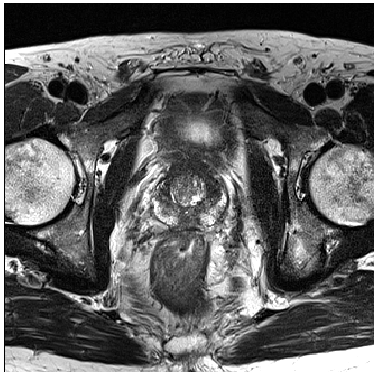}
    
  \end{minipage}
  \hfill
  \begin{minipage}[b]{0.495\textwidth}
    \includegraphics[width=\textwidth]{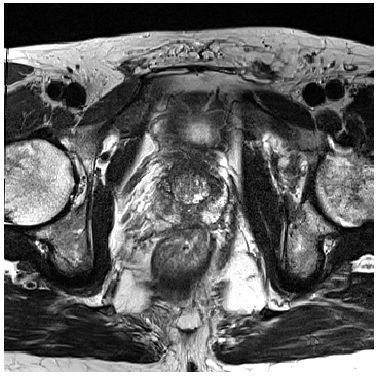}
  \end{minipage}
  
  \begin{minipage}[b]{0.495\textwidth}
    \includegraphics[width=\textwidth]{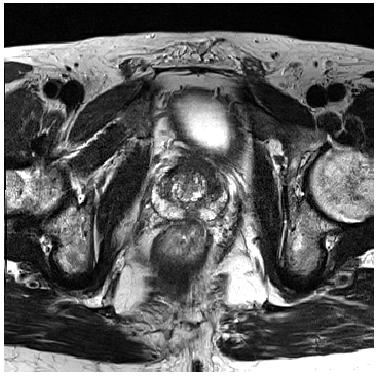}
    
  \end{minipage}
  \hfill
  \begin{minipage}[b]{0.495\textwidth}
    \includegraphics[width=\textwidth]{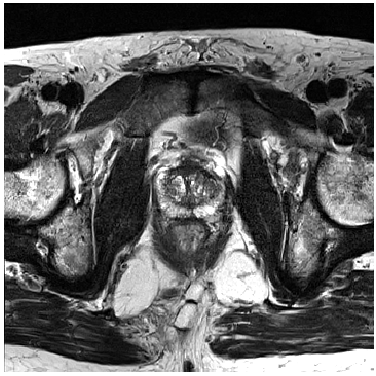}
  \end{minipage}
  
  \caption{Example of a elastic deformation augmentation applied to a T2W image. The original image is represented in the top left corner and the other are generated by deformations with a spline order of 3, an alpha of 1000, and a sigma of 40. The deformations are applied to the whole 3D, however, in this image only results in slice 13 are shown. This T2W sequence is from the IMPROD dataset. }
  \label{elastic_deformation}
  
\end{figure}

In Figure \ref{elastic_deformation} it is shown an example of elastic deformation, that was one of the augmentations used. The application of this transformation to the image and to the mask increases the variability of the data by giving slightly different shapes and sizes to the prostate cancer lesions. Other transformations were also applied, such as random contrast, random Poisson noise, random Gaussian noise, and random rotations. 

\subsection{Prostate segmentation}

The main goal of this problem is to predict the segmentation mask of the prostate from multi-parametric magnetic resonance images. This problem can be seen from two perspectives, first, it was attempted to achieve an overlapping between the ground truth and the predicted mask. And it can also be seen as the attempt to classify individually each pixel with a positive or negative label. 

Due to being present and visible in a considerable number of slices in each MRI, and also to the fact that the prostate occupies a considerable part of the slice when visible, the number of positive labels is abundant. Moreover, size and shape variations between patient prostates exist, but they are small and possible to learn by the model. The problem is, therefore, simpler than other versions, for example, lesion classification, and requires less tuning to attain the expected results. 

This problem was used as a baseline for the models, a test to the performance of the model, and the entire pipeline process that led to the segmentation.  The performance was evaluated in different sequences, variable resampling sizes, and data augmentation in order to gather details regarding the ability to solve this problem and the impact of these hyperparameters.

\subsection{Lesions segmentation}

Predicting a mask for prostate cancer lesion in multi-parametric magnetic resonance images is a problem considerably more difficult than the segmentation of the prostate. First, a lesion appears in fewer slices than the prostate and that combined with the smaller size the lesions results in much less positive pixels. Moreover, lesions are not mainly centered on the magnetic resonance image, and not only their position varies, but the shape and the slices where the lesion is visible vary too.  Finally, the number of lesions to be segmented vary between patients, and this increases the difficulty of the problem. This is also a far more difficult problem to humans than just segmenting the prostate.

Each patient might have several lesions, thus it will also have several masks, one for each prostate cancer lesion. The model, independently of which architecture is used, do not try to identify individual lesions. Instead, the model tries to segment all lesion voxels present in one mp-MRI. Therefore, it is necessary to create a mask composed by the union of all the individual lesion masks of a patient. Thus, to combine the masks, since the lesions do not overlap, a sum of all the masks is performed.

\chapter{Experiments and Results}

Once the methods are established, it is time to conduct several research experiments within the problem to be solved regarding the methods and the data. Experiments help to understand which techniques work and which ones do not work. Moreover, careful experimentation can improve already working models. For example, the data to be used in one experiment can be resized, the cropping can be performed differently, and the magnitude of data augmentations can be increased or decreased. Slight changes in any of these can have a considerable impact on the results, thus the experiments must be conducted thoroughly.

In spite of not being the only indicator of good research, results are crucial, and they should be a relevant part of any research. Frequently results are compared to different research projects due to the use of the same public dataset, however, since this thesis uses a private dataset, the results are compared between several different experiments, with considerable parameters changed between experiments.

In this chapter, the process used to conduct experiments is explored, the results for each experiment are given and results are carefully detailed and compared. Furthermore, an analysis of the methods is conducted, and the reasons for the different performance results are discussed with the advantages and disadvantages of the methods for each one of the problems presented initially in this thesis. Finally, some of the experiments are illustrated with visual examples of their performance, either using plots or segmentation masks side by side.

\section{Evaluation}

Deep learning problem evaluation is usually performed in unseen data, in order to also assess the capabilities of generalization from a specific model. In other words, if the model is overfitted to the training data. Splitting the dataset to construct a test set can be done with different ratios, such as 20\%/80\% or 25\%/75\%. However, the performance of the algorithm is frequently biased by the selected split, and results can appear to be better or worse depending on the data present in each set. In this problem a cross-validation technique with five different folds is used.  

\begin{figure}[h!]
    \centering
	\includegraphics[width=8cm]{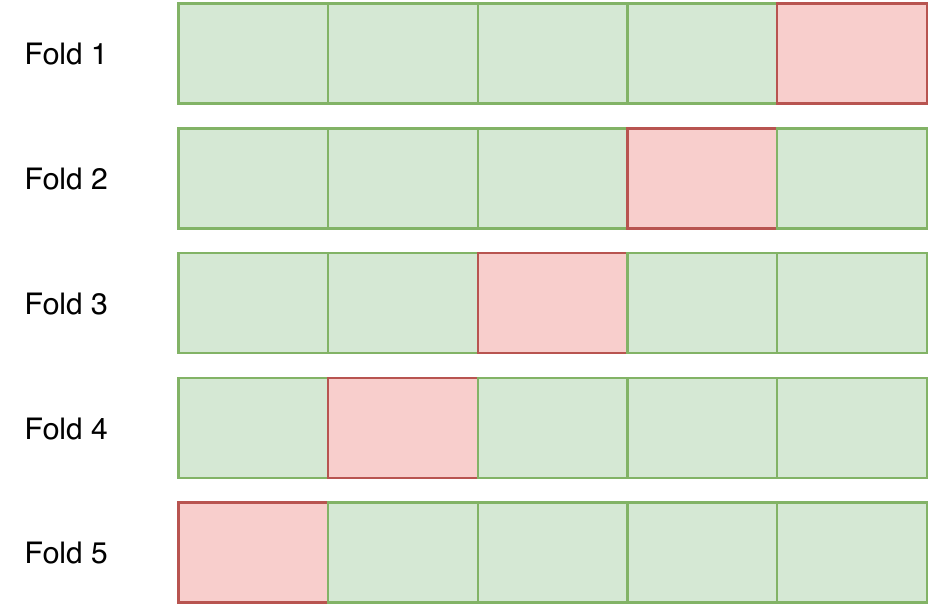} 
	\caption{Structure of the five folds. Each rectangle represents a subset of the dataset containing 20\% of the data. Green rectangles indicate that the data contained is part of the training set. On the other hand the red rectangle indicate that the data is used to test and evaluate the model. } 
    \label{folds_divisio}
\end{figure}

The construction of the folds can be seen in Image \ref{folds_divisio}, with each rectangle corresponding to 20\% of the data present in the dataset. To accurately analyze the performance in a problem, a model is trained for each fold on the 80\% training data (green rectangles) and tested afterward on the data present in the red rectangle of each fold. Finally, the results of all folds are averaged. It is also important to note that this technique works with different performance metrics, the data present in one rectangle is the same in all the folds and the split is performed patient-wise.

\section{Lesion classification}
\label{Sec:lesion_class_experiment}

Experiments on the classification of lesions are conducted independently for both problems with different losses, models, and metrics. However, some elements are common to experiments in both problems. Both work with two-dimensional images of lesions cropped from the slice where the approximated area of the lesion is bigger. The cropping of this lesion can be performed with a bounding box with a fixed size centered in the center of the lesion or an adjustable bounding box that resizes to have the exact size of the lesion that is resized afterward to the desired size. The cropping is performed at 64 and 32 pixels (with an adjusted bounding box the lesion is resized to have this dimension), while the model receives a 56x56 and 28x28 image respectively. Moreover, the augmentations applied were the same in both experiments with the images being subject to the following random operations:  crop with the size of the input of the model, horizontal flip, vertical flip, and an affine transformation with 15 degrees of rotation, a scaling between 0.50 and 1.50, a shear of 15, and translations of at most 0.07 in each direction. Whenever data augmentations are not used, the image is simply centrally cropped to the size of the input of the model.

\subsection{Clinical significance classification}
\label{clinical_classification_section}

To perform binary classification of prostate cancer lesion experiments, it was necessary to define some configurations for the experiments, in other words, a specific combination of techniques that can be applied in several experiments. The experiments can vary in the type of cropping technique to be used, the sizes of the cropping and the input size also vary, as well as if data augmentations are used or not. Since all three of these parameters have two different options each it results in $2^3 = 8$ different configurations to be used and evaluated.

\begin{table}[h]
    \centering
\caption{Different configurations of the experiments. Includes details of the cropping technique used, if data augmentation techniques were used and what were the values used for cropping the lesions and the input size for the model.}
\label{tab:experiments_binary_classification_setup}
\begin{center}
 \begin{tabular}{|c c c c c|} 
 \hline
 Config. & Crop Size & Input Size & Crop Type & Augmentations \\ [0.5ex] 
 \hline
 A & 64 & 56 & Adjusted Box & No\\ 
 \hline
 B & 64 & 56 & Adjusted Box & Yes\\ 
 \hline
 C & 64 & 56 & Fixed Box & No \\ 
 \hline
 D & 64 & 56 & Fixed Box & Yes\\ 
 \hline
 E & 32 & 28 & Adjusted Box & No\\ 
 \hline
 F & 32 & 28 & Adjusted Box & Yes\\ 
 \hline
 G & 32 & 28 & Fixed Box & No\\ 
 \hline
 H & 32 & 28 & Fixed Box & Yes\\ 
 \hline
\end{tabular}
\end{center}

\end{table}

The impact of the parameters is further analyzed individually in order to understand not only which combinations work, and which do not, but also to understand the independent impact of the parameter. Table \ref{tab:experiments_binary_classification_setup} shows the eight different configurations. Each configuration is associated with a letter, and it will be referred through that letter during the rest of this section. All the experiments' configurations described in the initial phase of the experiment, before retrieving the lesions, resample the MRI to 224x224 while keeping the original number of slices. An analysis of the effect of the resampling was also conducted and the results discussed.

\subsubsection{Cropping the lesions}

The impact of the cropping technique in the performance of the model was not clear, therefore, some experiments comparing both techniques described in this thesis were conducted. 

\begin{table}[h]
    \centering
\caption{Comparison of the performance of the experiments using different sequences types that used distinct sizes and different cropping techniques. For this comparison all the selected experiments configurations do not include data augmentations. Bold represents the best performance configuration in that sequence type. All the values in this table represent AUC scores that were computed in a cross-validation setup for each of the five folds, and averaged across them.}
\label{tab:binary_classification_experiments_cropping_comparison}
\begin{center}
 \begin{tabular}{|c c c c c|} 
 \hline
 Config. & ADC & T2W & ADC+T2W & Average \\ [0.5ex] 
 \hline
 A & \textbf{0.863} & 0.730 & 0.839 & 0.811\\ 
 \hline
 C & 0.737 & 0.592 & 0.701 & 0.677\\ 
 \hline
 E & 0.856 & \textbf{0.776} & \textbf{0.844} & \textbf{0.825}\\ 
 \hline
 G & 0.762 & 0.706 & 0.760 & 0.742\\ 
 \hline
\end{tabular}
\end{center}
\end{table}

Table \ref{tab:binary_classification_experiments_cropping_comparison} illustrates the results of each experiment by the sequence used as input of the model. It is possible to observe that configurations, which crop the lesion with a bounding box adjusted to the lesion size and resized afterward to the crop size, show better results regardless of this size (A and E). Both these experiments demonstrated considerable performance gains when compared to the use of a bounding box with a fixed size. Moreover, despite the slightly better performance of configuration A with ADC sequences when compared to the configuration E, the latter displays substantially better results with T2W and with the combination of both sequences, averaging also in a better AUC for the three different inputs. Regarding the other configurations, G is by far superior to C in all the inputs. Thus, it is possible to infer that a crop size of 32 with an input size of 28 further improves the performance regardless of the cropping technique used. 

\subsubsection{Data augmentation}

Similarly to the cropping, the effect of data augmentations in the performance of the model was also analyzed. The experiments to study its impacts were divided into two different groups in order to establish a fair comparison between configurations since, as seen before, the cropping type impacts the performance.  Therefore, the cropping technique was the criterion to divide configurations into groups. 

\begin{table}[h]
    \centering
\caption{Analysis of the effects of data augmentation techniques in experiments that used an adjusted box as cropping technique. Bold represents the best performance configuration in that sequence type. All the values in this table represent AUC scores that were computed in a cross-validation setup for each of the five folds, and averaged across them.}
\label{tab:adjusted_box_experiments_data_augmentation}
\begin{center}
 \begin{tabular}{|c c c c c|} 
 \hline
 Config. & ADC & T2W & ADC+T2W & Average \\ [0.5ex] 
 \hline
 A & \textbf{0.863} & 0.730 & 0.839 & 0.811\\ 
 \hline
 B & 0.840 & 0.772 & 0.830 & 0.814\\ 
 \hline
 E & 0.856 & 0.776 & \textbf{0.844} & 0.825\\ 
 \hline
 F & 0.855 & \textbf{0.815} & 0.828 & \textbf{0.833}\\ 
 \hline
\end{tabular}
\end{center}
\end{table}

Table \ref{tab:adjusted_box_experiments_data_augmentation} displays the results for the experiments that used the best performing cropping technique. Differently from what was expected, the data augmentations do not seem to positively impact all the different sequences used as input. For instance, on ADC sequences, the impact seems to be negative, and while it can be neglected when comparing E and F, it is somewhat larger when comparing A and B. In contrast, using T2W sequences as the input seems to greatly impact the results in a positive manner and this can be observed by direct comparison of A with B and E with F which show improvements of approximately 0.04. However, when combining both sequences, the impact is clearly negative in all the experiments analyzed. In spite of having some negative impact, data augmentations have improved the average AUC of a configuration when compared to a similar one without data augmentations.

\begin{table}[h]
    \centering
\caption{Analysis of the effects of data augmentation techniques in experiments that used a fixed box as cropping technique. Bold represents the best performance configuration in that sequence type. All the values in this table represent AUC scores that were computed in a cross-validation setup for each of the five folds, and averaged across them.}
\label{tab:fixed_box_experiments_data_augmentation}
\begin{center}
 \begin{tabular}{|c c c c c|} 
 \hline
 Config. & ADC & T2W & ADC+T2W & Average \\ [0.5ex] 
 
 \hline
 C & 0.737 & 0.592 & 0.701 & 0.677\\ 
 \hline
 D & 0.725 & 0.679 & 0.749 & 0.718\\ 
 \hline
 G & 0.762 & \textbf{0.706} & \textbf{0.760} & \textbf{0.742}\\ 
 \hline
 H & \textbf{0.818} & 0.640 & 0.757 & 0.738\\ 
 \hline
\end{tabular}
\end{center}
\end{table}

The impact of data augmentations on the experiments with different cropping techniques is also different and distinct as seen in Table \ref{tab:fixed_box_experiments_data_augmentation}. Furthermore, the impact also seems to be based on the cropping and input sizes used. For instance, on larger sizes (C and D) the impact on ADC seems to be negative with T2W and the combination of sequences benefiting greatly from the augmentations. Whereas on smaller crop sizes (G and H), ADC has a positive impact when using augmentations while the other two different inputs have a decreased performance.  It is also worth noting that different from the results in Table  \ref{tab:adjusted_box_experiments_data_augmentation}, the augmentations do not improve the average AUC of all the configurations, instead, it decreases the area under the curve in smaller input sizes. 

\subsubsection{Resampling}

All the previous experiment configurations started by resampling the magnetic resonance images to a 224x224 dimension while keeping the number of slices untouched. However, this resampling undoubtedly affects the performance, either negatively or positively. In order to properly assess this impact, all the previous experiments were replicated without the initial resampling and the results compared. Since in the previous section it was shown that both data augmentations and the cropping technique impact the results, the comparisons were performed between examples that had these parameters in comparison, while the size of the crop and the input size were the only variants within a results table.

\begin{table}[h]
    \centering
\caption{Comparison of the performance of the model if no resampling for 224x224 is used, between configurations that use no augmentations and use a fixed box crop technique. Bold represents the best performance configuration in that sequence type. All the values in this table represent AUC scores that were computed in a cross-validation setup for each of the five folds, and averaged across them. WR stands for "With Resampling" whereas NR stands for "No Resampling"}
\label{tab:resampling_experiments_fixed_no_augmentation}
\begin{center}
 \begin{tabular}{|c c c c c|} 
 \hline
 Config. & ADC & T2W & ADC+T2W & Average \\ [0.5ex] 
 \hline
 C - WR & 0.737 & 0.592 & 0.701 & 0.677\\ 
 \hline
 C - NR & 0.620 & 0.627 & 0.676 & 0.641\\ 
 \hline
 G - WR& \textbf{0.762} & \textbf{0.706} & \textbf{0.760} & \textbf{0.742}\\ 
 \hline
 G - NR & 0.631 & 0.656 & 0.655 & 0.647\\ 
 \hline
\end{tabular}
\end{center}
\end{table}

Table \ref{tab:resampling_experiments_adjusted_no_augmentation} shows the results of the configuration with a fixed box with worse performance, C, and the one with the best performance, G. Neither include data augmentations, and the size of their crop varies with G having the smaller value. In these cases, not resampling the images hurt the average AUC score, especially of ADC sequences. This, however, can be explained with the fact that the original size of ADC sequences is rather small, thereby the lesions are also small, meaning that it may be difficult to spot. On the other hand, while T2W performance decreased on configuration G, it improved on C. The reason behind this is the fact that the original size of T2W is considerably larger than the resampling size, meaning that a larger crop size might better capture the lesions while a small one might crop important parts of lesions. 

\begin{table}[h]
    \centering
\caption{Comparison of the performance of the model if no resampling for 224x224 is used, between configurations that use no augmentations and use a adjusted box crop technique. Bold represents the best performance configuration in that sequence type. All the values in this table represent AUC scores that were computed in a cross-validation setup for each of the five folds, and averaged across them. WR stands for "With Resampling" whereas NR stands for "No Resampling"}
\label{tab:resampling_experiments_adjusted_no_augmentation}
\begin{center}
 \begin{tabular}{|c c c c c|} 
 \hline
 Config. & ADC & T2W & ADC+T2W & Average \\ [0.5ex] 
 \hline
 A - WR& 0.863 & 0.730 & 0.839 & 0.811\\ 
 \hline
 A - NR & 0.834 & 0.741 & 0.818 & 0.798\\ 
 \hline
 E - WR & 0.856 & \textbf{0.776} & \textbf{0.844} & \textbf{0.825}\\ 
 \hline
 E - NR & \textbf{0.870} & 0.750 & 0.818 & 0.813\\ 
 \hline
\end{tabular}
\end{center}
\end{table}

Analyzing Table \ref{tab:resampling_experiments_adjusted_no_augmentation} where the results of two configurations with an adjusted box and no augmentations are shown, A and E, it is possible to observe similar patterns with lower average scores on experiments without resampling, and improved results on the experiment with larger crop size without resampling. However, in configuration E, the results for ADC show an improvement to 0.87, which means that this configuration is the best performing.

\begin{table}[h]
    \centering
\caption{Comparison of the performance of the model if no resampling for 224x224 is used, between configurations that have data augmentations and use a adjusted box crop technique. Bold represents the best performance configuration in that sequence type. All the values in this table represent AUC scores that were computed in a cross-validation setup for each of the five folds, and averaged across them. WR stands for "With Resampling" whereas NR stands for "No Resampling"}
\label{tab:resampling_experiments_adjusted_augmentation}
\begin{center}
 \begin{tabular}{|c c c c c|} 
 \hline
 Config. & ADC & T2W & ADC+T2W & Average \\ [0.5ex] 
 \hline
 B - WR & 0.840 & 0.772 & \textbf{0.830} & 0.814\\ 
 \hline
 B - NR & 0.839 & \textbf{0.825} & \textbf{0.830} & 0.831\\ 
 \hline
 F - WR & \textbf{0.855} & 0.815 & 0.828 & \textbf{0.833}\\ 
 \hline
 F - NR& 0.820 & 0.803 & 0.814 & 0.812\\ 
 \hline
\end{tabular}
\end{center}
\end{table}

Finally, Table \ref{tab:resampling_experiments_adjusted_augmentation} shows a comparison of the equivalent configurations shown in Table \ref{tab:resampling_experiments_adjusted_no_augmentation} but with data augmentations. One interesting observation from this table is that the experiment B without resampling attains a performance boost that allows it to outperform in the average AUC column all the other configurations besides F with resampling. This happens due to a major improvement in the performance using T2W sequences, that is in a majority of the experiments below 0.80, and achieves a higher maximum with this configuration. This also allows configuration B to be the only experiment to have a better performance without resampling than with resampling. Arguably, this can be associated with the combination of data augmentations that attenuated the degradation of the performance of ADC sequences and further improved the performance of T2W sequences.

\begin{figure}[!h]
  \centering
  \begin{minipage}[b]{0.8\textwidth}
    \includegraphics[width=\textwidth]{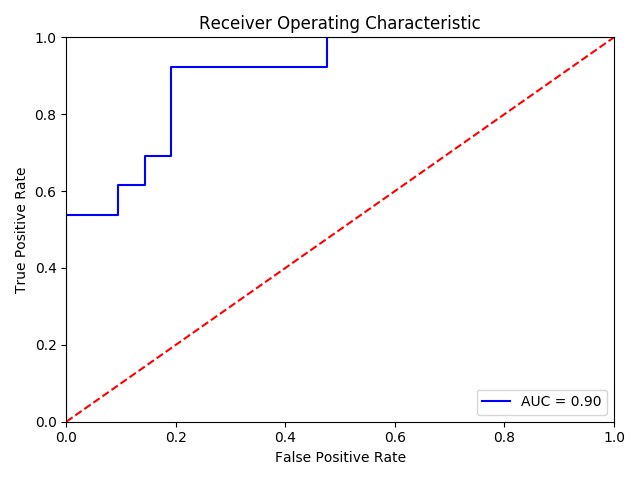}
  \end{minipage}
  \begin{minipage}[b]{0.8\textwidth}
    \includegraphics[width=\textwidth]{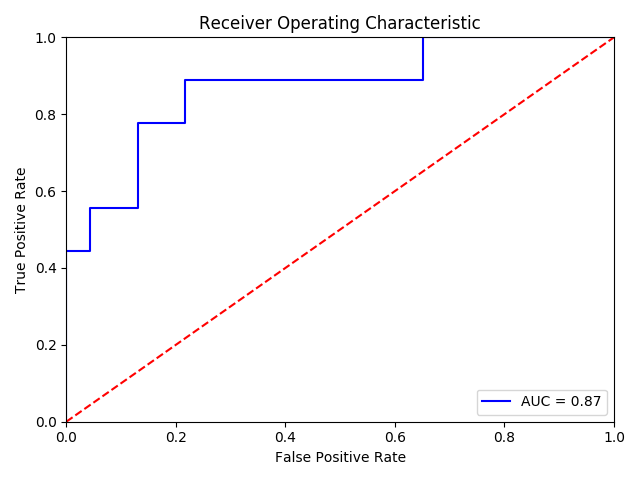}
  \end{minipage}
  \begin{minipage}[b]{0.8\textwidth}
    \includegraphics[width=\textwidth]{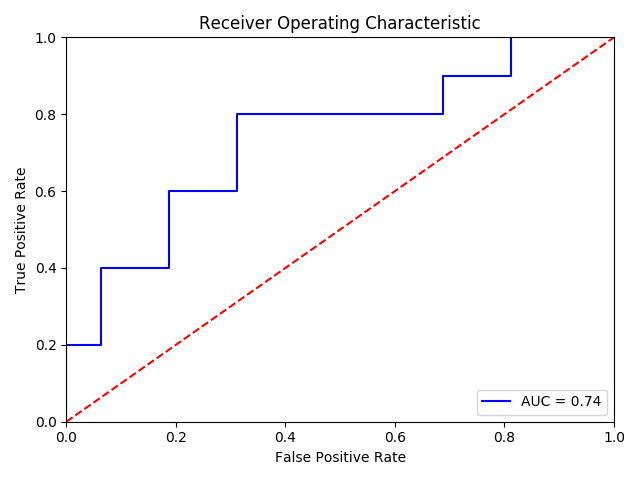}
  \end{minipage}

  \caption{Receiver operating characteristic (ROC) curve of the best performing model for some selected folds out of the five folds present in the problem. Folds are respectively from top to bottom numerated as 1, 3 and 4. }
  \label{best_performing_auc_binary_classification}
\end{figure}

For each fold used in the cross-validation, a Receiver operating characteristic (ROC) curve was generated and the AUC for that fold calculated. In Figure \ref{best_performing_auc_binary_classification}, it is possible to see these curves for some of the folds of the experiment configuration E without the resampling on ADC sequences. This was the configuration to achieve the best AUC score overall when used on ADC sequences. The folds represented in the image are the folds 1, 3, and 4. These were selected because fold 1 is, alongside with following 2 and 5, the best performing fold, while 3 and 4 are the worst performing. Despite having an inferior performance, fold 3 is close to the performance of the others, whereas fold 4 is distant by 0.16. Moreover, the latter fold has performed poorly in all the configurations, in some cases being below 0.60.

\begin{figure}[h!]
    \centering
	\includegraphics[width=12cm]{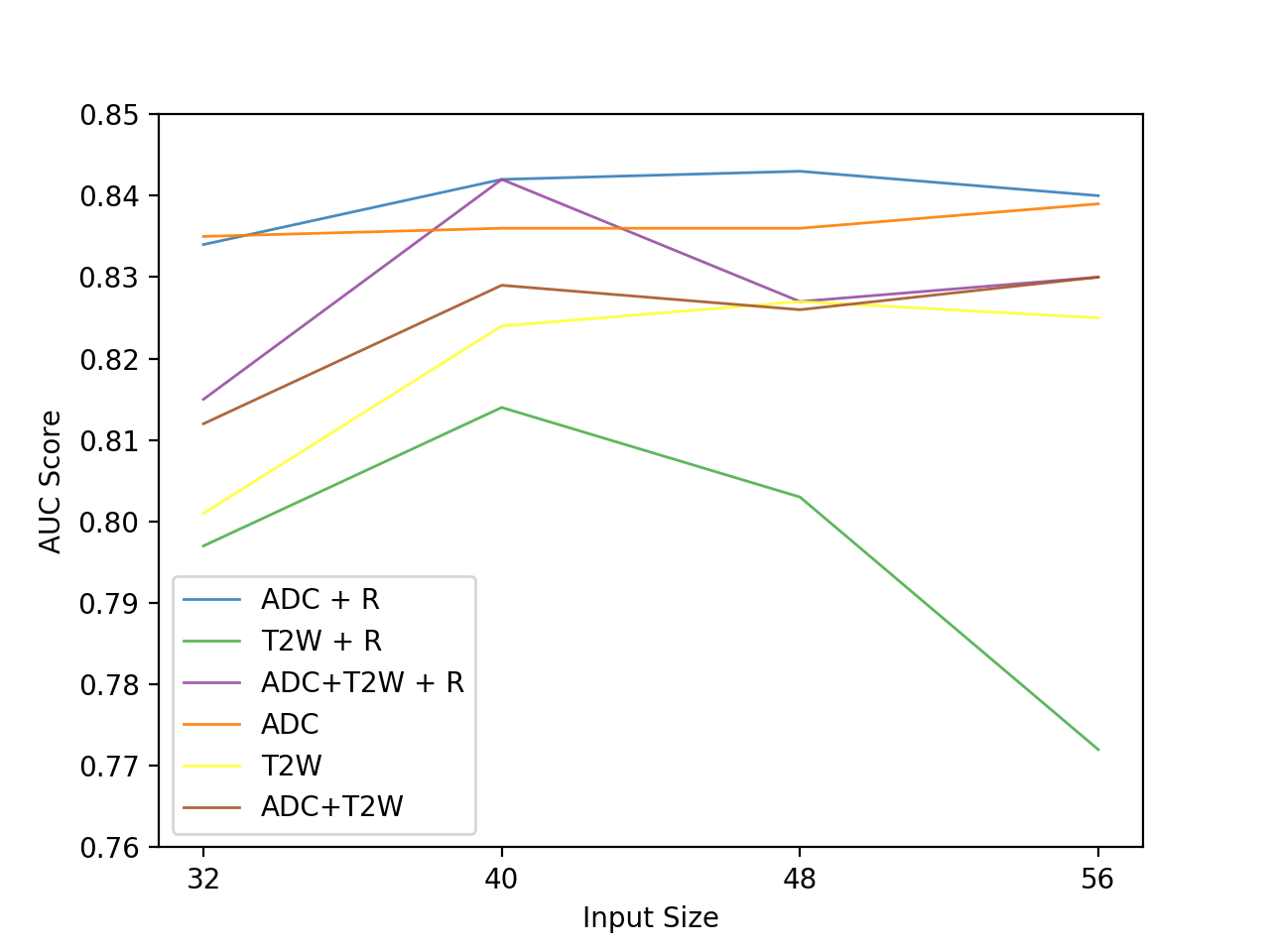} 
	\caption{Comparison of the AUC score with the variation of the input size from a adjusted box crop with a size of 64. The results are shown for four different input sizes: 32, 40, 48 and 56. Although the selected size for performance comparison is 56, it is possible to see that some sizes show better performance, and that the selected size leads to a decreased performance of the T2W with resampling. The letter R in front of the sequence means that before cropping the lesion was resampled.  } 
    \label{input_comparision_plot}
\end{figure}

The final experiments regarding the binary classification of lesions were conducted to analyze variation in the performance of experiments with a crop size of 64 over different input sizes as seen in Figure \ref{input_comparision_plot}. No data augmentations were used, and the crop was performed by adjusting the bounding box to the size of the lesion. Evaluations were performed in each of the three different input types (ADC, T2W, and ADC+T2W), with and without resampling for the following input sizes: 32, 40, 48, and 56. The results are interesting, and while it is impossible to find a common input size that maximizes the AUC in every experiment, it is possible to infer that 32 is an ineffective size overall. Moreover, all the experiments using T2W, either alone or combined with ADC, have a peak in performance when using an input of 40. The performance does not suffer significant variations when the selected size is greater or equal to 40 in most of the experiments, however, the experiment using T2W sequences as input with resampling suffered a considerable degradation in performance with larger input sizes.

\subsection{PI-RADS classification}

For the classification of the lesions accordingly with their PIRADS score, six different experiments were performed to assess the performance of the model in this problem. Before the experiments, it was also verified that only one of the lesions that included mask had a PIRADS score below 3. It was problematic and difficult to assess if the model can perform and learn how to predict the score of lesions with low scores. This also increased the difficulty of the training and the optimization of the model for the existent lesions.  Results for these experiments are given in terms of accuracy and confusion matrices. Since in this problem the classes are not independent and can be ordered in terms of proximity to each other, these matrices were constructed to indicate if the target class was missed by one or two classes. Moreover, the performance was also verified if the results aggregated both scores 4 and 5.

\begin{table}[h]
    \centering
\caption{Table with the results of the accuracy of the experiments for all the input sequences and with or without data augmentations. Bold highlights the best performance column-wise.  This results were calculated for separated scores 4 and 5. All the values in this table represent AUC scores that were computed in a cross-validation setup for each of the five folds, and averaged across them.}
\label{tab:accuracy_multi_class_classification}
\begin{center}
 \begin{tabular}{| c c c c c|} 
 \hline
  Augmentations & ADC & T2W & ADC+T2W & Average \\ [0.5ex] 
 \hline
 No & 0.631 & 0.658 & 0.644 & 0.644\\ 
 \hline
 Yes & \textbf{0.644} & \textbf{0.664} & \textbf{0.651} & \textbf{0.653}\\ 
 \hline
\end{tabular}
\end{center}
\end{table}

The first evaluation of the results was performed across all the five different PIRADS scores. The results are shown in Table \ref{tab:accuracy_multi_class_classification} and it is also possible to analyze the impact of data augmentations. And despite the improvements provided by those, the results are below what was expected. This can be explained based on the skewness of the dataset that does not include any sample for lesions with a score of 2, and only one lesion has as a score of 1. Differently from what was seen in the previous classification problem, the performance of using T2W as input is superior to the performance of using only ADC or combining both.

\begin{figure}[!h]
  \centering
  \begin{minipage}[b]{0.49\textwidth}
   \includegraphics[width=6.5cm]{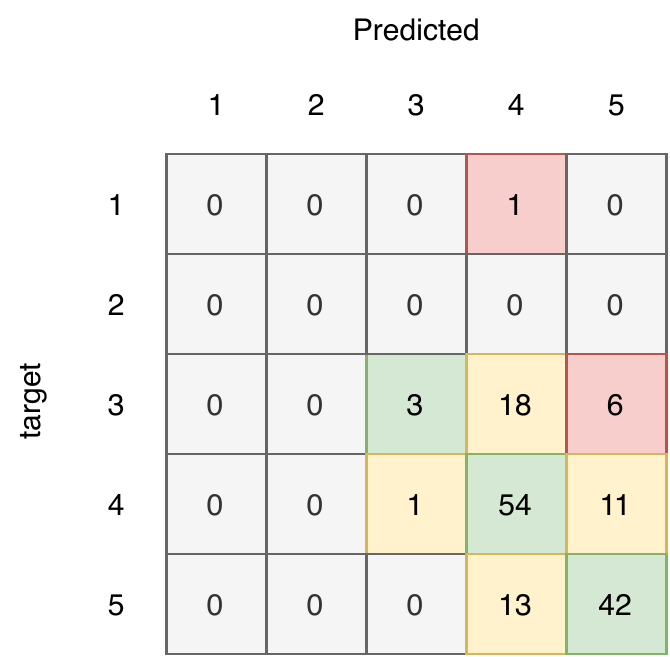} 

  \end{minipage}
  \hfill
  \begin{minipage}[b]{0.49\textwidth}
    \includegraphics[width=6.5cm]{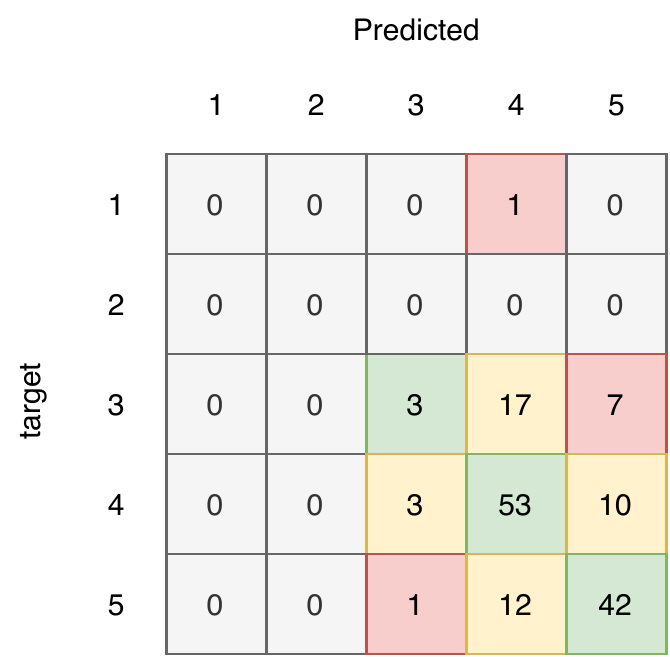} 
  \end{minipage}
\caption{ Confusion matrix for experiments to predict the PIRADS score using resampling, adjusted box cropping with a size of 32 and 28 input size on T2W sequences. Left image included data augmentation and had an accuracy of 66.4\% whereas the other did not included data augmentations with an accuracy of 64.8\%.
Green - Correct, Yellow - Missed by 1 class , Red - Missed by more than one class} 
\label{confusion_matrix_multi_class_no_aggregation}
\end{figure}

Figure \ref{confusion_matrix_multi_class_no_aggregation} displays the results of Table \ref{tab:accuracy_multi_class_classification} as confusion matrices. Both images represent experiments that used T2W sequences as input, however, the left image includes data augmentations, and as seen before a better performance. It is also possible to observe that there is an overestimation of lesions with a ground truth score of 3, being in a majority of the cases classified as 4 or 5. As mentioned before, the skewness of the dataset is most likely the reason for this overestimation.

\begin{table}[h]
    \centering
\caption{Table with the results of the accuracy of the experiments for all the input sequences and with or without data augmentations. Bold highlights the best performance column-wise.  Results were calculated for classes 4 and 5 as a single class. All the values in this table represent AUC scores that were computed in a cross-validation setup for each of the five folds, and averaged across them.}
\label{tab:accuracy_multi_class_classification_aggregated}
\begin{center}
 \begin{tabular}{| c c c c c|} 
 \hline
  Augmentations & ADC & T2W & ADC+T2W & Average \\ [0.5ex] 
 \hline
  No & \textbf{0.852} & 0.805 & \textbf{0.846} & \textbf{0.834}\\ 
 \hline
  Yes & 0.826 & \textbf{0.826} & 0.832 & 0.828\\ 
 \hline
\end{tabular}
\end{center}
\end{table}

In order to explore the effects of this overestimation, scores 4 and 5 were grouped in the final result. Not only overestimation, but also underestimation is studied, however, with less impact. The results in Table \ref{tab:accuracy_multi_class_classification_aggregated} reflect the impact of using this joint group which for both ADC and combined sequences as the input changes the best performing model.

\begin{figure}[!h]
  \centering
  \begin{minipage}[b]{0.49\textwidth}
    \includegraphics[width=6.5cm]{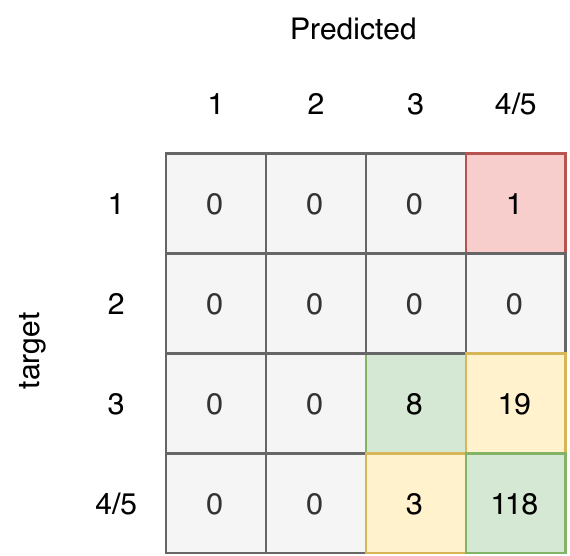} 
	
  \end{minipage}
  \hfill
  \begin{minipage}[b]{0.49\textwidth}
    \includegraphics[width=6.5cm]{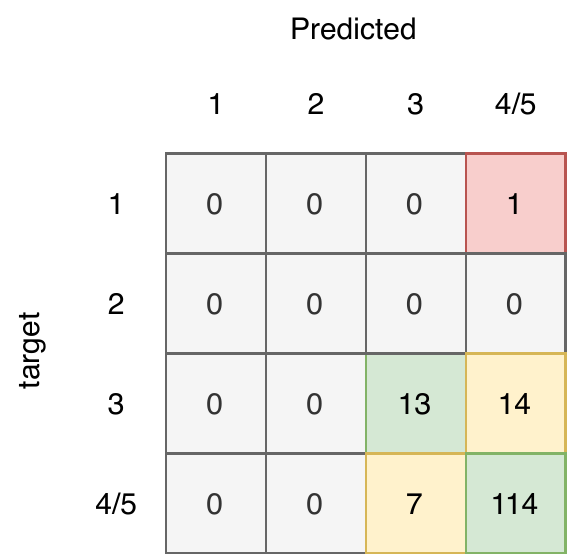} 
	
  \end{minipage}
  \caption{ Confusion matrix for experiments to predict the PIRADS score using resampling, adjusted box cropping with a size of 32 and 28 input size without data augmentations. Left image combined both T2W and ADC and had an accuracy of 84.6\% whereas the other did not included T2W sequences with an accuracy of 85.2\%. Green - Correct, Yellow - Missed by 1 class , Red - Missed by more than one class } 
  \label{confusion_matrix_multi_class_aggregation}
\end{figure}

Despite the best results shown in Table \ref{tab:accuracy_multi_class_classification_aggregated}, in Figure \ref{confusion_matrix_multi_class_aggregation} it is possible to observe a deficiency of the model in classifying lesions with a score of 3. In the left image, 70\% of lesions with label 3 are overestimated and 52\% in the right image. These high values represent the effect of the distribution of the dataset in the results. 

\section{Detection and segmentation}
\label{Sec:lesion_segm_experiment}

The experiments to detect and segment prostate cancer lesions and to detect and segment the prostate were executed similarly. The same preprocessing techniques, the same models, and the same resampling sizes were all used in experiments for both problems. In this section the 3D U-Net and the 3D ResNet-18 models are experimented with different input sizes and data augmentations.

\begin{table}[h]
    \centering
\caption{Different configurations of the experiments. Includes the dimensions of the input to the model as well as the loss function to be minimized while optimizing the model. }
\label{tab:experiments_configurations_prostate_segmentaiton}
\begin{center}
 \begin{tabular}{|c c c |} 
 \hline
 Config.   & Size & Loss Function\\ [0.5ex] 
 \hline
 A & 128x128x16 & 3D Dice\\ 
 \hline
 B &  128x128x16 & 3D Dice+BCE\\
 \hline
 C & 224x224x16 & 3D Dice\\ 
 \hline
 D &   224x224x16 & 3D Dice+BCE\\[1ex] 
 \hline
\end{tabular}
\end{center}
\end{table}

 Table \ref{tab:experiments_configurations_prostate_segmentaiton} lists the dimensions used and the input sizes for each configuration. The first size 128x128x16 was selected because it is the original size of ADC sequences regarding width and height. While 224x224x16 was used since the original width and height (360x360) of T2W sequences was too large to even fit one batch of size one in the GPU memory (RTX 2080ti 11 Gb). Models for these problems ran in parallel with two GPUs, which means that half of the batch was run in one GPU while the other half ran in the other. Moreover, the use of 16 slices is due to the architecture of the 3D U-Net model that reduces every dimension by half at each downsample block. Furthermore, the loss function used in each configuration varies also, with some configurations using exclusively a 3D dice loss, whereas the others use a combination of this latter loss with the binary cross-entropy loss applied pixel-wise. The evaluation of the experiments was performed with the resized ground truth mask to the size of the respective configuration.

\subsection{Prostate segmentation}
\label{prostate_segmentatiion_section}
Regarding the process of segmenting the prostate, the experiments conducted had their results compared based on the resampling size used, based on the models and finally the impact of data augmentations was also analyzed and carefully discussed. The experiments using 3D U-Net ran considerably faster than the 3D ResNet-18 experiments. While the former used a learning rate of $5*10^{-4}$ the latter used a learning rate of $2*10^{-4}$. Both versions of the experiments ran for 200 epochs and used the Adam optimizer with a weight decay of $10^{-4}$. Regarding the size of the batches used, configurations A and B used a batch size of four while C and D used two as batch size. This difference is due to GPU memory limitations.

\subsubsection{Input size and loss function}

Both the size of the input used to feed the model and the loss function used to calculate the error and backpropagate it in order to optimize the network are important factors in the performance of a model. Thus, before comparing the models against each other, it is important to compare and understand the effects of these parameters on the performance of each model.

\begin{table}[h]
    \centering
\caption{Experiments on the segmentation of the prostate using a 3D U-Net model in the different configurations. All the values in this table represent DICE scores that were computed in a cross-validation setup for each of the five folds, and averaged across them. The average column represents the mean of both sequences. Bold highlights the best performance column-wise. }
\label{tab:experiments_prostate_segmentation_resampling_uned}
\begin{center}
 \begin{tabular}{|c c c c c|} 
 \hline
 Config.  & Model & ADC & T2W & Average\\ [0.5ex] 
 \hline
 A & 3D U-Net &  0.893 & 0.889 & 0.891\\ 
 \hline
 B &  3D U-Net &  \textbf{0.898} & \textbf{0.911} & \textbf{0.905}\\
 \hline
 C & 3D U-Net &  0.873 & 0.882 & 0.878\\ 
 \hline
 D &  3D U-Net &  0.893 & 0.908 & 0.901\\[1ex] 
 \hline
\end{tabular}
\end{center}
\end{table}

Table \ref{tab:experiments_prostate_segmentation_resampling_uned} shows the performance of the 3D U-Net for all four configurations and the two different input sizes. It is possible to observe three different patterns. First, the performance using T2W sequences is always superior to the performance of using ADC sequences in the same configuration. Secondly, in configurations with the same loss, the performance is always superior, for all the input sequences in the configuration with a smaller input size. Finally, the performance improves when using a combination of loss functions.


Despite the improvements that the loss function brought to the 3D U-Net, Table \ref{tab:experiments_prostate_segmentation_resampling_3d_resnet} shows that the same is not reflected in the 3D ResNet-18 model. Not only is it difficult to analyze and detect any pattern, but also the use of different losses seemed to have a residual impact on the performance of the model. While having slightly more impact on ADC sequences, the input size impact can also be neglected when T2W sequences are used. 

\begin{table}[h]
    \centering
\caption{Experiments on the segmentation of the prostate using a 3D ResNet-18 as model in the different configurations. All the values in this table represent DICE scores that were computed in a cross-validation setup for each of the five folds, and averaged across them. The average column represents the mean of both sequences. Bold highlights the best performance column-wise. }
\label{tab:experiments_prostate_segmentation_resampling_3d_resnet}
\begin{center}
 \begin{tabular}{|c c c c c|} 
 \hline
 Config.  & Model & ADC & T2W & Average\\ [0.5ex] 
 \hline
 A & 3D ResNet-18 &  \textbf{0.895} & 0.886 & \textbf{0.891}\\ 
 \hline
 B &  3D ResNet-18 &  0.888 & 0.886 & 0.887\\
 \hline
 C & 3D ResNet-18 &  0.876 & 0.888 & 0.882\\ 
 \hline
 D &  3D ResNet-18 &  0.872 & \textbf{0.891} & 0.882\\[1ex] 
 \hline
\end{tabular}
\end{center}
\end{table}

\subsubsection{Models}

Comparison between models is also useful in order to understand how they perform and in which conditions should a model be used. The comparison of these models was only performed on configurations that had an input size of 128x128x16, configurations A and B.


\begin{table}[h]
    \centering
\caption{Comparison of the models 3D ResNet-18 and 3D U-Net in the configurations A and B, the ones that gave better results overall in the segmentation of the prostate.  All the values in this table represent DICE scores that were computed in a cross-validation setup for each of the five folds, and averaged across them. The average column represents the mean of both sequences. Bold highlights the best performance column-wise.}
\label{tab:experiments_prostate_segmentation_model_comparision}
\begin{center}
 \begin{tabular}{|c c c c c|} 
 \hline
 Config.  & Model & ADC & T2W & Average\\ [0.5ex] 
 \hline
 A & 3D U-Net &  0.893 & 0.889 & 0.891\\ 
 \hline
 A & 3D ResNet-18 &  0.895 & 0.886 & 0.891\\ 
 \hline
 B &  3D U-Net &  \textbf{0.898} & \textbf{0.911} & \textbf{0.905}\\
 \hline
 B &  3D ResNet-18 &  0.888 & 0.886 & 0.887\\
 \hline
\end{tabular}
\end{center}
\end{table}

On Table \ref{tab:experiments_prostate_segmentation_model_comparision} it is possible to observe that the performance of the models varies, especially when a combination of loss functions is used. While having a quite similar performance with the 3D dice loss, the 3D U-Net outperforms the other model when the binary cross-entropy is also used. These results are different from what would be expected considering the complexity difference of both models. However, the particular U-shaped architecture has shown to achieve slightly better results, and since it is also faster to train, further experiments in the segmentation of the prostate use the 3D U-Net as the model.

\subsubsection{Data Augmentations}

In an attempt to further improve the performance of the experiments, a couple of variations of these were performed using data augmentation techniques. The data augmentations used did not, however, have a high magnitude, since it is important to restrain the distortions that may go beyond the distribution of the data in the real-world.

\begin{table}[h]
    \centering
\caption{Analysis of the impact of data augmentation techniques in the performance of the 3D U-Net model in the configurations A and B in segmenting the prostate.  All the values in this table represent DICE scores that were computed in a cross-validation setup for each of the five folds, and averaged across them. The average column represents the mean of both sequences. Bold highlights the best performance column-wise. }
\label{tab:experiments_prostate_segmentation_augmentations_uned}
\begin{center}
 \begin{tabular}{|c c c c c c|} 
 \hline
 Config.  & Model  & Augmentation & ADC & T2W & Average\\ [0.5ex] 
 \hline
 A & 3D U-Net & No &  0.893 & 0.889 & 0.891\\ 
 \hline
 A & 3D U-Net & Yes &  0.902 & 0.900 & 0.901\\ 
 \hline
 B &  3D U-Net & No &  0.898 & \textbf{0.911} & 0.905\\
 \hline
 B &  3D U-Net & Yes &  \textbf{0.915} & 0.910 & \textbf{0.913}\\[1ex] 
 \hline
\end{tabular}
\end{center}
\end{table}

As expected, the use of these augmentations slightly improved the performance of the model for both the loss functions used. The improvement margin was similar in both cases, however, due to the initial better performance of the combination of both losses, the use of augmentations in that experiment improves the DICE score to a value of 0.915. Another particularity is that these new experiments show better performance in ADC images when compared to T2W images, which is the opposite of previous experiments.

\subsubsection{Final Results}

Configuration B, including data augmentations and using a 3D U-Net as the model, was the best experiment of this problem. Thus, the experiment was selected to be analyzed fold-wise. Table \ref{tab:experiments_prostate_segmentation_best_model} shows the performance of the model on the different folds, and it is possible to observe that the results do not vary significantly with fold 3 being the only one to be below a 0.910 DICE score.

\begin{table}[h]
    \centering
\caption{Performance in each fold of the model that performed the best in each sequence in the table \ref{tab:experiments_prostate_segmentation_augmentations_uned}.  All the values in this table represent DICE scores . The average column represents the mean of both sequences. Bold highlights the best performance column-wise. }
\label{tab:experiments_prostate_segmentation_best_model}
\begin{center}
 \begin{tabular}{| c c c c |} 
 \hline
  Fold & ADC & T2W & Average \\ [0.5ex] 
 \hline
  1 &  0.9182 & \textbf{0.9164} & \textbf{0.9173}\\ 
 \hline
  2 &  0.9136 & 0.9113 & 0.9125 \\ 
 \hline
  3 &  0.9086  & 0.9023 & 0.9055\\
 \hline
  4 &  0.9146 & 0.9103 & 0.9125 \\
 \hline
  5 &  \textbf{0.9188} & 0.9116 & 0.9152\\[1ex] 
 \hline
\end{tabular}
\end{center}
\end{table}

\begin{figure}[!h]
  \centering
  \begin{minipage}[b]{0.496\textwidth}
    \includegraphics[width=\textwidth]{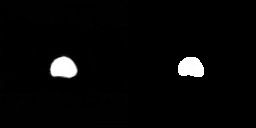}
  \end{minipage}
  \hfill
  \begin{minipage}[b]{0.496\textwidth}
    \includegraphics[width=\textwidth]{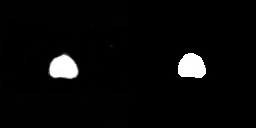}
  \end{minipage}
  \hfill
  \begin{minipage}[b]{0.496\textwidth}
    \includegraphics[width=\textwidth]{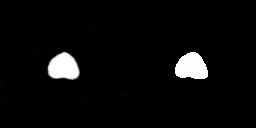}
  \end{minipage}
  \hfill
  \begin{minipage}[b]{0.496\textwidth}
    \includegraphics[width=\textwidth]{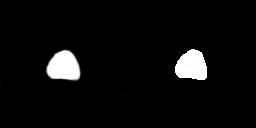}
  \end{minipage}
  \caption{Sample predictions made by a 3D U-Net model from the fold 5 of experiment configuration B with data augmentations using ADC as input. Images represent slices with mask predictions on the same patient compared to the ground truth. Left side of each image contains the prediction whereas the right size contains the label.} 
  \label{results_best_prostate_segmentation_images}
\end{figure}

The results in this problem indicate that the segmentation is being performed with significant quality and the visualizations of the predictions are also similar to the visualizations of the ground truth. Figure \ref{results_best_prostate_segmentation_images} shows predictions for consecutive slices in the fold 5 of this experiment using ADC sequences as input.  

Table \ref{tab:experiments_prostate_segmentation_comparision methods} shows a comparison between the methods presented in this paper and other methods presented in literature. It is important to note that this is not an objective comparison of methods, since a different dataset was used. However, these results denote that despite performing in different dataset, the presented method achieves and perhaps surpasses the expected performance.

\begin{table}[h]
    \centering
\caption{Comparison of the results presented in other papers that attempted to approach prostate segmentation with deep learning methods. Bold highlights the best performance column-wise.  All the values in this table represent DICE scores. }
\label{tab:experiments_prostate_segmentation_comparision methods}
\begin{center}
 \begin{tabular}{| c c c c | c|} 
 \hline
  Method & ADC & T2W & ADC+T2W  & Best\\ [0.5ex] 
 \hline
  Ours &  \textbf{0.92} & \textbf{0.91} & - & \textbf{0.92} \\ 
 \hline
  \cite{Schelb2019} &  0.86 & 0.86 & \textbf{0.88} & 0.88 \\ 
 \hline
  \cite{To2018} &  -  & 0.85 & - & 0.85\\
 \hline
  \cite{Karimi2019} &  - & \textbf{0.91} & - & 0.91 \\
 \hline
  \cite{Feldman2019} &  - & - & 0.87 & 0.87\\[1ex] 
 \hline
\end{tabular}
\end{center}
\end{table}

\subsection{Lesion segmentation}



The segmentation of prostate cancer lesions had similar experiments to the previous problem regarding the segmentation of the prostate.  Similar analyses were conducted on the input size, data augmentations, and different models. Despite these similarities that also include the same Adam optimizer, the same weight decay value of $10^{-4}$ and the same batch size for the different configurations (A and B have a batch size of four while C and D have a batch size of two in order to fit in the GPU) there are some differences. Due to being a slightly more complicated problem, the models trained for 250 epochs in this problem, and the learning rates were $2*10^{-4}$ for the 3D U-Net Model and $7*10^{-5}$ for the 3D ResNet-18 model. The results in the experiments regarding this problem reflect the difficulty of the task at hand.

\subsubsection{Input size and loss function}

Studying the loss function and the input size that optimizes the results of each model is important in order to understand the limitations of the model and how they can be explored.

\begin{table}[h!]
    \centering
\caption{Experiments on the segmentation of the prostate cancer lesions using a 3D U-Net as model in the different configurations.  All the values in this table represent DICE scores that were computed in a cross-validation setup for each of the five folds, and averaged across them. The average column represents the mean of both sequences. Bold highlights the best performance column-wise. }
\label{tab:experiments_lesions_segmentation_resampling_uned}
\begin{center}
 \begin{tabular}{|c c c c c|} 
 \hline
 Config.  & Model & ADC & T2W & Average\\ [0.5ex] 
 \hline
 A & 3D U-Net &  0.500 & 0.500 & 0.500\\ 
 \hline
 B &  3D U-Net &  \textbf{0.674} & \textbf{0.530} & \textbf{0.602}\\
 \hline
 C & 3D U-Net &  0.630 & 0.511 & 0.571\\ 
 \hline
 D &  3D U-Net &  0.652 & 0.516 & 0.584\\[1ex] 
 \hline
\end{tabular}
\end{center}
\end{table}

An increase in the size of the input dramatically increases the time needed in the backward and forward pass of the training, and also the inference time. Therefore, increased training and inference can only be justified if these larger sizes represent significant performance improvements. Table \ref{tab:experiments_lesions_segmentation_resampling_uned} illustrates the results for the experiments using a 3D U-Net model. It is possible to observe two patterns in the table. Firstly, once more, when no data augmentations are being used, the combination of 3D dice loss and binary cross-entropy shows significant improvements regardless of the sequence given as input or the size of the sequence. Secondly, the performance of ADC sequences is usually superior to the performance of T2W sequences, and in most cases, there is a considerable difference between the performance of both. 

\begin{table}[h]
    \centering
\caption{Experiments on the segmentation of prostate cancer lesions using a 3D ResNet-18 as model in the different configurations.  All the values in this table represent DICE scores that were computed in a cross-validation setup for each of the five folds, and averaged across them. The average column represents the mean of both sequences. Bold highlights the best performance column-wise. }
\label{tab:experiments_lesions_segmentation_resampling_3d_resnet}
\begin{center}
 \begin{tabular}{|c c c c c|} 
 \hline
 Config.  & Model & ADC & T2W & Average\\ [0.5ex] 
 \hline
 A & 3D ResNet-18 &  0.500 & 0.385 & 0.443\\ 
 \hline
 B &  3D ResNet-18 &  \textbf{0.501} & 0.438 & 0.470\\
 \hline
 C & 3D ResNet-18 &  0.500 & 0.501 & 0.501\\ 
 \hline
 D &  3D ResNet-18 &  0.500 & \textbf{0.503} & \textbf{0.502}\\[1ex] 
 \hline
\end{tabular}
\end{center}
\end{table}

The results observed in Table \ref{tab:experiments_lesions_segmentation_resampling_3d_resnet} are rather poor. Despite the slight performance increase on T2W sequences when the size is increased, neither 3D dice nor the combined loss seem to be enough to optimize the model. Smaller sizes show a performance degradation of T2W sequences, thus an inferior performance compared to the ADC sequences.

\subsubsection{Models}

Differently from what was seen in the prostate segmentation problem, the models show a completely distinct performance when compared against each other.

\begin{table}[h]
    \centering
\caption{Comparison of the models 3D ResNet-18 and 3D U-Net in the configurations A and B, the ones that gave better results overall in the segmentation of prostate cancer lesions. All the values in this table represent DICE scores that were computed in a cross-validation setup for each of the five folds, and averaged across them. The average column represents the mean of both sequences. Bold highlights the best performance column-wise. }
\label{tab:experiments_lesions_segmentation_model_comparision}
\begin{center}
 \begin{tabular}{|c c c c c|} 
 \hline
 Config.  & Model & ADC & T2W & Average\\ [0.5ex] 
 \hline
 A & 3D U-Net &  0.500 & 0.500 & 0.500\\ 
 \hline
 A & 3D ResNet-18 &  0.500 & 0.385 & 0.443\\ 
 \hline
 B &  3D U-Net &  \textbf{0.674} & \textbf{0.530} & \textbf{0.602}\\
 \hline
 B &  3D ResNet-18 &  0.501 & 0.438 & 0.470\\
 \hline
\end{tabular}
\end{center}
\end{table}

Table \ref{tab:experiments_lesions_segmentation_model_comparision} displays the values for these experiments, and it is possible to observe that not only the performance of the 3D ResNet-18 model does not go beyond 0.501, but it gets near that value with ADC sequences since T2W performs always poorly. The results of the best 3D U-Net model are 34.5\% and 21.0\% higher for ADC and T2W sequences respectively when compared to the other model in the same configuration. Once more the performance of the 3D U-Net model outperforms the 3D ResNet-18, this time by a significant margin. Further experiments use the 3D U-Net as the model.

\subsubsection{Data Augmentations}

Data augmentation will once more be used to see how much further the performance of the model can be improved. The previous two best-performing experiments in the comparison of the models were used in two new experiments that included data augmentations in the Data loading stage.

\begin{table}[h]
    \centering
\caption{Analysis of the impact of data augmentation techniques in the performance of the 3D U-Net model in the configurations A and B to segment prostate cancer lesions.  All the values in this table represent DICE scores that were computed in a cross-validation setup for each of the five folds, and averaged across them. The average column represents the mean of both sequences. Bold highlights the best performance column-wise. }
\label{tab:experiments_lesions_segmentation_augmentations_uned}
\begin{center}
 \begin{tabular}{|c c c c c c|} 
 \hline
 Config.  & Model  & Augmentation & ADC & T2W & Average\\ [0.5ex] 
 \hline
 A & 3D U-Net & No &  0.500 & 0.500 & 0.500\\ 
 \hline
 A & 3D U-Net & Yes &  \textbf{0.690} & 0.491 & 0.591\\ 
 \hline
 B &  3D U-Net & No &  0.674 & \textbf{0.530} & 0.602\\
 \hline
 B &  3D U-Net & Yes &  0.681 & 0.526 & \textbf{0.604}\\[1ex] 
 \hline
\end{tabular}
\end{center}
\end{table}

The impact of the augmentations in the performance of the model did not seem to be positive when T2W sequences were given as input with results slightly worse than the ones without data augmentations. However, the performance in ADC sequences greatly benefited from the inclusion of such transformations. Observing the data presented in Table \ref{tab:experiments_lesions_segmentation_augmentations_uned}, it is possible to observe a small 1\% improvement when ADC sequences were given to a model optimized with the combined loss function, yet the performance of this model was already far superior to previous models without data augmentations. The largest performance boost is with the use of the 3D dice and ADC sequences, where the augmentations help the model to achieve a performance 38\% higher when compared to the model without augmentations. This new value is the best DICE score in all the experiments.

\subsubsection{Final Results}

Using data augmentation in configuration A gave the best results for ADC inputs, while for T2W, configuration B without augmentations had the best results. Both these experiments can be seen in Table \ref{tab:experiments_lesions_segmentation_best_model} with the results in each fold discriminated. 

\begin{table}[h]
    \centering
\caption{Performance in each fold of the model that performed the best in each sequence in the table \ref{tab:experiments_lesions_segmentation_augmentations_uned}.  All the values in this table represent DICE scores. The average column represents the mean of both sequences. Bold highlights the best performance column-wise. }
\label{tab:experiments_lesions_segmentation_best_model}
\begin{center}
 \begin{tabular}{| c c c c|} 
 \hline
  Fold & ADC & T2W & Average\\ [0.5ex] 
 \hline
  1 &  0.4686 & 0.2964 & 0.3825  \\ 
 \hline
  2 &  0.6722 & 0.5214 & 0.5968\\ 
 \hline
  3 &  0.7003  & 0.5832 & 0.6418\\
 \hline
  4 &  \textbf{0.8302} & \textbf{0.6564} & \textbf{0.7433} \\
 \hline
  5 &  0.7896 & 0.6005 & 0.6951\\[1ex] 
 \hline
\end{tabular}
\end{center}
\end{table}

It is important to note that the performance varies significantly between folds, especially between folds 1 and 4 with the latter having 77.1\% and 121.5\% higher performance in ADC and T2W sequences respectively when compared to the performance of the former. It is also worth noting that while T2W results are relatively poorer, the variation fold-wise is similar to the variation that occurs with ADC sequences. The best performing fold for ADC sequences displays somewhat promising results.

\begin{figure}[!h]
  \centering
  \begin{minipage}[b]{0.496\textwidth}
    \includegraphics[width=\textwidth]{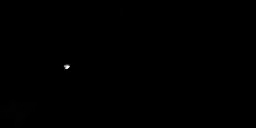}
  \end{minipage}
  \hfill
  \begin{minipage}[b]{0.496\textwidth}
    \includegraphics[width=\textwidth]{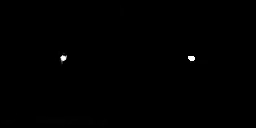}
  \end{minipage}
  \hfill
  \begin{minipage}[b]{0.496\textwidth}
    \includegraphics[width=\textwidth]{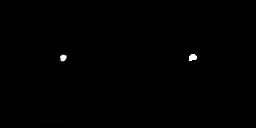}
  \end{minipage}
  \hfill
  \begin{minipage}[b]{0.496\textwidth}
    \includegraphics[width=\textwidth]{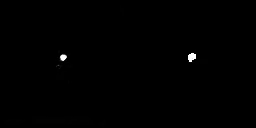}
  \end{minipage}
  \caption{Sample predictions made by a 3D U-Net model from the fold 4 of experiment configuration A with data augmentations using ADC as input. Images represent slices with mask predictions on the same patient compared to the ground truth. Left side of each image contains the prediction whereas the right size contains the label.} 
  \label{results_best_lesion_segmentation_images}
\end{figure}

 Figure \ref{results_best_lesion_segmentation_images} displays some predictions given of the model for the fold 4 of ADC sequences. The predictions have some distance from the ground truth, either by being smaller, having a somewhat more simplistic shape, or simply by appearing one or two slices before expected.

\clearpage

\chapter{Conclusion and Future Work}
\label{Chap:conclusion}
This research approached deep learning methods for prostate cancer diagnosis problems based on multi-parametric magnetic resonance images. From the information contained in these images, the purpose was to understand which problems could be solved or addressed using computer vision methods and deep learning. Some of the problems required lesions to be classified. Others required the prostate or prostate cancer lesions to be detected in the MRI and segmented accordingly to the annotations given by experts.  

Deep learning approaches dwell in a world that is unreachable for some domains, the world of big data and massive volumes of information. This is, for now, the case of medical use cases. The information is very limited, not only due to the requirements of experts, but also due to privacy matters. Hence, it was a challenge that required a thorough analysis of the problem, the preprocessing techniques, the models and the hyperparameters to use. Overall, a deep analysis of the performance of each problem was conducted and the results discussed. 

The image quality and the annotations carefully made by experts supported the learning process of the model. However, other characteristics were considerably less positive. First of all, the small number of samples hurt the performance. Secondly, the distribution of the PIRADS score of lesions with mask increased the difficulty of the problem. Moreover, the overall number of patients was rather small. This affected the performance in some tasks, such as the segmentation of lesions. The latter task due to its nature requires significantly more data in order to achieve better results. 

Regarding the classification of lesions, the performance varied greatly between tasks. Results in the cross-validation score presented in Section \ref{clinical_classification_section} for the clinical significance of lesions were superior to the 0.84 AUC obtained by XmasNet in the ProstateX challenge \cite{Liu2017}. This while using fewer sequences as input, thus also using fewer parameters \cite{Liu2017}. And while the challenge dataset had poor data quality, the number of images, from the 346 patients, was considerably higher. Hence, the results not only matched the expectations, but they can also be considered promising. It is expected that if the number of samples increases while keeping the quality of the data, the performance may also increase. Despite these results, the  distribution of the PIRADS score from lesion with masks did not allow the performance in this task to be the expected. The performance was significantly worse than the binary classification and was not nearly enough to be adopted in a clinical situation or the problem to be considered solved.

The process of detecting those lesions and segment them is rather complicated. Thus, in order to test and validate the research methodology to be followed, one experiment was conducted before segmenting lesions. The extra prostate mask annotations given in the dataset were used to detect and segment the prostate itself. The results shown in Section \ref{prostate_segmentatiion_section} exceeded expectations. Not only both the models perform as required on both sizes and with both losses. But experiments were further improved with data augmentation techniques. The results were not only promising, but they also surpassed performance in previous research \cite{Karimi2019,To2018,Feldman2019,Schelb2019}. Therefore, all the experiments were repeated for lesion segmentation, yet, without achieving results close to the previous problem. The results indicate, however, some promising details. Such as the performance increase in the experiments that used data augmentations. These details may indicate that increasing the data in the dataset is also crucial to improve upon the 0.69 DICE score obtained in this task. And while some folds performed poorly, others presented results that achieved a DICE score up to 0.83. Understanding the reason behind these distinct results might in the future lead to gains in the performance.  And despite the results in this task being inferior to the ones in the previous segmentation task, it is important to note that the results are somewhat slightly better than the ones presented by previous researchers \cite{Feldman2019, Schelb2019}.

The relevance of the work presented in this thesis goes beyond the simple analysis of the results. The thesis demonstrates that deep learning methods can be applied to solve computer vision tasks on multi-parametric magnetic resonance images regarding the prostate cancer diagnosis. Despite only being able to achieve the desired results on two of the four tasks. The remaining tasks have shown a potential that can be achieved by addressing some of the flaws regarding the data previously mentioned. 

Further research is still needed to study other plausibly relevant questions. Studies on the performance of the model in classifications tasks can be conducted by averaging the prediction of two models trained in different sequences accordingly to different configurations. It can also be analyzed if the Gleason score can be better predicted through this data compared to the PIRADS score. Regarding the segmentation, the registration of both sequences or the addition of more augmentations can be empirically studied. All these future work is based on the results given in this thesis that conducted and presented more than 112 different experiments. More experiments were conducted in the background and were not presented in the thesis but they were essential to tune the necessary hyperparameters for the final experiments. The goals of the thesis were therefore achieved, and the ground for future work on this dataset established.

\bibliographystyle{unsrt}
\bibliography{masters}

\begin{thebibliography}{100}

\bibitem{Stewart2014}
Bernard~W Stewart and Christopher~P Wild.
\newblock {\em {World Cancer Report 2014}}.
\newblock International Agency for Research on Cancer, 2014.

\bibitem{newcases}
{Key Statistics for Prostate Cancer | Prostate Cancer Facts}.

\bibitem{Romer1977}
Alfred~Sherwood Romer and Thomas S. (Thomas~Sturges) Parsons.
\newblock {\em {The vertebrate body}}.
\newblock Saunders, 1977.

\bibitem{Gallagher1998}
Richard~P Gallagher and Neil Fleshner.
\newblock {Prostate cancer: 3. Individual risk factors}.
\newblock Technical Report~7, 1998.

\bibitem{10.1002/cncr.11262}
Maurice P~A Zeegers, Annemarie Jellema, and Harry Ostrer.
\newblock {Empiric risk of prostate carcinoma for relatives of patients with
  prostate carcinoma}.
\newblock {\em Cancer}, 97(8):1894--1903, 2003.

\bibitem{10.1002/cncr.11635}
David~C Miller, Khaled~S Hafez, Andrew Stewart, James~E Montie, and John~T Wei.
\newblock {Prostate carcinoma presentation, diagnosis, and staging}.
\newblock {\em Cancer}, 98(6):1169--1178, 2003.

\bibitem{TheAmericanCancerSociety2019}
{The American Cancer Society}.
\newblock {Can Prostate Cancer Be Found Early?}, 2019.

\bibitem{PDQScreeningandPreventionEditorialBoard2002}
{PDQ Screening and Prevention Editorial Board}.
\newblock {\em {Prostate Cancer Screening (PDQ{\textregistered}): Health
  Professional Version}}.
\newblock 2002.

\bibitem{Gomella2011}
Leonard~G. Gomella, Xiaolong~S. Liu, Edouard~J. Trabulsi, Wm~Kevin Kelly,
  Ronald Myers, Timothy Showalter, Adam Dicker, and Richard Wender.
\newblock {Screening for prostate cancer: The current evidence and guidelines
  controversy}.
\newblock {\em Canadian Journal of Urology}, 18(5):5875--5883, oct 2011.

\bibitem{Naji2018}
Leen Naji, Harkanwal Randhawa, Zahra Sohani, Brittany Dennis, Deanna
  Lautenbach, Owen Kavanagh, Monica Bawor, Laura Banfield, and Jason Profetto.
\newblock {Digital Rectal Examination for Prostate Cancer Screening in Primary
  Care: A Systematic Review and Meta-Analysis.}
\newblock {\em Annals of family medicine}, 16(2):149--154, mar 2018.

\bibitem{Grossman2018}
David~C. Grossman, Susan~J. Curry, Douglas~K. Owens, Kirsten Bibbins-Domingo,
  Aaron~B. Caughey, Karina~W. Davidson, Chyke~A. Doubeni, Mark Ebell, John~W.
  Epling, Alex~R. Kemper, Alex~H. Krist, Martha Kubik, C.~{Seth Landefeld},
  Carol~M. Mangione, Michael Silverstein, Melissa~A. Simon, Albert~L. Siu, and
  Chien~Wen Tseng.
\newblock {Screening for prostate cancer USPreventive
  servicestaskforcerecommendation statement}.
\newblock {\em JAMA - Journal of the American Medical Association},
  319(18):1901--1913, may 2018.

\bibitem{MHoffman2020}
Richard {M Hoffman}.
\newblock {Screening for prostate cancer}, 2020.

\bibitem{Roberts2016}
Matthew~J Roberts, Harrison~Y Bennett, Patrick~N Harris, Michael Holmes, Jeremy
  Grummet, Kurt Naber, and Florian M~E Wagenlehner.
\newblock {Prostate Biopsy Related Infection: a Systematic Review of Risk
  Factors, Prevention Strategies and Management Approaches.}
\newblock {\em Urology}, 2016.

\bibitem{El-ShaterBosaily2015}
A.~{El-Shater Bosaily}, C.~Parker, L.~C. Brown, R.~Gabe, R.~G. Hindley,
  R.~Kaplan, M.~Emberton, H.~U. Ahmed, Mark Emberton, Hashim Ahmed, Ahmed
  El~Shater Bosaily, Alex Kirkham, Alex Freeman, Charles Jameson, Richard
  Hindley, Christopher Parker, Colin Cooper, Robert Oldroyd, Richard Kaplan,
  Louise Brown, {Rhian Gabe}, Yolanda Collaco-Moraes, Cybil Adusei, Katie Ward,
  Sophie Stewart, Katie Thompson~Claire Mulrenan, Hannah Gardner, Carlos
  Diaz-Montana, Chris Coyle, Mark Sculpher, Rita Faria, {David Guthrie}, John
  Chester, Richard Cowan, Michael Jewitt, H.~Ahmed, J.~Coe, A.~{El-Shater
  Bosaily}, M.~Emberton, A.~Freeman, M.~Hung, C.~Jameson, A.~Kirkham,
  S.~Punwani, R.~Scott, Richard Hindley, A.~Edwards, H.~El-Mahallawi,
  D.~Peppercorn, J.~Smith, A.~Thrower, M.~Winkler, K.~Ansu, T.~Barwick,
  S.~Edwards, L.~Honeyfield, N.~Qazi, B.~Statton, V.~Stewart, E.~Temple,
  N.~Burns-Cox, P.~Burn, K.~Gordon, H.~Routley, A.~Maccormick, D.~Paterson,
  A.~Henderson, E.~Bernsten, R.~Casey, D.~Day, S.~Ghosh, J.~James, P.~J.
  McMillan, G.~Russell, R.~Persad, J.~Ash-Miles, M.~Elmahdy, S.~Pandian,
  C.~Shiridzinomwa, M.~Sohail, A.~Treasure, M.~Ghei, V.~Conteh, L.~Harbin,
  R.~Katz, J.~Kumaradevan, A.~Trinidade, A.~Verjee, T.~Dudderidge, J.~Smart,
  D.~Rosario, J.~Catto, F.~Selem, I.~Shergill, and S.~Agarwal.
\newblock {PROMIS - Prostate MR imaging study: A paired validating cohort study
  evaluating the role of multi-parametric MRI in men with clinical suspicion of
  prostate cancer}.
\newblock {\em Contemporary Clinical Trials}, 42:26--40, may 2015.

\bibitem{Patel2009}
Amit~R Patel and J~Stephen Jones.
\newblock {Optimal biopsy strategies for the diagnosis and staging of prostate
  cancer.}
\newblock {\em Current opinion in urology}, 19(3):232--7, may 2009.

\bibitem{Bennett2016}
H.~Y. Bennett, M.~J. Roberts, S.~A.R. Doi, and R.~A. Gardiner.
\newblock {The global burden of major infectious complications following
  prostate biopsy}, jun 2016.

\bibitem{Bonekamp2011}
David Bonekamp, Michael~A. Jacobs, Riham El-Khouli, Dan Stoianovici, and
  Katarzyna~J. Macura.
\newblock {Advancements in MR imaging of the prostate: From diagnosis to
  interventions}.
\newblock {\em Radiographics}, 31(3):677--704, may 2011.

\bibitem{Sonn2013}
Geoffrey~A. Sonn, Shyam Natarajan, Daniel~J.A. Margolis, Malu MacAiran,
  Patricia Lieu, Jiaoti Huang, Frederick~J. Dorey, and Leonard~S. Marks.
\newblock {Targeted biopsy in the detection of prostate cancer using an office
  based magnetic resonance ultrasound fusion device}.
\newblock {\em Journal of Urology}, 189(1):86--92, 2013.

\bibitem{Marks2013}
Leonard Marks, Shelena Young, and Shyam Natarajan.
\newblock {MRI-ultrasound fusion for guidance of targeted prostate biopsy}, jan
  2013.

\bibitem{Kuru2013}
Timur~H. Kuru, Matthias~C. Roethke, Jonas Seidenader, Tobias
  Simpfend{\"{o}}rfer, Silvan Boxler, Khalid Alammar, Philip Rieker,
  Valentin~I. Popeneciu, Wilfried Roth, Sascha Pahernik, Heinz~Peter Schlemmer,
  Markus Hohenfellner, and Boris~A. Hadaschik.
\newblock {Critical evaluation of magnetic resonance imaging targeted,
  transrectal ultrasound guided transperineal fusion biopsy for detection of
  prostate cancer}.
\newblock {\em Journal of Urology}, 190(4):1380--1386, oct 2013.

\bibitem{Pokorny2014}
Morgan~R. Pokorny, Maarten {De Rooij}, Earl Duncan, Fritz~H. Schr{\"{o}}der,
  Robert Parkinson, Jelle~O. Barentsz, and Leslie~C. Thompson.
\newblock {Prospective study of diagnostic accuracy comparing prostate cancer
  detection by transrectal ultrasound-guided biopsy versus magnetic resonance
  (MR) imaging with subsequent mr-guided biopsy in men without previous
  prostate biopsies}.
\newblock {\em European Urology}, 66(1):22--29, 2014.

\bibitem{ProstateX}
{PROSTATEx - Overview}.

\bibitem{ProstateX2}
{PROSTATEx-2 Challenge}.

\bibitem{McKinney2020}
Scott~Mayer McKinney, Marcin Sieniek, Varun Godbole, Jonathan Godwin, Natasha
  Antropova, Hutan Ashrafian, Trevor Back, Mary Chesus, Greg~C Corrado, Ara
  Darzi, Mozziyar Etemadi, Florencia Garcia-Vicente, Fiona~J Gilbert, Mark
  Halling-Brown, Demis Hassabis, Sunny Jansen, Alan Karthikesalingam,
  Christopher~J Kelly, Dominic King, Joseph~R Ledsam, David Melnick, Hormuz
  Mostofi, Lily Peng, Joshua~Jay Reicher, Bernardino Romera-Paredes, Richard
  Sidebottom, Mustafa Suleyman, Daniel Tse, Kenneth~C Young, Jeffrey {De Fauw},
  and Shravya Shetty.
\newblock {International evaluation of an AI system for breast cancer
  screening}.
\newblock {\em Nature}, 577(7788):89--94, 2020.

\bibitem{JAronen}
Hannu {J Aronen}.
\newblock {Improved Prostate Cancer Diagnosis - Combination of Magnetic
  Resonance Imaging and Biomarkers - Full Text View - ClinicalTrials.gov}.

\bibitem{Merisaari2019}
Harri Merisaari, Ivan Jambor, Otto Ettala, Peter~J. Bostr{\"{o}}m, Ileana
  {Montoya Perez}, Janne Verho, Aida Kiviniemi, Kari Syv{\"{a}}nen, Esa
  K{\"{a}}hk{\"{o}}nen, Lauri Eklund, Tapio Pahikkala, Paula Vainio, Jani
  Saunavaara, Hannu~J. Aronen, and Pekka Taimen.
\newblock {IMPROD biparametric MRI in men with a clinical suspicion of prostate
  cancer (IMPROD Trial): Sensitivity for prostate cancer detection in
  correlation with whole-mount prostatectomy sections and implications for
  focal therapy}.
\newblock {\em Journal of Magnetic Resonance Imaging}, 50(5):1641--1650, nov
  2019.

\bibitem{Liu2017}
Saifeng Liu, Huaixiu Zheng, Yesu Feng, and Wei Li.
\newblock {Prostate Cancer Diagnosis using Deep Learning with 3D
  Multiparametric MRI}.
\newblock mar 2017.

\bibitem{Cicek2016}
{\"{O}}zg{\"{u}}n {\c{C}}i{\c{c}}ek, Ahmed Abdulkadir, Soeren~S. Lienkamp,
  Thomas Brox, and Olaf Ronneberger.
\newblock {3D U-Net: Learning Dense Volumetric Segmentation from Sparse
  Annotation}.
\newblock jun 2016.

\bibitem{chen2019med3d}
Sihong Chen, Kai Ma, and Yefeng Zheng.
\newblock Med3d: Transfer learning for 3d medical image analysis.
\newblock {\em arXiv preprint arXiv:1904.00625}, 2019.

\bibitem{Basch2012}
Ethan Basch, Thomas~K Oliver, Andrew Vickers, Ian Thompson, Philip Kantoff,
  Howard Parnes, D~Andrew Loblaw, Bruce Roth, James Williams, and Robert~K Nam.
\newblock {Screening for prostate cancer with prostate-specific antigen
  testing: American Society of Clinical Oncology Provisional Clinical Opinion.}
\newblock {\em Journal of clinical oncology : official journal of the American
  Society of Clinical Oncology}, 30(24):3020--5, aug 2012.

\bibitem{Wolf2010}
A.~M.~D. Wolf, R.~C. Wender, R.~B. Etzioni, I.~M. Thompson, A.~V. D'Amico,
  R.~J. Volk, D.~D. Brooks, C.~Dash, I.~Guessous, K.~Andrews, C.~DeSantis, and
  R.~A. Smith.
\newblock {American Cancer Society Guideline for the Early Detection of
  Prostate Cancer: Update 2010}.
\newblock {\em CA: A Cancer Journal for Clinicians}, 60(2):70--98, mar 2010.

\bibitem{Greene2013}
Kirsten~L. Greene, Peter~C. Albertsen, Richard~J. Babaian, H.~Ballentine
  Carter, Peter~H. Gann, Misop Han, Deborah~Ann Kuban, A.~Oliver Sartor,
  Janet~L. Stanford, Anthony Zietman, and Peter Carroll.
\newblock {Prostate Specific Antigen Best Practice Statement: 2009 Update}.
\newblock {\em The Journal of Urology}, 2013.

\bibitem{prostate_guidelines_webmed}
{Prostate Cancer Guidelines: Guidelines Summary, Genetic Testing,
  Multiparametric Magnetic Resonance Imaging}.

\bibitem{dre_secondary}
{Prostate Cancer: Early Detection Guideline - American Urological Association}.

\bibitem{Catalona1994}
William~J. Catalona, Jerome~P. Richie, Frederick~R. Ahmann, M'Liss~A. Hudson,
  Peter~T. Scardino, Robert~C. Flanigan, Jean~B. Dekernion, Timothy~L. Ratliff,
  Louis~R. Kavoussi, Bruce~L. Dalkin, W.~Bedford Waters, Michael~T. Macfarlane,
  and Paula~C. Southwick.
\newblock {Comparison of Digital Rectal Examination and Serum Prostate Specific
  Antigen in the Early Detection of Prostate Cancer: Results of a Multicenter
  Clinical Trial of 6,630 Men}.
\newblock {\em Journal of Urology}, 151(5):1283--1290, may 1994.

\bibitem{Velonas2013}
Vicki Velonas, Henry Woo, Cristobal Remedios, and Stephen Assinder.
\newblock {Current Status of Biomarkers for Prostate Cancer}.
\newblock {\em International Journal of Molecular Sciences},
  14(6):11034--11060, may 2013.

\bibitem{Cooperberg2011}
Matthew~R. Cooperberg, Peter~R. Carroll, and Laurence Klotz.
\newblock {Active surveillance for prostate cancer: Progress and promise}, sep
  2011.

\bibitem{Ghei2005}
M~Ghei, S~Pericleous, A~Kumar, R~Miller, S~Nathan, and B~H Maraj.
\newblock {Finger-guided transrectal biopsy of the prostate: a modified, safer
  technique.}
\newblock {\em Annals of the Royal College of Surgeons of England},
  87(5):386--7, sep 2005.

\bibitem{Stuckey}
Mike Stuckey.
\newblock {Tales from a prostate biopsy - Health - Men's health - Low Blow |
  NBC News}.

\bibitem{Hambrock2012}
Thomas Hambrock, Caroline Hoeks, Christina {Hulsbergen-Van De Kaa}, Tom
  Scheenen, Jurgen F{\"{u}}tterer, Stefan Bouwense, Inge {Van Oort}, Fritz
  Schr{\"{o}}der, Henkjan Huisman, and Jelle Barentsz.
\newblock {Prospective assessment of prostate cancer aggressiveness using 3-T
  diffusion-weighted magnetic resonance imaging-guided biopsies versus a
  systematic 10-core transrectal ultrasound prostate biopsy cohort}.
\newblock {\em European Urology}, 61(1):177--184, jan 2012.

\bibitem{Bell2014}
Neil Bell, Sarah {Connor Gorber}, Amanda Shane, Michel Joffres, Harminder
  Singh, James Dickinson, Elizabeth Shaw, Lesley Dunfield, and Marcello
  Tonelli.
\newblock {Recommendations on screening for prostate cancer with the
  prostate-specific antigen test}.
\newblock {\em CMAJ}, 186(16):1225--1234, nov 2014.

\bibitem{Roberts2017}
Matthew~J. Roberts, Harrison~Y. Bennett, Patrick~N. Harris, Michael Holmes,
  Jeremy Grummet, Kurt Naber, and Florian~M.E. Wagenlehner.
\newblock {Prostate Biopsy-related Infection: A Systematic Review of Risk
  Factors, Prevention Strategies, and Management Approaches}, jun 2017.

\bibitem{Turkbey2018}
Baris Turkbey and Peter~L. Choyke.
\newblock {Future Perspectives and Challenges of Prostate MR Imaging}, mar
  2018.

\bibitem{Sartor2018}
Oliver Sartor and Johann~S. {De Bono}.
\newblock {Metastatic prostate cancer}, feb 2018.

\bibitem{Dhondt2019}
Bert Dhondt, Elise {De Bleser}, Tom Claeys, Sarah Buelens, Nicolaas Lumen,
  Jo~Vandesompele, Anneleen Beckers, Valerie Fonteyne, Kim {Van der Eecken},
  Aur{\'{e}}lie {De Bruycker}, J{\'{e}}r{\^{o}}me Paul, Pierre Gramme, and Piet
  Ost.
\newblock {Discovery and validation of a serum microRNA signature to
  characterize oligo- and polymetastatic prostate cancer: not ready for prime
  time}.
\newblock {\em World Journal of Urology}, 37(12):2557--2564, dec 2019.

\bibitem{Wallis2016}
Christopher~J.D. Wallis, Alyson~L. Mahar, Richard Choo, Sender Herschorn,
  Ronald~T. Kodama, Prakesh~S. Shah, Cyril Danjoux, Steven~A. Narod, and
  Robert~K. Nam.
\newblock {Second malignancies after radiotherapy for prostate cancer:
  Systematic review and meta-analysis}, mar 2016.

\bibitem{Mouraviev2006}
V.~Mouraviev, B.~Evans, and T.~J. Polascik.
\newblock {Salvage prostate cryoablation after primary interstitial
  brachytherapy failure: A feasible approach}.
\newblock {\em Prostate Cancer and Prostatic Diseases}, 9(1):99--101, mar 2006.

\bibitem{Wallis2018}
Christopher~J.D. Wallis, Adam Glaser, Jim~C. Hu, Hartwig Huland, Nathan
  Lawrentschuk, Daniel Moon, Declan~G. Murphy, Paul~L. Nguyen, Matthew~J.
  Resnick, and Robert~K. Nam.
\newblock {Survival and Complications Following Surgery and Radiation for
  Localized Prostate Cancer: An International Collaborative Review}, jan 2018.

\bibitem{surgery}
{Surgery for Prostate Cancer}.

\bibitem{Ilic2017}
Dragan Ilic, Sue~M. Evans, Christie~Ann Allan, Jae~Hung Jung, Declan Murphy,
  and Mark Frydenberg.
\newblock {Laparoscopic and robotic-assisted versus open radical prostatectomy
  for the treatment of localised prostate cancer}, sep 2017.

\bibitem{hopekinsdsd}
{Erectile Dysfunction After Prostate Cancer | Johns Hopkins Medicine}.

\bibitem{Hinton}
Geoffrey Hinton, Li~Deng, Dong Yu, George Dahl, Abdel-Rahman Mohamed, Navdeep
  Jaitly, Andrew Senior, Vincent Vanhoucke, Patrick Nguyen, Tara Sainath, and
  Brian Kingsbury.
\newblock {Deep Neural Networks for Acoustic Modeling in Speech Recognition}.
\newblock Technical report.

\bibitem{Krizhevsky}
Alex Krizhevsky, Ilya Sutskever, and Geoffrey~E Hinton.
\newblock {ImageNet Classification with Deep Convolutional Neural Networks}.
\newblock Technical report.

\bibitem{He}
Kaiming He, Xiangyu Zhang, Shaoqing Ren, and Jian Sun.
\newblock {Delving Deep into Rectifiers: Surpassing Human-Level Performance on
  ImageNet Classification}.
\newblock Technical report.

\bibitem{Ren2017}
Shaoqing Ren, Kaiming He, Ross Girshick, and Jian Sun.
\newblock {Faster R-CNN: Towards Real-Time Object Detection with Region
  Proposal Networks}.
\newblock {\em IEEE Transactions on Pattern Analysis and Machine Intelligence},
  39(6):1137--1149, jun 2017.

\bibitem{Redmon2018}
Joseph Redmon and Ali Farhadi.
\newblock {YOLOv3: An Incremental Improvement}.
\newblock apr 2018.

\bibitem{Ronneberger2015}
Olaf Ronneberger, Philipp Fischer, and Thomas Brox.
\newblock {U-net: Convolutional networks for biomedical image segmentation}.
\newblock In {\em Lecture Notes in Computer Science (including subseries
  Lecture Notes in Artificial Intelligence and Lecture Notes in
  Bioinformatics)}, volume 9351, pages 234--241. Springer Verlag, may 2015.

\bibitem{Li2019}
Peiliang Li, Xiaozhi Chen, and Shaojie Shen.
\newblock {Stereo R-CNN based 3D Object Detection for Autonomous Driving}.
\newblock {\em Proceedings of the IEEE Computer Society Conference on Computer
  Vision and Pattern Recognition}, 2019-June:7636--7644, feb 2019.

\bibitem{Aafaq2019}
Nayyer Aafaq, Ajmal Mian, Wei Liu, Syed~Zulqarnain Gilani, and Mubarak Shah.
\newblock {Video description: A survey of methods, datasets, and evaluation
  metrics}, oct 2019.

\bibitem{Yousefikamal2019}
Parvin Yousefikamal.
\newblock {Breast Tumor Classification and Segmentation using Convolutional
  Neural Networks}.
\newblock may 2019.

\bibitem{10.5555/3104322.3104425}
Vinod Nair and Geoffrey~E Hinton.
\newblock {Rectified Linear Units Improve Restricted Boltzmann Machines}.
\newblock In {\em Proceedings of the 27th International Conference on
  International Conference on Machine Learning}, ICML'10, pages 807--814,
  Madison, WI, USA, 2010. Omnipress.

\bibitem{misra2019mish}
Diganta Misra.
\newblock Mish: A self regularized non-monotonic neural activation function.
\newblock {\em arXiv preprint arXiv:1908.08681}, 2019.

\bibitem{Maas2013}
Andrew~L Maas, Awni~Y Hannun, and Andrew~Y Ng.
\newblock {Rectifier Nonlinearities Improve Neural Network Acoustic Models}.
\newblock Technical report, 2013.

\bibitem{Lecun2015}
Yann Lecun, Yoshua Bengio, and Geoffrey Hinton.
\newblock {Deep learning}, may 2015.

\bibitem{Jain2017}
Prateek Jain and Purushottam Kar.
\newblock {Non-convex Optimization for Machine Learning}.
\newblock {\em Foundations and Trends in Machine Learning}, 10(3-4):142--336,
  dec 2017.

\bibitem{Goodfellow-et-al-2016}
Ian Goodfellow, Yoshua Bengio, and Aaron Courville.
\newblock {\em Deep Learning}.
\newblock MIT Press, 2016.
\newblock \url{http://www.deeplearningbook.org}.

\bibitem{Lin2017}
Tsung-Yi Lin, Priya Goyal, Ross Girshick, Kaiming He, and Piotr Doll{\'{a}}r.
\newblock {Focal Loss for Dense Object Detection}.
\newblock {\em IEEE Transactions on Pattern Analysis and Machine Intelligence},
  42(2):318--327, aug 2017.

\bibitem{Mohri}
Mehryar Mohri, Afshin Rostamizadeh, and Ameet Talwalkar.
\newblock {\em {Foundations of machine learning}}.

\bibitem{Russel2012}
Stuart Russel and Peter Norvig.
\newblock {\em {Artificial intelligence—a modern approach 3rd Edition}}.
\newblock 2012.

\bibitem{Hinton1999}
Geoffrey~E. Hinton and Terrence J. (Terrence~Joseph) Sejnowski.
\newblock {\em {Unsupervised learning : foundations of neural computation}}.
\newblock MIT Press, 1999.

\bibitem{Taddy}
Matt Taddy.
\newblock {\em {Business data science : combining machine learning and
  economics to optimize, automate, and accelerate business decisions}}.

\bibitem{PaulH.Sra2011}
{Paul H.Sra}.
\newblock {\em {Optimization for Machine Learning (Neural Information
  Processing Series)}}.
\newblock 2011.

\bibitem{Hawkins2004}
Douglas~M. Hawkins.
\newblock {The Problem of Overfitting}, jan 2004.

\bibitem{Hubel1968}
D.~H. Hubel and T.~N. Wiesel.
\newblock {Receptive fields and functional architecture of monkey striate
  cortex}.
\newblock {\em The Journal of Physiology}, 195(1):215--243, mar 1968.

\bibitem{Fukushima1988}
Kunihiko Fukushima.
\newblock {Neocognitron: A hierarchical neural network capable of visual
  pattern recognition}.
\newblock {\em Neural Networks}, 1(2):119--130, jan 1988.

\bibitem{Marr1980}
D.~Marr and E.~Hildreth.
\newblock {Theory of edge detection}.
\newblock {\em Proceedings of the Royal Society of London - Biological
  Sciences}, 207(1167):187--217, 1980.

\bibitem{CS231n}
{CS231n Convolutional Neural Networks for Visual Recognition}.

\bibitem{Yu2015}
Fisher Yu and Vladlen Koltun.
\newblock {Multi-Scale Context Aggregation by Dilated Convolutions}.
\newblock {\em 4th International Conference on Learning Representations, ICLR
  2016 - Conference Track Proceedings}, nov 2015.

\bibitem{conv_size}
{torch.nn — PyTorch master documentation}.

\bibitem{HabibiAghdam2017}
Hamed {Habibi Aghdam} and Elnaz {Jahani Heravi}.
\newblock {\em {Guide to Convolutional Neural Networks}}.
\newblock Springer International Publishing, 2017.

\bibitem{LeCun1989}
Y.~LeCun, B.~Boser, J.~S. Denker, D.~Henderson, R.~E. Howard, W.~Hubbard, and
  L.~D. Jackel.
\newblock {Backpropagation Applied to Handwritten Zip Code Recognition}.
\newblock {\em Neural Computation}, 1(4):541--551, dec 1989.

\bibitem{Springenberg2015}
Jost~Tobias Springenberg, Alexey Dosovitskiy, Thomas Brox, and Martin
  Riedmiller.
\newblock {Striving for simplicity: The all convolutional net}.
\newblock In {\em 3rd International Conference on Learning Representations,
  ICLR 2015 - Workshop Track Proceedings}. International Conference on Learning
  Representations, ICLR, dec 2015.

\bibitem{Graham2014}
Benjamin Graham.
\newblock {Fractional Max-Pooling}.
\newblock dec 2014.

\bibitem{Shapiro2001}
Linda~G. Shapiro and George~C. Stockman.
\newblock {\em {Computer vision}}.
\newblock Prentice Hall, 2001.

\bibitem{Lee2003}
Lawrence~W Lee and San Francisco.
\newblock {United States US 2004OO59754A1 (12) Patent Application Publication}.
\newblock Technical Report~10, jul 2003.

\bibitem{Pham2000}
Dzung~L. Pham, Chenyang Xu, and Jerry~L. Prince.
\newblock {Current Methods in Medical Image Segmentation}.
\newblock {\em Annual Review of Biomedical Engineering}, 2(1):315--337, aug
  2000.

\bibitem{Forouzanfar2010}
Mohamad Forouzanfar, Nosratallah Forghani, and Mohammad Teshnehlab.
\newblock {Parameter optimization of improved fuzzy c-means clustering
  algorithm for brain MR image segmentation}.
\newblock {\em Engineering Applications of Artificial Intelligence},
  23(2):160--168, mar 2010.

\bibitem{BenGeorge}
E~{Ben George}.
\newblock {MR Brain Image Segmentation using Bacteria Foraging Optimization
  Algorithm}.
\newblock Technical report.

\bibitem{Kamalakannan2010}
Sridharan Kamalakannan, Arunkumar Gururajan, Hamed Sari-Sarraf, Rodney Long,
  and Sameer Antani.
\newblock {Double-edge detection of radiographic lumbar vertebrae images using
  pressurized open DGVF snakes}.
\newblock {\em IEEE Transactions on Biomedical Engineering}, 57(6):1325--1334,
  jun 2010.

\bibitem{Tan2012}
Nelly Tan, Daniel~J.A. Margolis, Timothy~D. McClure, Albert Thomas, David~S.
  Finley, Robert~E. Reiter, Jiaoti Huang, and Steven~S. Raman.
\newblock {Radical prostatectomy: Value of prostate MRI in surgical planning},
  aug 2012.

\bibitem{Tahmassebi2018}
Amirhessam Tahmassebi, Georg~J Wengert, Thomas~H Helbich, Zsuzsanna
  Bago-Horvath, Sousan Alaei, Rupert Bartsch, Peter Dubsky, Pascal Baltzer,
  Paola Clauser, Panagiotis Kapetas, Elizabeth~A Morris, Anke Meyer-Baese, and
  Katja Pinker.
\newblock {Impact of Machine Learning With Multiparametric Magnetic Resonance
  Imaging of the Breast for Early Prediction of Response to Neoadjuvant
  Chemotherapy and Survival Outcomes in Breast Cancer Patients}.
\newblock 2018.

\bibitem{Marino2018}
Maria~Adele Marino, Thomas Helbich, Pascal Baltzer, and Katja Pinker-Domenig.
\newblock {Multiparametric MRI of the breast: A review}.
\newblock {\em Journal of Magnetic Resonance Imaging}, 47(2):301--315, feb
  2018.

\bibitem{Barentsz2012}
Jelle~O. Barentsz, Jonathan Richenberg, Richard Clements, Peter Choyke, Sadhna
  Verma, Geert Villeirs, Olivier Rouviere, Vibeke Logager, and Jurgen~J.
  F{\"{u}}tterer.
\newblock {ESUR prostate MR guidelines 2012}.
\newblock {\em European Radiology}, 22(4):746--757, 2012.

\bibitem{Weinreb2016}
Jeffrey~C. Weinreb, Jelle~O. Barentsz, Peter~L. Choyke, Francois Cornud,
  Masoom~A. Haider, Katarzyna~J. Macura, Daniel Margolis, Mitchell~D. Schnall,
  Faina Shtern, Clare~M. Tempany, Harriet~C. Thoeny, and Sadna Verma.
\newblock {PI-RADS Prostate Imaging - Reporting and Data System: 2015, Version
  2}.
\newblock {\em European Urology}, 69(1):16--40, jan 2016.

\bibitem{Thompson2016}
J.~E. Thompson, P.~J. {Van Leeuwen}, D.~Moses, R.~Shnier, P.~Brenner,
  W.~Delprado, M.~Pulbrook, M.~B{\"{o}}hm, A.~M. Haynes, A.~Hayen, and P.~D.
  Stricker.
\newblock {The diagnostic performance of multiparametric magnetic resonance
  imaging to detect significant prostate cancer}.
\newblock {\em Journal of Urology}, 195(5):1428--1435, may 2016.

\bibitem{Vargas2016}
H.~A. Vargas, A.~M. H{\"{o}}tker, D.~A. Goldman, C.~S. Moskowitz, T.~Gondo,
  K.~Matsumoto, B.~Ehdaie, S.~Woo, S.~W. Fine, V.~E. Reuter, E.~Sala, and
  H.~Hricak.
\newblock {Updated prostate imaging reporting and data system (PIRADS v2)
  recommendations for the detection of clinically significant prostate cancer
  using multiparametric MRI: critical evaluation using whole-mount pathology as
  standard of reference}.
\newblock {\em European Radiology}, 26(6):1606--1612, jun 2016.

\bibitem{Schieda2015}
Nicola Schieda, Jeffrey~S. Quon, Christopher Lim, Mohammed El-Khodary, Wael
  Shabana, Vivek Singh, Christopher Morash, Rodney~H. Breau, Matthew~D.F.
  McInnes, and Trevor~A. Flood.
\newblock {Evaluation of the European Society of Urogenital Radiology (ESUR)
  PI-RADS scoring system for assessment of extra-prostatic extension in
  prostatic carcinoma}.
\newblock {\em European Journal of Radiology}, 84(10):1843--1848, oct 2015.

\bibitem{Abdi2015}
Hamidreza Abdi, Farshad Pourmalek, Homayoun Zargar, Triona Walshe, Alison~C.
  Harris, Silvia~D. Chang, Christopher Eddy, Alan~I. So, Martin~E. Gleave,
  Lindsay Machan, S.~Larry Goldenberg, and Peter~C. Black.
\newblock {Multiparametric magnetic resonance imaging enhances detection of
  significant tumor in patients on active surveillance for prostate cancer}.
\newblock {\em Urology}, 85(2):423--429, feb 2015.

\bibitem{Read_prostate}
{How to Read Your Prostate MRI Report}.

\bibitem{Abragam1983}
A.~Abragam.
\newblock {\em {Principles of Nuclear Magnetism (The International Series of
  Monographs on Physics)}}.
\newblock Clarendon Press, 1983.

\bibitem{Claridge2016}
Timothy~D.W. Claridge.
\newblock {\em {High-Resolution NMR Techniques in Organic Chemistry: Third
  Edition}}.
\newblock Elsevier Inc., may 2016.

\bibitem{UniversityofWisconsin}
{University of Wisconsin}.
\newblock {Magnetic Resonance Imaging}.

\bibitem{Johnson}
Keith~A. Johnson.
\newblock {basic MR imaging}.

\bibitem{Dunn2015}
Joel Dunn and Paul~Kenneth Marsden.
\newblock {Hyperpolarised 13C NMR Spectroscopy for studying cardiac metabolism
  View project PET/MRI Reconstruction View project}.
\newblock {\em Physics in Medicine {\&} Biology}, 56(13):6441, 2015.

\bibitem{Merboldt1985}
Klaus~Dietmar Merboldt, Wolfgang Hanicke, and Jens Frahm.
\newblock {Self-diffusion NMR imaging using stimulated echoes}.
\newblock {\em Journal of Magnetic Resonance (1969)}, 64(3):479--486, oct 1985.

\bibitem{Koh2007}
Dow-Mu Koh and David~J. Collins.
\newblock {Diffusion-Weighted MRI in the Body: Applications and Challenges in
  Oncology}.
\newblock {\em American Journal of Roentgenology}, 188(6):1622--1635, jun 2007.

\bibitem{Hammer}
Mark Hammer.
\newblock {MRI Physics: Diffusion-Weighted Imaging - XRayPhysics}.

\bibitem{Elster}
Allen~D. Elster.
\newblock {Causes of restricted diffusion - Questions and Answers in MRI}.

\bibitem{Yoo2019}
Sunghwan Yoo, Isha Gujrathi, Masoom~A. Haider, and Farzad Khalvati.
\newblock {Prostate Cancer Detection using Deep Convolutional Neural Networks}.
\newblock {\em Scientific Reports}, 9(1):1--10, dec 2019.

\bibitem{Karimi2019}
Davood Karimi, Golnoosh Samei, Yanan Shao, and Septimiu Salcudean.
\newblock {A deep learning-based method for prostate segmentation in
  T2-weighted magnetic resonance imaging}.
\newblock jan 2019.

\bibitem{To2018}
Minh Nguyen~Nhat To, Dang~Quoc Vu, Baris Turkbey, Peter~L. Choyke, and Jin~Tae
  Kwak.
\newblock {Deep dense multi-path neural network for prostate segmentation in
  magnetic resonance imaging}.
\newblock {\em International Journal of Computer Assisted Radiology and
  Surgery}, 13(11):1687--1696, nov 2018.

\bibitem{Feldman2019}
Aharon Feldman, Zhenzhen Dai, Eric Carver, Chang Liu, Joon Lee, Milan Pantelic,
  Mohamed Elshaikh, and Ning Wen.
\newblock {Prostate and Prostate Cancer Segmentation Using a Deep
  Learning-Based Object Detection Algorithm}.
\newblock {\em Clinical Research}, may 2019.

\bibitem{Schelb2019}
Patrick Schelb, Simon Kohl, Jan~Philipp Radtke, Manuel Wiesenfarth, Philipp
  Kickingereder, Sebastian Bickelhaupt, Tristan~Anselm Kuder, Albrecht
  Stenzinger, Markus Hohenfellner, Heinz-Peter Schlemmer, Klaus~H. Maier-Hein,
  and David Bonekamp.
\newblock {Classification of Cancer at Prostate MRI: Deep Learning versus
  Clinical PI-RADS Assessment}.
\newblock {\em Radiology}, 293(3):607--617, dec 2019.

\bibitem{Hara}
Kensho Hara, Hirokatsu Kataoka, and Yutaka Satoh.
\newblock {Can Spatiotemporal 3D CNNs Retrace the History of 2D CNNs and
  ImageNet?}
\newblock Technical report.

\bibitem{Hara2}
Kensho Hara, Hirokatsu Kataoka, and Yutaka Satoh.
\newblock {Learning Spatio-Temporal Features with 3D Residual Networks for
  Action Recognition}.
\newblock Technical report.

\end{thebibliography}

\appendix

\chapter{Prostate Segmentation}

\begin{figure}[ht]
  \centering
  \begin{minipage}[b]{0.325\textwidth}
    \includegraphics[width=\textwidth]{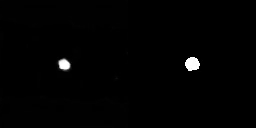}
  \end{minipage}
  \hfill
  \begin{minipage}[b]{0.325\textwidth}
    \includegraphics[width=\textwidth]{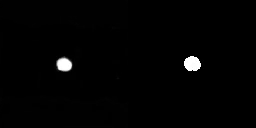}
  \end{minipage}
  \hfill
  \begin{minipage}[b]{0.325\textwidth}
    \includegraphics[width=\textwidth]{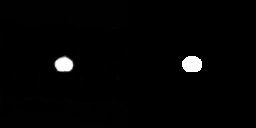}
  \end{minipage}
  \hfill
  \begin{minipage}[b]{0.325\textwidth}
    \includegraphics[width=\textwidth]{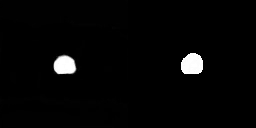}
  \end{minipage}
  \hfill
  \begin{minipage}[b]{0.325\textwidth}
    \includegraphics[width=\textwidth]{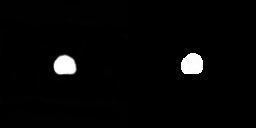}
  \end{minipage}
  \hfill
  \begin{minipage}[b]{0.325\textwidth}
    \includegraphics[width=\textwidth]{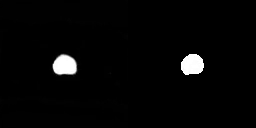}
  \end{minipage}
  \hfill
  \begin{minipage}[b]{0.325\textwidth}
    \includegraphics[width=\textwidth]{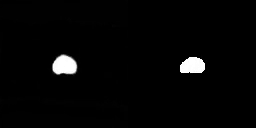}
  \end{minipage}
  \hfill
  \begin{minipage}[b]{0.325\textwidth}
    \includegraphics[width=\textwidth]{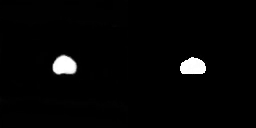}
  \end{minipage}
  \hfill
  \begin{minipage}[b]{0.325\textwidth}
    \includegraphics[width=\textwidth]{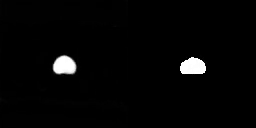}
  \end{minipage}
  \hfill
  \begin{minipage}[b]{0.325\textwidth}
    \includegraphics[width=\textwidth]{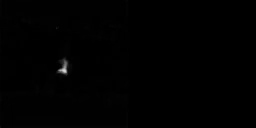}
  \end{minipage}
  \hfill
  \begin{minipage}[b]{0.325\textwidth}
    \includegraphics[width=\textwidth]{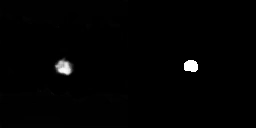}
  \end{minipage}
  \hfill
  \begin{minipage}[b]{0.325\textwidth}
    \includegraphics[width=\textwidth]{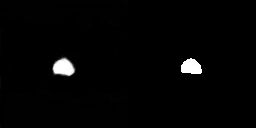}
  \end{minipage}
  \hfill
  \begin{minipage}[b]{0.325\textwidth}
    \includegraphics[width=\textwidth]{images/results/prostate/prostate_seg00057.png}
  \end{minipage}
  \hfill
  \begin{minipage}[b]{0.325\textwidth}
    \includegraphics[width=\textwidth]{images/results/prostate/prostate_seg00061.png}
  \end{minipage}
  \hfill
  \begin{minipage}[b]{0.325\textwidth}
    \includegraphics[width=\textwidth]{images/results/prostate/prostate_seg00065.png}
  \end{minipage}
  \hfill
  \begin{minipage}[b]{0.325\textwidth}
    \includegraphics[width=\textwidth]{images/results/prostate/prostate_seg00069.png}
  \end{minipage}
  \hfill
  \begin{minipage}[b]{0.325\textwidth}
    \includegraphics[width=\textwidth]{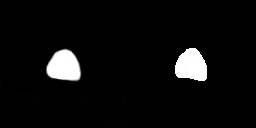}
  \end{minipage}
  \hfill
  \begin{minipage}[b]{0.325\textwidth}
    \includegraphics[width=\textwidth]{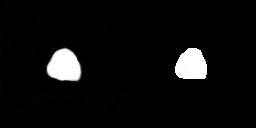}
  \end{minipage}
  
  \caption{Sample predictions made by a 3D U-Net model from the fold 5 of experiment configuration B with data augmentations using ADC as input. Images represent slices with mask predictions on the same patient compared to the ground truth. Left side of each image contains the prediction whereas the right size contains the label.}
\end{figure}

\begin{figure}[ht]
  \centering
  
  \begin{minipage}[b]{0.325\textwidth}
    \includegraphics[width=\textwidth]{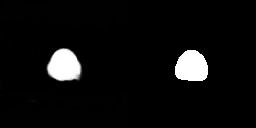}
  \end{minipage}
  \hfill
  \begin{minipage}[b]{0.325\textwidth}
    \includegraphics[width=\textwidth]{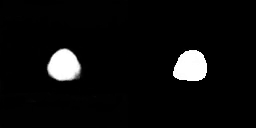}
  \end{minipage}
  \hfill
  \begin{minipage}[b]{0.325\textwidth}
    \includegraphics[width=\textwidth]{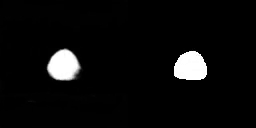}
  \end{minipage}
  \hfill
  \begin{minipage}[b]{0.325\textwidth}
    \includegraphics[width=\textwidth]{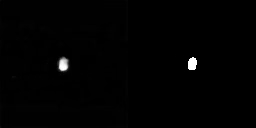}
  \end{minipage}
  \hfill
  \begin{minipage}[b]{0.325\textwidth}
    \includegraphics[width=\textwidth]{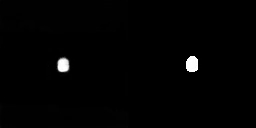}
  \end{minipage}
  \hfill
  \begin{minipage}[b]{0.325\textwidth}
    \includegraphics[width=\textwidth]{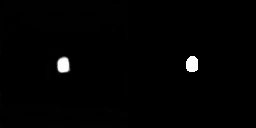}
  \end{minipage}
  \hfill
  \begin{minipage}[b]{0.325\textwidth}
    \includegraphics[width=\textwidth]{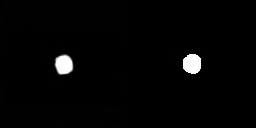}
  \end{minipage}
  \hfill
  \begin{minipage}[b]{0.325\textwidth}
    \includegraphics[width=\textwidth]{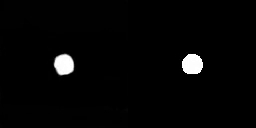}
  \end{minipage}
  \hfill
  \begin{minipage}[b]{0.325\textwidth}
    \includegraphics[width=\textwidth]{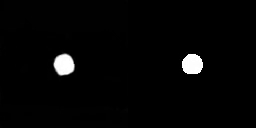}
  \end{minipage}
  \hfill
  \begin{minipage}[b]{0.325\textwidth}
    \includegraphics[width=\textwidth]{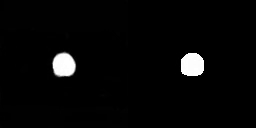}
  \end{minipage}
  \hfill
  \begin{minipage}[b]{0.325\textwidth}
    \includegraphics[width=\textwidth]{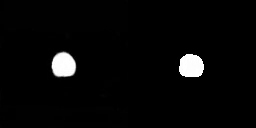}
  \end{minipage}
  \hfill
  \begin{minipage}[b]{0.325\textwidth}
    \includegraphics[width=\textwidth]{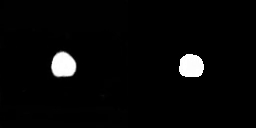}
  \end{minipage}
  \hfill
  \begin{minipage}[b]{0.325\textwidth}
    \includegraphics[width=\textwidth]{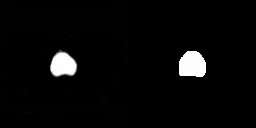}
  \end{minipage}
  \hfill
  \begin{minipage}[b]{0.325\textwidth}
    \includegraphics[width=\textwidth]{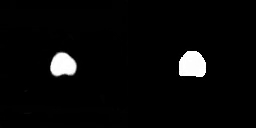}
  \end{minipage}
  \hfill
  \begin{minipage}[b]{0.325\textwidth}
    \includegraphics[width=\textwidth]{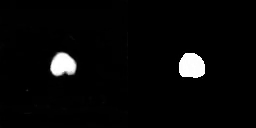}
  \end{minipage}
  \hfill
  \begin{minipage}[b]{0.325\textwidth}
    \includegraphics[width=\textwidth]{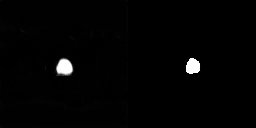}
  \end{minipage}
  \hfill
  \begin{minipage}[b]{0.325\textwidth}
    \includegraphics[width=\textwidth]{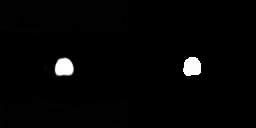}
  \end{minipage}
  \hfill
  \begin{minipage}[b]{0.325\textwidth}
    \includegraphics[width=\textwidth]{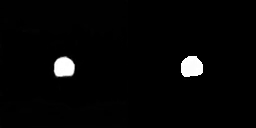}
  \end{minipage}
  \hfill
  \begin{minipage}[b]{0.325\textwidth}
    \includegraphics[width=\textwidth]{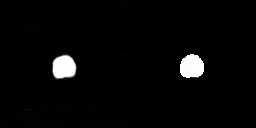}
  \end{minipage}
  \hfill
  \begin{minipage}[b]{0.325\textwidth}
    \includegraphics[width=\textwidth]{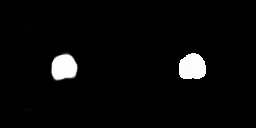}
  \end{minipage}
  \hfill
  \begin{minipage}[b]{0.325\textwidth}
    \includegraphics[width=\textwidth]{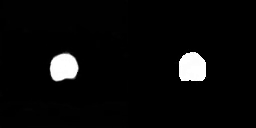}
  \end{minipage}
  \hfill
  \begin{minipage}[b]{0.325\textwidth}
    \includegraphics[width=\textwidth]{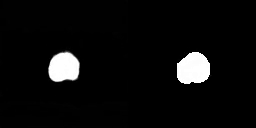}
  \end{minipage}
  \caption{Sample predictions made by a 3D U-Net model from the fold 5 of experiment configuration B with data augmentations using ADC as input. Images represent slices with mask predictions on the same patient compared to the ground truth. Left side of each image contains the prediction whereas the right size contains the label.}
\end{figure}

\clearpage

\chapter{Lesions Segmentation }

\begin{figure}[ht]
  \centering
  \begin{minipage}[b]{0.325\textwidth}
    \includegraphics[width=\textwidth]{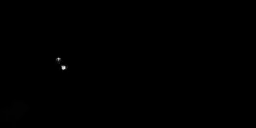}
  \end{minipage}
  \hfill
  \begin{minipage}[b]{0.325\textwidth}
    \includegraphics[width=\textwidth]{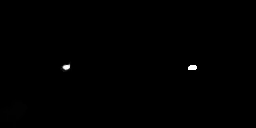}
  \end{minipage}
  \hfill
  \begin{minipage}[b]{0.325\textwidth}
    \includegraphics[width=\textwidth]{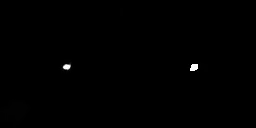}
  \end{minipage}
  \hfill
  \begin{minipage}[b]{0.325\textwidth}
    \includegraphics[width=\textwidth]{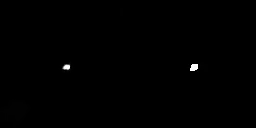}
  \end{minipage}
  \hfill
  \begin{minipage}[b]{0.325\textwidth}
    \includegraphics[width=\textwidth]{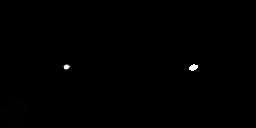}
  \end{minipage}
  \hfill
  \begin{minipage}[b]{0.325\textwidth}
    \includegraphics[width=\textwidth]{images/results/lesions/lesion_seg00037.png}
  \end{minipage}
  \hfill
  \begin{minipage}[b]{0.325\textwidth}
    \includegraphics[width=\textwidth]{images/results/lesions/lesion_seg00113.png}
  \end{minipage}
  \hfill
  \begin{minipage}[b]{0.325\textwidth}
    \includegraphics[width=\textwidth]{images/results/lesions/lesion_seg00117.png}
  \end{minipage}
  \hfill
  \begin{minipage}[b]{0.325\textwidth}
    \includegraphics[width=\textwidth]{images/results/lesions/lesion_seg00121.png}
  \end{minipage}
  \hfill
  \begin{minipage}[b]{0.325\textwidth}
    \includegraphics[width=\textwidth]{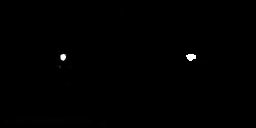}
  \end{minipage}
  \hfill
  \begin{minipage}[b]{0.325\textwidth}
    \includegraphics[width=\textwidth]{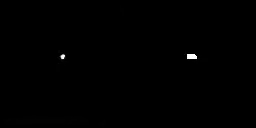}
  \end{minipage}
  \hfill
  \begin{minipage}[b]{0.325\textwidth}
    \includegraphics[width=\textwidth]{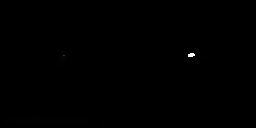}
  \end{minipage}
  \hfill
  \begin{minipage}[b]{0.325\textwidth}
    \includegraphics[width=\textwidth]{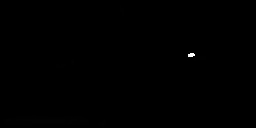}
  \end{minipage}
  \hfill
  \begin{minipage}[b]{0.325\textwidth}
    \includegraphics[width=\textwidth]{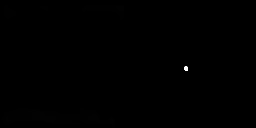}
  \end{minipage}
  \hfill
  \begin{minipage}[b]{0.325\textwidth}
    \includegraphics[width=\textwidth]{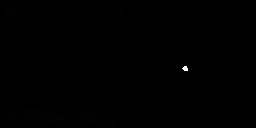}
  \end{minipage}
  \hfill
  \begin{minipage}[b]{0.325\textwidth}
    \includegraphics[width=\textwidth]{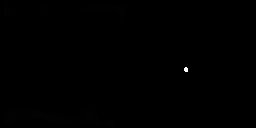}
  \end{minipage}
  \hfill
  \begin{minipage}[b]{0.325\textwidth}
    \includegraphics[width=\textwidth]{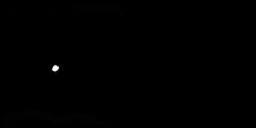}
  \end{minipage}
  \hfill
  \begin{minipage}[b]{0.325\textwidth}
    \includegraphics[width=\textwidth]{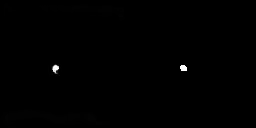}
  \end{minipage}
  
  \caption{Sample predictions made by a 3D U-Net model from the fold 4 of experiment configuration A with data augmentations using ADC as input. Images represent slices with mask predictions on the same patient compared to the ground truth. Left side of each image contains the prediction whereas the right size contains the label.}
\end{figure}

\begin{figure}[ht]
  \centering
  
  \begin{minipage}[b]{0.325\textwidth}
    \includegraphics[width=\textwidth]{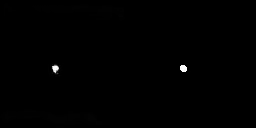}
  \end{minipage}
  \hfill
  \begin{minipage}[b]{0.325\textwidth}
    \includegraphics[width=\textwidth]{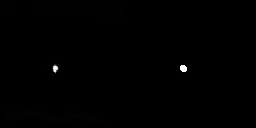}
  \end{minipage}
  \hfill
  \begin{minipage}[b]{0.325\textwidth}
    \includegraphics[width=\textwidth]{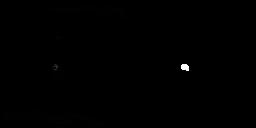}
  \end{minipage}
  \hfill
  \begin{minipage}[b]{0.325\textwidth}
    \includegraphics[width=\textwidth]{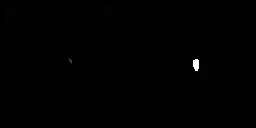}
  \end{minipage}
  \hfill
  \begin{minipage}[b]{0.325\textwidth}
    \includegraphics[width=\textwidth]{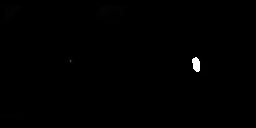}
  \end{minipage}
  \hfill
  \begin{minipage}[b]{0.325\textwidth}
    \includegraphics[width=\textwidth]{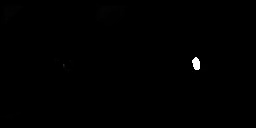}
  \end{minipage}
  \hfill
  \begin{minipage}[b]{0.325\textwidth}
    \includegraphics[width=\textwidth]{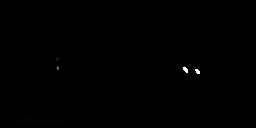}
  \end{minipage}
  \hfill
  \begin{minipage}[b]{0.325\textwidth}
    \includegraphics[width=\textwidth]{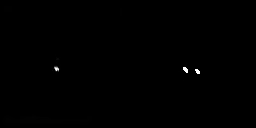}
  \end{minipage}
  \hfill
  \begin{minipage}[b]{0.325\textwidth}
    \includegraphics[width=\textwidth]{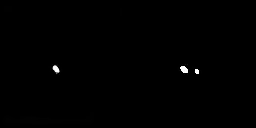}
  \end{minipage}
  \hfill
  \begin{minipage}[b]{0.325\textwidth}
    \includegraphics[width=\textwidth]{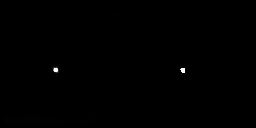}
  \end{minipage}
  \hfill
  \begin{minipage}[b]{0.325\textwidth}
    \includegraphics[width=\textwidth]{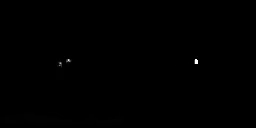}
  \end{minipage}
  \hfill
  \begin{minipage}[b]{0.325\textwidth}
    \includegraphics[width=\textwidth]{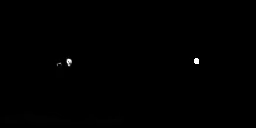}
  \end{minipage}
  \hfill
  \begin{minipage}[b]{0.325\textwidth}
    \includegraphics[width=\textwidth]{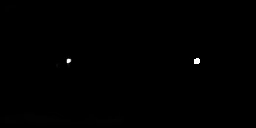}
  \end{minipage}
  \hfill
  \begin{minipage}[b]{0.325\textwidth}
    \includegraphics[width=\textwidth]{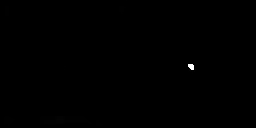}
  \end{minipage}
  \hfill
  \begin{minipage}[b]{0.325\textwidth}
    \includegraphics[width=\textwidth]{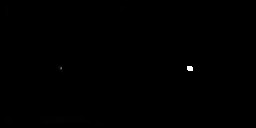}
  \end{minipage}
  \hfill
  \begin{minipage}[b]{0.325\textwidth}
    \includegraphics[width=\textwidth]{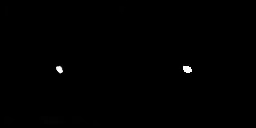}
  \end{minipage}
  \hfill
  \begin{minipage}[b]{0.325\textwidth}
    \includegraphics[width=\textwidth]{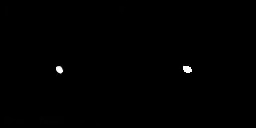}
  \end{minipage}
  \hfill
  \begin{minipage}[b]{0.325\textwidth}
    \includegraphics[width=\textwidth]{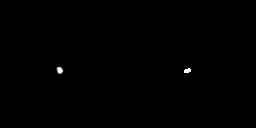}
  \end{minipage}
  \hfill
  \begin{minipage}[b]{0.325\textwidth}
    \includegraphics[width=\textwidth]{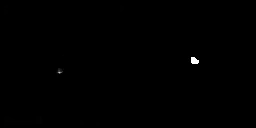}
  \end{minipage}
  \hfill
  \begin{minipage}[b]{0.325\textwidth}
    \includegraphics[width=\textwidth]{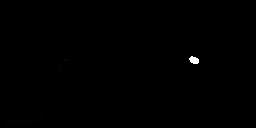}
  \end{minipage}
  \hfill
  \begin{minipage}[b]{0.325\textwidth}
    \includegraphics[width=\textwidth]{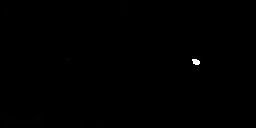}
  \end{minipage}
  \hfill
  \begin{minipage}[b]{0.325\textwidth}
    \includegraphics[width=\textwidth]{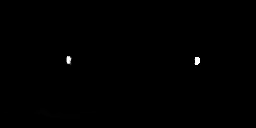}
  \end{minipage}
  \caption{Sample predictions made by a 3D U-Net model from the fold 4 of experiment configuration A with data augmentations using ADC as input. Images represent slices with mask predictions on the same patient compared to the ground truth. Left side of each image contains the prediction whereas the right size contains the label.}
\end{figure}

\end{document}